\documentclass[reqno]{article}

\usepackage{amsmath}
\usepackage{amsthm}
\usepackage{amsfonts}
\usepackage{amssymb}
\usepackage{times,macros}

\newcommand{\al}{\alpha}

\newcommand{\de}{\delta}
\newcommand{\ve}{\varepsilon}
\newcommand{\fsl}{{\mathfrak s}{\mathfrak l}}
\newcommand{\Uq}{{\cal U}_q(\fsl(2,\BR))}
\newcommand{\nn}{\nonumber}
\DeclareMathOperator*{\Res}{Res}
\newcommand{\0}{{\mathfrak 0}}
\newcommand{\1}{{\mathfrak 1}}
\newcommand{\2}{{\mathfrak 2}}
\newcommand{\3}{{\mathfrak 3}}
\newcommand{\FU}{{\mathfrak U}}
\newcommand{\rf}[1]{(\ref{#1})}
\newcommand{\ti}{\times}
\newcommand{\ra}{\to}
\newcommand{\BS}{{\mathbb S}}
\newcommand{\CO}{{\mathcal O}}
\newcommand{\CC}{{\mathcal C}}
\newcommand{\CQ}{{\mathcal Q}}
\newcommand{\CA}{{\mathcal A}}
\newcommand{\CB}{{\mathcal B}}

\newcommand{\CT}{{\mathcal T}}
\newcommand{\CK}{{\mathcal K}}
\newcommand{\CM}{{\mathcal M}}
\newcommand{\SA}{{\mathsf A}}

\newcommand{\SD}{{\mathsf D}}
\newcommand{\la}{\lambda}
\newcommand{\SRN}{{\rm N}}

\newcommand{\sy}{{\mathsf y}}

\newcommand{\by}{{\mathbf y}}
\newcommand{\bx}{{\mathbf x}}
\newcommand{\bb}{{\mathbf b}}
\newcommand{\bk}{{\mathbf k}}

\newcommand{\SOmega}{{\mathsf \Omega}} 
\newcommand{\somega}{{\scriptstyle{\mathsf\Omega}}}

\newcommand{\ot}{\otimes}
\newcommand{\fr}[2]{{\textstyle \frac{#1}{#2} }}
\newcommand{\klein}{\scriptscriptstyle}
\newcommand{\kt}{{\klein T}}
\newcommand{\XXZ}{{\rm\klein XXZ}}
\newcommand{\SG}{{\rm\klein SG}}
\newcommand{\SGo}{{\rm\klein SGo}}
\renewcommand{\=}[1]{\stackrel{(\ref{#1})}{=}}
\newcommand{\+}{{+}}
\newcommand{\ppm}{\pm}
\newcommand{\xx}{{X}}  
\renewcommand{\fsl}{{\mathfrak s}{\mathfrak l}}
\newcommand{\BR}{{\mathbb R}}
\newcommand{\BN}{{\mathbb N}}
\newcommand{\BC}{{\mathbb C}}
\newcommand{\BZ}{{\mathbb Z}}
\newcommand{\CS}{{\mathcal S}}
\newcommand{\CD}{{\mathcal D}}
\newcommand{\CH}{{\mathcal H}}
\newcommand{\CU}{{\mathcal U}}
\newcommand{\SR}{{\mathsf R}}
\newcommand{\SE}{{\mathsf E}}
\newcommand{\SF}{{\mathsf F}}
\newcommand{\SK}{{\mathsf K}}
\newcommand{\SC}{{\mathsf C}}
\newcommand{\SO}{{\mathsf O}}
\newcommand{\SB}{{\mathsf B}}
\newcommand{\SH}{{\mathsf H}}
\newcommand{\SP}{{\mathsf P}}

\newcommand{\SM}{{\mathsf M}}
\newcommand{\ST}{{\mathsf T}}
\newcommand{\SX}{{\mathsf X}}
\newcommand{\SQ}{{\mathsf Q}}
\newcommand{\SY}{{\mathsf Y}}
\newcommand{\SZ}{{\mathsf Z}}
\newcommand{\SV}{{\mathsf V}}
\newcommand{\SW}{{\mathsf W}}
\newcommand{\SU}{{\mathsf U}}
\newcommand{\SJ}{{\mathsf J}}
\newcommand{\sx}{{\mathsf x}}
\newcommand{\spp}{{\mathsf p}}
\newcommand{\sS}{{\mathsf s}}
\newcommand{\sj}{{\mathsf j}}

\newcommand{\su}{{\mathsf u}}

\newcommand{\sr}{{\mathsf r}}
\newcommand{\sh}{{\mathsf h}}
\newcommand{\se}{{\mathsf e}}
\newcommand{\sff}{{\mathsf f}}

\newcommand{\sk}{{\mathsf k}}
\newcommand{\De}{\Delta}
\newcommand{\CP}{{\mathcal P}}
\newcommand{\bra}{\langle}
\newcommand{\ket}{\rangle}
\theoremstyle{plain}
\newtheorem{thm}{Theorem}

\newtheorem{propn}{Proposition}
\newtheorem{lem}{Lemma}

\newtheorem{conj}{Conjecture}
\newtheorem{defn}{Definition}
\theoremstyle{remark}
\newtheorem{rem}{Remark}
\begin{document}
\thispagestyle{empty}
\title{Quantization of models with non--compact quantum group symmetry.\\
Modular XXZ magnet and lattice sinh--Gordon model.}

\author{Andrei G. Bytsko$^{1}$ \ and \ 
	J{\"o}rg Teschner$^{2}$}

\address{$^1$ Steklov Mathematics Institute,
 Fontanka 27, 191023, St.~Petersburg, Russia  \\[1mm]
 $^2$ DESY Theory Group, Notkestrasse 85, 
 D-22603, Hamburg, Germany\\[2ex]
February 2006}

\maketitle

\footnotetext[1]{Supported in part by a Humboldt Stiftung fellowship,
 INTAS grant YS--03-55-962, the Russian Foundation for Fundamental 
 Research grant 05--01--00922, and a grant from the Russian Science 
 Support Foundation}
\footnotetext[2]{Part of this work was supported by a 
 DFG Heisenberg fellowship}

\section{Introduction} 

\subsection{Motivation}
One of the main motivations for studying integrable lattice models is 
their role as a lattice regularization of quantum field theories in 
continuous space--time. Integrable nonlinear sigma models are of 
particular interest, and recently there has been a growing interest in 
nonlinear sigma models with {\em non--compact} target spaces. 
This interest is motivated by possible applications to string
theory on curved space--times in general, and to gauge theories
via the AdS--CFT correspondence in particular. 

However,
the quantization and the solution of such non--compact nonlinear sigma
models still represents a major challenge for the field of 
integrable models. Compared to the better understood 
nonlinear sigma models with {compact} target spaces one may expect 
important qualitative differences, which make it problematic to
apply the known techniques from the compact cases to the sigma
models with non--compact target spaces. This point is exemplified by
the relation between the Wess--Zumino--Novikov--Witten (WZNW)
models associated to compact and non--compact symmetric spaces respectively.
The solution of the latter is possible \cite{T1}, but it is considerably 
more difficult than the solution of WZNW models associated to 
compact groups.

In more general sigma models one can not hope to find the powerful 
Kac--Moody symmetries of the WZNW models but the integrable structure 
may still survive. In order to enter the next level of complexity one may 
therefore try to exploit the integrability of some of these models.
Turning to a new class of models it is always advisable
to look for the simplest member which still exhibits most of
the new qualitative features. In the case of the conformal 
WZNW models it has turned out that Liouville theory 
already displays many of the 
relevant differences which distinguish the non--compact WZNW
models from rational conformal field theories \cite{T2}. Moving outside
of the class of sigma models soluble thanks to Kac--Moody or similarly
powerful {\em chiral} symmetries it seems 
natural to look for a useful  counterpart of Liouville theory
within this larger class of models. 

A natural candidate for such a model exists: the sinh--Gordon model. 
Indeed, there is some evidence \cite{ZZ,Lu} that the sinh--Gordon model 
can be seen as a ``deformation'' of Liouville theory which 
preserves its integrable structure when the conformal symmetry is lost.
While there certainly exists a good basis for the study of the 
sinh--Gordon model in infinite volume --- S-matrix and the form factors
are known \cite{VG,FMS,KMu,BL,Le} and the basic ingredients of
the QISM approach were developed \cite{S1} --- there does not seem 
to exist a sytematic approach to
the quantization and solution of the sinh--Gordon model in {\em finite}
spatial volume yet. Part of the problem is due to the usual 
divergencies and ordering problems of quantum (field) theory.
But the other part of the problem seems to be closely related
to the non--compactness of the target space in the sinh--Gordon model.

Our main motivation behind the present project was therefore
to find an integrable
lattice regularization for the sinh--Gordon model. This not only tames the
usual short distance singularities, it will also allow us to take
care of the troubles from non--compactness of the target space in 
a mathematically well--defined framework. One particular feature
that directly follows from the non--compactness of the target space will 
be the failure of the usual algebraic Bethe ansatz method \cite{F1}
for the model at hand. This failure 
means that we will have to use the more general separation of 
variables method \cite{Sk2,Sk3,Sm} instead.

\subsection{Lattice sinh--Gordon and the modular XXZ magnet}

A quantum integrable system is a quantum system $(\CH,\CA,\SH)$, with
Hilbert space $\CH$, algebra of observables $\CA$, Hamiltonian $\SH$,
in which there exists a set $\CQ=\{\ST_0,\ST_1,\dots\}$ of self--adjoint
operators such that
\begin{align*}
{\rm (A)} \quad &  [\ST,\ST']=0\quad\forall\ \ST,\, \ST'\in\CQ,\\
{\rm (B)} \quad &  [\ST,\SH]=0\quad\forall\ \ST\in\CQ,\\
{\rm (C)} \quad &  {\rm if}\;\,[\ST,\SO]=0 
\;\,{\rm for~all}\;\,\ST\in\CQ,\;\,
{\rm then}\;\,\SO=\SO(\CQ).
\end{align*}
Property (C) expresses completeness of the set $\CQ$ of integrals of 
motion. It is equivalent to the statement that the spectrum of $\CQ$ is 
non--degenerate, i.e., that simultaneous eigenstates of $\ST_k$, 
$k\in\BZ^{\geq 0}$ are uniquely determined by the tuple of their 
eigenvalues. We will consider the so--called one--dimensional lattice
models for which one has 
\begin{equation}
\CH=\CK^{\ot \SRN}, \qquad \CA=\CB^{\ot \SRN},
\end{equation}
with one copy of Hilbert space $\CK$ and algebra of local observables 
$\CB$ being associated to each of the $\SRN$ sites
of a one--dimensional lattice.

The quantum inverse scattering method (QISM) \cite{FST,F1} goes a long 
way towards the construction of large classes of 
quantum integrable models of this type. In this framework one 
usually characterizes $\CK$ as a representation of a Hopf algebra 
$\FU$ of ``symmetries'', and $\CB$ is generated from the
operators which represent the elements of $\FU$ on $\CK$.
It is clear that the representation theoretic properties 
of $\CK$ will influence the physical properties of the 
resulting integrable model decisively. Good control
over these properties will be crucial in the construction and
solution of such models.

In general it is a highly nontrivial problem to find the ``right" 
representation $\CK$ which leads to a useful lattice regularization of 
a particular quantum field theory. We will here propose a particular 
choice for $\CK$ which will lead to a lattice model with particularly 
nice properties, and which will be shown to yield the sinh--Gordon 
Hamiltonian density in the continuum limit of 
the corresponding classical lattice model. The representations
in question will be representations of the non--compact real 
form $\Uq$ of $\CU_q(\fsl_2)$ which 
have been studied in \cite{PT1,F3,PT2,BT}. 

The non--compactness of 
the target space will be reflected in the infinite--dimensionality of 
the representation~$\CK$. It is furthermore worth noting that
the same representations were previously found to reflect a key
internal structure of Liouville theory~\cite{PT1,T2}. 
In view of the existing evidence~\cite{ZZ,Lu} for the connection between
Liouville theory and the sinh--Gordon model, it is quite natural that 
the same class of representations appears in our
lattice version of the sinh--Gordon model as well.

The corresponding
representations possess a remarkable {\em duality}~--- they are
simultaneously representations of ${\cal U}_q(\fsl_2)$ and 
${\cal U}_{\tilde{q}}(\fsl_2)$, where $q=e^{i \pi b^2}$ and 
$\tilde{q}=e^{i \pi b^{-2}}$. One may therefore view them \cite{F3} 
as representations of the {\em modular double} 
${\cal U}_q(\fsl_2)\ot{\cal U}_{\tilde{q}}(\fsl_2)$ (see 
also~\cite{KLS,BT}). The parameter $b$ turns out to be proportional to
the coupling constant $\beta$ of the sinh--Gordon model. The 
self--duality of our representations will be directly related to 
the self--duality of the sinh--Gordon model under $b\ra b^{-1}$ 
which was previously observed in its scattering theory. The importance
of this self--duality for our analysis can hardly be over--emphasized.

It turns out that there is a close relative of our lattice sinh--Gordon
model which is simpler in some respects. This integrable
lattice model can be seen as a non--compact counterpart of the
XXZ model with spins in infinite--dimensional representations 
of the modular double. We will refer to this model as the 
{\em modular} XXZ magnet. As some technical issues are simpler in 
the case of the modular XXZ magnet, we will first construct
the latter model before we turn to the lattice sinh--Gordon model. 
In any case, it seems to us that the study of the modular XXZ magnet
is of interest in its own right.  We note in particular that despite 
the different underlying representation theory, our model has many 
structural similarities with the non--compact XXX type magnet based
on infinite--dimensional highest weight representations
of~$\fsl_2$ which was studied in~\cite{DKM,KM}. The latter model plays 
an important role in high energy QCD~\cite{Li,FK1}.

\subsection{Plan of the paper}

To make our paper accessible to a reasonably wide audience, we 
presented the general description of our approach, the main definitions 
and results in the main body of the paper and collected more technical 
developments in Appendices.
The article is organized as follows. 

In Section~2 we define
the modular XXZ magnet in terms of the representations $\CP_s$
describe its Hilbert space of states, construct the corresponding 
fundamental R--operator $\SR(u)$, and discuss the construction of its 
Hamiltonian and the set of integrals of motion~$\CQ$. 

The same is done for the lattice sinh--Gordon model in Section~3. 
We show that the Hamiltonian density of the sinh--Gordon model is 
recovered in the continuum limit of the corresponding classical model. 
We also show that the algebraic Bethe ansatz fails due to the 
non--compactness of the target space in the sinh--Gordon model.

An important first step towards the solution of these models is
taken in Section 4. We refine the spectral problem for the integrals 
of motion $\ST_k$ by constructing the Q--operator $\SQ(u)$ which is 
related to the $\ST_k$ via the so--called Baxter equation.
Analyzing the properties of $\SQ(u)$,
we derive a set of conditions for its eigenvalues $q_t(u)$ which can 
be seen as quantization conditions and which replace the usual Bethe 
ansatz equations in our models. The self--duality of 
our representations furthermore allows us to derive the so--called quantum 
Wronskian relation for the sinh--Gordon model with odd~$\SRN$, 
which encodes valuable additional 
information about the spectrum.

In order to show that the conditions found in Section 4 are also 
sufficient to characterize the spectrum we apply the separation
of variables approach to our models in Section 5. 

Section 6 contains concluding remarks on the conditions which characterize
the spectrum of our models, the continuum limit, and the relation
with the lattice and continuum versions of Liouville theory. 
We observe in particular that our results
are consistent with the results and conjectures of \cite{Za,Lu}
on the continuum sinh--Gordon model in a nontrivial way. 

Appendices contain necessary technical details.
Appendix~A collects the relevant information on the special functions
that we use.
Appendix~B discusses the precise mathematical nature of the self--duality
$b\ra b^{-1}$ of the representations that we use.
Appendic~C contains some important technical results on the 
structure of the monodromy matrix.
Appendix~D is devoted to the construction of the fundamental R--operator, 
the key object for the construction of local integrals of motion for the 
lattice models. Appendix~E contains details on the derivation of the 
properties of the 
Q--operator~$\SQ(u)$.

\vspace*{1.5mm}
{\par\small
{\em Acknowledgements.} We thank  S.\,Derkachov, L.\,Faddeev, S.\,Lukyanov,
N.\,Reshetikhin, and especially F.\,Smirnov for stimulating discussions. 
 A.B. is grateful to J.T. and R.\,Schrader for hospitality during 
his visits to the Institute for Theoretical Physics, FU--Berlin
and to V.\,Schomerus for hospitality during a visit to DESY, Hamburg.
\par}

\newpage
\renewcommand{\contentsname}{\large\bf Contents}
{\small
\tableofcontents
}
\newpage


\section{Modular XXZ magnet}\label{XXZ}   

In this section we will begin to develop the QISM for the {\em modular} 
XXZ magnet --- an XXZ type non--compact spin chain, which has $\Uq$ as 
a quantum symmetry. 

\subsection{Quantum group symmetry  $\Uq$}

Let $q=e^{i\gamma}$, $\gamma=\pi b^2$, $b\in (0,1)$. We will also
use the notation $Q=b+b^{-1}$.

The quantum group $\CU_q(\fsl_2)$ is a Hopf algebra with generators
$E$, $F$, $K$, $K^{-1}$ satisfying the relations
\begin{equation}\label{def}
 KE=qEK,\qquad KF=q^{-1}FK, \qquad 
 [E,F]=\fr{1}{q-q^{-1}}(K^2-K^{-2}) \,
\end{equation}
and equipped with the following co--product:
\begin{equation}\label{De} 
\begin{aligned}
 \De(E)=&\, E\ot K+K^{-1}\ot E \,, \\ 
\De(F)=&\, F\ot K+K^{-1}\ot F \,,
\end{aligned}
 \qquad \De(K)=K\ot K \,. 
\end{equation}
The relevant real form of $\CU_q(\fsl_2)$ is $\Uq$, which is defined 
by the following star--structure:
\begin{equation}
\label{star}
 K^*=K, \qquad E^*=E, \qquad F^*=F \,. 
\end{equation} 
The center of $\Uq$ is generated by the $q$--Casimir element:
\begin{equation}\label{Cas}
 C = (2 \sin\gamma)^2 \, FE - qK^2 - q^{-1}K^{-2} + 2 \,,
 \qquad C^*=C \,.
\end{equation}

\subsection{Representations ${\cal P}_s$ --- algebra of observables}

A one--parameter family of unitary representations 
$\CP_s$ of $\Uq$ can be constructed from a
pair of self--adjoint operators $\spp$ and $\sx$ on $L^2(\BR)$ 
which satisfy
\hbox{$[\spp,\sx]=(2\pi i)^{-1}$} as follows:
\begin{equation}\label{EFK1}
\begin{aligned}
\pi_{s}(E)\;\equiv\;\SE_{s}\;=\;&e^{+\pi b \sx} \,
 \frac{\cosh\pi b(\spp-s)}{\sin\gamma} \, e^{+\pi b \sx} \,, \\
 \pi_{s}(F)\;\equiv\;\SF_{s}\;=\;&
 e^{-\pi b \sx} \, \frac{\cosh\pi b(\spp+s)}{\sin\gamma} \,
 e^{-\pi b \sx} \,,
\end{aligned}
 \qquad \pi_{s}(K)\;\equiv\SK_s\;=\;e^{-\pi b\spp} \,.
\end{equation}
For this representation
\begin{equation}\label{cps}
\SC_s\;\equiv\; \pi_{s} ( C ) =  4 \cosh^2 \pi b s  \,.
\end{equation}
It is remarkable and important that the operators $\SE_s$, $\SF_s$ and 
$\SK_s$ are {\em positive} self--adjoint. 
Indeed, the representations  $\CP_s$ are 
the {\em only} ``reasonable" representations of $\Uq$ which 
have this property. 
This property will play a key role in much of 
the following developments. It will in particular ensure 
seld--adjointness of operators such as the Hamiltonian and
the integrals of motion. It is also the mathematical basis
for the self--duality of the representations $\CP_s$, as
shown in Appendix~\ref{posapp} (see, in particular, eq.~\rf{bdual}).

The lattice model that we are about to define will have 
one of the representations $\CP_s$ attached to each site of the
one--dimensional lattice.
This means that we take
\begin{equation}
\CH\,=\,\bigl( L^2(\BR) \bigr)^{\ot \SRN}
\end{equation}
as the Hilbert space of our model, and let
\begin{equation}
\widehat{\CA}\,=\,\big(\pi_s(\CU)\big)^{\ot \SRN},\quad \CU\equiv\Uq
\end{equation}
be a set of generators for
our algebra of observables. Note that the operators in 
$\widehat{\CA}$ are all unbounded, but 
there exists a basis for $\widehat{\CA}$
whose elements are positive self--adjoint (see Appendix~\ref{posapp}). 
The latter fact allows us to construct large classes of 
{\em non--polynomial} operator
functions of the generators in $\widehat{\CA}$ via standard  
functional calculus for self--adjoint operators and/or
pseudo--differential operator calculus. 

\subsection{Integrals of motion}

As the next step we shall introduce our main ansatz for the
set $\CQ$ of integrals of motion using the usual scheme
of the QISM. To this aim let us assemble the generators of 
$\widehat{\CA}$ into the following 
L--matrix acting on ${\mathbb C}^2 \ot {\cal P}_s$:
\begin{equation}\label{Lxxz}
 L^{\XXZ} (u)   
  = \left( \begin{array}{cc}  
  e^{\pi b u} \sk_s - e^{-\pi b u} \sk_s^{-1} &  
     i \,  e^{\pi b u} \, \sff_s \\
  i \,  e^{-\pi b u} \, \se_s  & 
 e^{\pi b u} \sk_s^{-1} - e^{-\pi b u} \sk_s 
    \end{array} \right) \,, \qquad u\in {\mathbb C}.
\end{equation}
In the definition of $L^{\XXZ} (u)$ we have used the rescaled generators
$\se_s$, $\sff_s$, $\sk_s$ which are defined by
\begin{equation}\label{efk}
 \se_s = (2 \sin\gamma) \, \SE_s \,, \quad   
 \sff_s = (2 \sin\gamma) \, \SF_s \,, \quad 
 \sk_s = \SK_s \,.
\end{equation}
Occasionally we will
omit the  superscript $\XXZ$ for the sake of brevity.
The defining relations~(\ref{def}) and (\ref{De})
of $\Uq$ are equivalent to 
\begin{eqnarray}
 \label{rLL} 
 R_{12}(u) \, L_{13} (u + v) \, L_{23} (v) &=&
 L_{23} (v) \, L_{13} (u + v) \, R_{12}(u) \,,\\[0.5mm]
 \label{dLL}
 \bigl(\mathrm{id} \otimes  \Delta \bigr) L^{\ppm} &=& 
 L^{\ppm}_{13} \, L^{\ppm}_{12}  \,,
\end{eqnarray}
where $L^{\ppm}$ arise in the decomposition 
\begin{equation}\label{Lpm}
 L (u) = e^{\pi b u} \, L^{+} - e^{-\pi b u} \, L^{-} \,,
\end{equation}
and the  auxiliary R--matrix is given by
\begin{equation}\label{R12}
  R(u) = 
 \left( \begin{array}{cccc}
  \sinh \pi b(u + i b) & & & \\ [-1mm]
   &  \sinh \pi b u  & 
   i\sin\pi b^2 \,  e^{ \pi b u} & \\ 
   & i\sin\pi b^2 \, e^{- \pi b u}
   &  \sinh \pi b u & \\ [-1mm]
   & & & \sinh\pi b(u + i b)
 \end{array} \right) \,.
\end{equation}
Out of the L--matrices we may then construct the monodromy~$\SM(u)$,
\begin{equation}\label{Mono}
 \SM(u) \,\equiv\, \left(\begin{matrix} \SA(u) & \SB(u) \\
 \SC(u) & \SD(u) \end{matrix}
 \right)\,\equiv\,L_\SRN(u) \cdot \ldots \cdot L_2(u) \cdot L_1(u) \,.
\end{equation}
Of particular importance is the one--parameter family of operators:
\begin{equation}\label{Monotr}
 \ST(u) \,= \,\mathrm{tr}\, \bigl(\SM(u)\bigr)=  
	\SA(u)+\SD(u)\,.
\end{equation}
The trace in (\ref{Monotr}) is taken over the auxiliary space,
which is~${\mathbb C}^2$ for the models we consider. 
\begin{lem} The operators $\ST_m$ which appear in the expansion
\begin{equation}\label{Texp}
 \ST(u)\,=\,e^{\pi b\SRN u}\,
	\sum_{m=0}^{\SRN}(-e^{-2\pi b u})^m\,\ST_m,
\end{equation}
are positive self--adjoint and mutually commuting, $[\ST_m,\ST_n]=0$.
\end{lem}
Commutativity of $\ST_m$ follows by the standard argument from
the relation (\ref{rLL}); the proof of their positivity and
self--adjointness is given in Appendix~\ref{monoapp}.

\begin{defn}
Let us define the set of commuting charges 
as $\CQ=\{\ST_0,\ST_1,\dots,\ST_\SRN\}$. 
\end{defn}

\begin{rem}\label{posadj}
Self--adjointness of the operators $\ST_m$, $m=0,\dots, \SRN$
ensures the existence of a joint spectral decomposition
for the family~$\CQ$.

Let us emphasize that the crucial positivity
of the operators
$\ST_m$ is a direct consequence of the fact that 
the generators $E$, $F$ and $K$ of $\Uq$ are 
represented by {\em positive} operators in the representations
$\CP_s$. This makes clear why these representations are particularly 
well--suited for defining non--compact analogues of the XXZ spin chains.
We will later make a similar observation in the lattice sinh--Gordon model.
\end{rem}

\subsection{Fundamental R--operator and Hamiltonian}
 
Our next aim is to construct a {\em local} Hamiltonian which commutes 
with the elements of~$\CQ$. We will adapt the approach from \cite{FTT} 
to the case at hand. The main ingredient of this approach is the 
so--called fundamental R--operator corresponding to (\ref{Lxxz}). This
operator, $\SR^\XXZ_{s_\2 s_\1}(u)$, acts on 
$\CP_{s_\2} \otimes \CP_{s_\1}$ and is supposed
to satisfy the commutation relations
\begin{equation}\label{RLL}
 \SR_{\2\3}(u) \, L_{\1\3} (u + v) \, L_{\1\2} (v) =
 L_{\1\2} (v) \, L_{\1\3} (u + v) \, \SR_{\2\3}(u) \,.
\end{equation}
For our purposes it will be sufficient\footnote{The general 
solution $\SR^\XXZ_{s_\2 s_\1}(u)$ 
is needed if we wish to construct an inhomogeneous spin
chain, for instance the one with alternating spins (see,
e.g.,~\cite{BD}).}
to deal with ${\SR}(u) \equiv {\SR}_{s s}^\XXZ(u)$
acting on~$\CP_{s} \otimes \CP_{s}$. 

\begin{defn}
Let the operator $\SR(u)$ be defined by the formula
\begin{equation}\label{Rsimpledef}
 {\SR}(u) = \SP \, w_b(u + \sS) \, w_b(u - \sS) =\SP \, D_{u}(\sS) \,,
\end{equation}
where $\SP$ is the operator which just permutes the
two tensor factors in $\CP_{s} \otimes \CP_{s}$, and $\sS$ 
is the unique positive self--adjoint operator such that
\begin{equation}\label{tensC}
4 \cosh^2 \pi b \, \sS \,=\,
	(\pi_{s} \ot \pi_{s}) \Delta (C) \,
\end{equation}
The special functions $w_b(x)$ and $D_\al(x)$ are defined in 
Appendix~\ref{Qdil}.
\end{defn}

\begin{thm} \label{Rthm} The operator $\SR(u)$ satisfies the
equation \rf{RLL} where $L(u)$ is  given by~(\ref{Lxxz}).
\end{thm}
The proof of this theorem is given in Appendix~\ref{proofSH},
where the construction of the operator $\SR^\XXZ_{s_\2 s_\1}(u)$ 
is presented for the
general case, $s_\1 \neq s_\2$, see equation~\rf{Rxxz}.

The operator $\SR(u)$ has the following further properties
\begin{align} 
\label{RGGa}
 \text{regularity} \quad&  {\SR}(0) = \SP \,,\\
\label{RGGb}
 \text{reflection property} \quad&
 {\SR}(-u)  =\SP \, {\SR}^{-1}(u) \, \SP \,,\\
\label{RGGc}
 \text{unitarity} \quad&  
 {\SR}^* (u) = {\SR}^{-1} (u) \quad \text{for } u\in\BR\,,
\end{align}
which follow from the properties of $w_b(u)$ 
and $D_\al(x)$ listed in Appendix~\ref{Qdil}.

The regularity condition (\ref{RGGa}) allows us to apply the standard 
recipe \cite{FTT} of the QISM in order to construct a Hamiltonian with 
local (nearest neighbour) interaction of sites: 
\begin{equation}\label{HRxxz}
\begin{aligned}
 \SH^\XXZ &= 
 \frac{i}{\pi b} \, \SU^{-1} \, \left[
\frac{\partial}{\partial u}  \mathrm{tr}_a \, 
 \bigl( \SR_{a \SRN}(u) \cdot \ldots \cdot \SR_{a 2}(u) 
 \cdot \SR_{a 1}(u) \bigl) \right]_{u=0} \\
 &= \sum_{n=1}^{\SRN}  \frac{i}{\pi b} \partial_u 
 D_u(\sS_{n,n+\1}) \Bigm|_{u=0} = \sum_{n=1}^{\SRN} 
 H^\XXZ_{n,n+1} \,.
\end{aligned}
\end{equation}
We are using the following notation: We 
identify $\sS_{\SRN,\SRN+1}\equiv \sS_{\SRN,1}$,
the cyclic shift operator $\SU$ is defined by 
$\SU\, f(x_\1,x_\2,\ldots,x_\SRN) = f(x_\2,\ldots,x_\SRN,x_\1)$, 
and 
the subscript $a$ stands for an auxiliary copy of the 
space ${\cal P}_s$. The trace operation is defined for 
an operator $\SO : {\cal P}_s \mapsto {\cal P}_s$ in the 
usual way: if the integral kernel of $\SO$ in the momentum
representation is given by $O(k|k')$, then 
$\mathrm{tr}\, \SO = \int_{-\infty}^\infty dk\, O(k|k)$.
According to this definition we have 
\hbox{$\mathrm{tr}_a \, \SP_{ab} = {\sf 1}_b$}.

Substituting the integral representation (\ref{Dint}) for 
$D_u(x)$ into (\ref{HRxxz}), we obtain the following local 
Hamiltonian density 
\begin{equation}\label{Hphi}
 H^\XXZ_{n,n+1} = 
  - \frac{1}{\pi} \, \int\limits_{\BR+i0} dt\,
 \frac{\cos (2 b t\, \sS_{n,n+1} )}{\sinh t \, \sinh{b^2 t}  } \,.
\end{equation}

It may then be shown in the usual manner \cite{FTT} that $\SH$ commutes 
with the trace of the monodromy matrix, $\ST(u)$, which means that
\begin{equation}\label{HTcomm}
[\,\SH\,,\,\ST_k\,]\,=\,0,\;\;{\rm for}\;\;k=0,\dots,\SRN.
\end{equation}

As in any quantum mechanical system, the fundamental problem that
we would like to solve is the problem to determine the spectrum of~$\SH$. 
However, thanks to the commutativity \rf{HTcomm} it seems promising to 
first solve the following\\[1ex]
\noindent{\bf Auxiliary Spectral Problem:} {\em Find the spectrum of the 
operator $\ST(u)$, i.e., the joint spectral decomposition for the family 
of operators} $\CQ=\{\ST_0,\dots,\ST_\SRN\}$.\\[1ex]
Simple counting of the degrees of freedom suggests that the spectrum of 
$\CQ$ may be simple, i.e., that an eigenstate $\Psi_t$ of~$\ST(u)$, 
\[
 \ST(u)\, \Psi_t \,=\, t(u) \, \Psi_t \,,
\]
is uniquely characterized  by the eigenvalue 
$t(u)\,=\,e^{\pi b\SRN u}\,\sum_{m=0}^{\SRN}(-e^{-2\pi bu})^m\,t_m$. 
This would imply that $\SH=\SH(\CQ)$, so that the solution to the Auxiliary 
Spectral Problem also yields the spectral decomposition of~$\SH$.

\subsection{Classical limit}\label{Hxxz}

Let us discuss the classical limit of the quantum 
Hamiltonian~(\ref{Hphi}).
So far we have been working in the units where the Planck 
constant $\hbar$ was chosen to be unity. In order to recover it 
explicitly, we have to make the following rescaling 
\begin{equation}\label{hbar}
  b^2 \to \hbar \, b^2 \,,\quad 
 \spp \to \hbar^{-\frac{1}{2}} \spp \,, \quad
 \sx  \to \hbar^{-\frac{1}{2}} \sx \,, \quad
 \sS \to \hbar^{-\frac{1}{2}} \sS \,, \quad 
 s \to \hbar^{-\frac{1}{2}} s \,, 
\end{equation}
so that we have $q=e^{i \hbar \gamma}$, $\gamma= \pi b^2$.
The operators $\se_s$, $\sff_s$, $\sk_s$ are not affected by
the procedure~(\ref{hbar}). In the limit $\hbar \to 0$ they
become classical variables $\se$, $\sff$, $\sk$ with the following
Poisson brackets obtained by the correspondence 
principle, $[\,,] \to -i \hbar \{\,,\}$
\begin{equation}\label{Pbr}
 \{ \se , \sk \}= \gamma \,\sk \se \,, \quad
 \{ \sff , \sk \}= - \gamma \,\sk \sff \,, \quad
 2 \gamma  \{ \se , \sff \} = \sk^2 -\sk^{-2} \,.
\end{equation}

Using the asymptotics (computed by means of contour integration)
\begin{equation}\label{lim}
\begin{aligned}
 \lim\limits_{\hbar \to 0}  \frac{\hbar}{2}  
 \int_{\BR+i0} \! dt\,
 \frac{e^{-i t z}}{\sinh t \, \sinh{\hbar t}  }  & =
 \frac{1}{2} 
 \int_{\BR+i0} \! dt\, \frac{e^{-i t z}}{t \, \sinh t}\\
 &= \sum_{n\geq 1}^{\infty} \frac{ (-1)^n \, e^{\pi n z}}{n} 
 = -\log (1+e^{\pi z}) \,,
\end{aligned}
\end{equation}
we obtain from (\ref{Hphi}) the corresponding classical 
lattice Hamiltonian density,
\begin{equation}\label{Hcl1}
\begin{aligned}
 {}  H^{\XXZ, \rm cl}_{n,n+1} & = 
 \lim_{h\to 0} \hbar H^\XXZ_{n,n+1}
 = \fr{1}{\gamma} \log (4 \cosh^2 \pi b \, \sS^{\rm cl}_{n,n+1}) =
 \fr{1}{\gamma} \log (\SC^{\rm cl}_{n,n+1}) \\
 {}& = \fr{1}{\gamma} \log \bigl( (\se_n \sff_{n+1} + 
 \sff_n \se_{n+1}) \, \sk_n^{-1} \sk_{n+1} \\
 & \qquad \qquad  + 2 \sk_n^{-2} \sk_{n+1}^2 + 
 2 \cosh(2\pi b s) (\sk_n^{-2} + \sk_{n+1}^2) +2  \bigr)  \,.
\end{aligned}
\end{equation}
Here $\SC^{\rm cl}_{n,n+1}$ is the classical limit
of the tensor Casimir operator given by~(\ref{tensC}).

\subsection{Comparison with similar models}\label{CSM}

L--matrix (\ref{Lxxz}) and R--matrix (\ref{R12}) are suitable
for the usual XXZ model as well. The only (but essential)
difference is that in the latter case matrix coefficients of
the L--matrix act on a highest weight module of~$\CU_q(\fsl_2)$.
We also remark that (\ref{Lxxz}) differs from the most commonly
used ``standard'' L--matrix in that it contains extra 
factors~$e^{\pm \pi b u}$ in the off--diagonal elements. 
The ``standard'' L--matrix does not satisfy (\ref{Lpm})
but is symmetric (if the matrix transposition ${}^\kt$ is combined
with the operator transposition ${}^t$ such that $\sff^t_s=\se_s$ 
and $\sk^t_s=\sk_s$) and corresponds to the symmetric 
auxiliary R--matrix~(\ref{Rlsg}).

Let $\psi_b(x)$ denote the logarithmic derivative of
the function $S_b(x)$ defined by~(\ref{Sbx}). Properties
(\ref{Sb1})--(\ref{Sb3}) show that $S_b(x)$ can be regarded 
as a b--analogue of the gamma function. The Hamiltonian density
(\ref{Hphi}) rewritten in terms of $\psi_b(x)$ looks as follows
\begin{equation}\label{Hpsi}
 H^\XXZ_{n,n+1} = \fr{2}{\pi b} \, \psi_b(\fr{Q}{2} + i\sS)=
 \fr{2}{\pi b} \, \psi_b(\fr{Q}{2} - i\sS) =
 \fr{1}{\pi b} \bigl( \psi_b(\fr{Q}{2} + i\sS) +
  \psi_b(\fr{Q}{2} - i\sS) \bigr) \,.
\end{equation}
Equivalence of these expressions is due to~(\ref{Sb3}). The last
of them resembles the form of the Hamiltonian density of
the non--compact XXX magnet \cite{DKM} expressed in terms of 
the ordinary $\psi$--function.

The special function $w_b(u)$ is closely related (cf.~Eq.~(\ref{gw})) 
to the non--compact quantum dilogarithm~$g_b(u)$. 
Counterparts of (\ref{Rsimpledef}) and (\ref{Rxxz}) for the compact 
XXZ magnet look similar in terms of the q--gamma function which, 
in turn, is closely related to the compact analogue of 
$g_b(u)$ given by $s_q(t)=\prod_{n=0}^{\infty}(1+tq^{2n+1})$.

It is also worth noticing that the R--operator (\ref{Rsimpledef}) 
resembles the fundamental R--operator $r(\sS,\lambda)$ found 
in~\cite{FV} for a simpler L--matrix related to the Volterra 
model~\cite{V1}. The main difference is that the operator argument 
$\sS$ of $r(\sS,\lambda)$ has a much simpler structure in terms of 
the variables $\spp$ and~$\sx$. It would be interesting to clarify 
the connection between these two R--operators.

\section{Lattice sinh--Gordon model}\label{LSG}  
\label{sectSG}

\subsection{Definition of the model}

In this section we will begin to develop the QISM for a lattice version 
of the sinh--Gordon model, which has $\Uq$ as a quantum symmetry. We are 
going to keep much of the set--up from Section~\ref{XXZ}, but we will now 
be using the following L--matrix acting on ${\mathbb C}^2 \ot {\cal P}_s$
\begin{equation}\label{lSG}
 L^{\SG}(u) 
  = \fr{1}{i} e^{-\pi b s}\, \left( \begin{array}{cc}  
  i \, \se_s  & 
 e^{\pi b u} \sk_s^{-1} - e^{-\pi b u} \sk_s  \\  
 e^{\pi b u} \sk_s - e^{-\pi b u} \sk_s^{-1} &
  i \, \sff_s   
 \end{array} \right) \,.
\end{equation}

L--matrix (\ref{lSG}) satisfies the intertwining relation
(\ref{rLL}) where the auxiliary R--matrix is now given~by
\begin{equation}\label{Rlsg}
 R(u) = 
 \left( \begin{array}{cccc} 
 \sinh \pi b ( u+ ib) & & & \\ [-1mm]
 & \sinh \pi b u & i \sin\gamma & \\ [-1mm]
 & i \sin\gamma & \sinh \pi b u & \\ [-1mm]
 & & & \sinh \pi b ( u+ ib) 
 \end{array} \right) \,.
\end{equation}
This R--matrix possesses the following symmetry
\begin{equation}\label{symR}
  [ R(u) , \sigma_a \otimes  \sigma_a] = 0
 \,, 
\end{equation}
where $\sigma_a$, $a=1,2,3$, are the Pauli matrices. We may then proceed 
the same way as in Section~\ref{XXZ} to define the operator $\ST(u)$ and 
the family $\CQ=\{\ST_0,\dots,\ST_{\SRN}\}$ of commuting observables.
We again find (see Appendix~\ref{monoapp})
that the corresponding operators $\ST_m$, $m=0,\dots,\SRN$ 
are  positive self--adjoint as a direct consequence of the positivity
of $\se_s$, $\sff_s$, $\sk_s$. The existence of  a joint spectral
decomposition for the family $\CQ$ is thereby ensured by the spectral
theorem (cf. Remark~\ref{posadj}).

\subsection{Fundamental R--operator and Hamiltonian}\label{FRSG}

Now our aim is to find the fundamental R--operator
$\SR_{s_\2 s_\1}^{\SG} (u)$ corresponding to
L--matrix~(\ref{lSG}).  Fortunately, it turns out that it can be 
constructed from the R--operator of the XXZ chain in a simple way.
To demonstrate this, we first introduce an automorphism $\theta$
such that
\begin{equation}\label{theta}
 \theta(\spp)=-\spp\,, \qquad\theta(\sx)=-\sx\,.
\end{equation}
It is useful to notice that $\theta$ can be realzied as
an inner automorphism, 
$\theta(\SO)= \somega \, \SO \, \somega^{-1}$,
where $\somega$ is the parity operator whose action in the momentum 
representation is defined by $(\somega f)(k)=f(-k)$. Notice that
$\somega$ is unitary and satisfies $\somega^{-1}=\somega$.
Observe that for the representation $\CP_s$ we have (cf.~(\ref{EFK1}))
\begin{equation}\label{thetaEFK}
 \theta (\se)  = \sff \,, \qquad 
 \theta (\sff)  = \se \,, \qquad 
 \theta (\sk)  = {\sk}^{-1} \,.
\end{equation}

\begin{defn}
Let the operator $\SR_{s_\2 s_\1}^{\SG} (u)$ be defined by the 
formula
\begin{equation}\label{Rsg2'}
    \SR_{s_\2 s_\1}^{\SG} (u) \,=\, 
	(\sk \otimes \sk)^{- i \frac{u}{2b}} \cdot (\somega\ot 1)\cdot
	 {\SR}^{\XXZ}_{s_\2 s_\1}(u)\cdot(1\ot\somega)\cdot 
	(\sk \otimes \sk)^{-  i \frac{u}{2b}} \,,
\end{equation}
where $\SR^{\XXZ}_{s_\2 s_\1}(u)$ is given by \rf{Rsimpledef}
if $s_\1=s_\2$ and by~(\ref{Rxxz}) otherwise.
\end{defn}

\begin{propn}\label{Rfsg}
The operator $\SR_{s_\2 s_\1}^{\SG} (u)$ satisfies the 
equation \rf{RLL} where $L(u)$ is given by~(\ref{lSG}).
\end{propn}

\begin{proof}
It will be convenient to consider 
$\check\SR_{\1\2}^{\SG}(u)\equiv\SP_{\!\!\2\1}\SR_{s_\2 s_\1}^{\SG}(u):
	\CP_{s_\2}\ot\CP_{s_\1}\ra\CP_{s_\1}\ot\CP_{s_\2}$
and analogously defined
$\check{\SR}_{\1\2}^\XXZ(u) \equiv 
	\SP_{\!\!\2\1}\SR_{s_\2 s_\1}^{\XXZ}(u)$.
Equation \rf{RLL} for $\SR_{s_\2 s_\1}^{\SG}(u)$ is then 
equivalent~to 
\begin{equation}\label{cRLL}
 \check\SR_{\2\3}^{\SG}(u) \, L_{\1\3}^{\SG} (u + v) \, 
	L_{\1\2}^{\SG} (v) = L_{\1\3}^{\SG} (v) \, 
 L_{\1\2}^{\SG} (u + v) \, \check\SR_{\2\3}^{\SG}(u) \,.
\end{equation}
Let us also note that, by using
$(\sk\ot\sk)^{-1}\check\SR^\XXZ_{\1\2}(u)(\sk\ot\sk)=
	\check\SR^\XXZ_{\1\2}(u)$,
we may rewrite the expression for $\check\SR_{\1\2}^{\SG}(u)$ which 
follows {}from \rf{Rsg2'} as
\begin{equation}\label{Rsg2}
\begin{aligned}
 \check{\SR}_{\1 \2}^{\SG} (u) &\,= \,
	\big({\rm id}\ot\theta\big)\big(\check{\SR}'_{\1\2}(u)\big),\\
 \check{\SR}'_{\1\2}(u)&\,\equiv\,  
	(\sk \otimes 1)^{- i \frac{u}{b}} \cdot 
\check{\SR}_{\1 \2}^{\XXZ}(u)\cdot
 (1 \otimes \sk)^{ i\frac{u}{b}} \,. 
\end{aligned}
\end{equation} 
The key to the proof of the Proposition will then be the following 
relation between the L--matrices $L^{\SG}(u)$ and $L^\XXZ(u)$
\begin{equation}\label{Lsg2}
 L^{\SG}(u)\,=\, -i e^{-\pi b s} \,
 \sigma_1 \,\sk^{-i\frac{u}{b}}\,L^\XXZ(u)\, 
	\sk^{i\frac{u}{b}}\,\equiv\, \sigma_1\,{L}'(u) \,.
\end{equation}
Inserting \rf{Lsg2} into \rf{cRLL}, we get an 
expression which contains $ \sigma_1 \, {L}_{\1\3}'(u) \, \sigma_1$. 
Observe that (cf.~\rf{thetaEFK})
\begin{equation}\label{Lt}
 \sigma_1 \, {L}'(u) \, \sigma_1 =
 (id \otimes \theta) \, {L}'(u) \,.
\end{equation}
Therefore, by using $\theta(\SO_\1\SO_\2)=\theta(\SO_\1)\theta(\SO_\2)$, 
one finds that \rf{cRLL} is equivalent to
\begin{equation}
 \check\SR_{\2\3}'(u) \, L_{\1\3}' (u + v) \, L_{\1\2}' (v) =
 L_{\1\3}' (v) \, L_{\1\2}'(u + v) \, \check\SR_{\2\3}'(u) \,,
\end{equation}
which is now easily reduced to the Theorem~\ref{Rthm} (its general 
case for $\SR^{\XXZ}_{s_\2 s_\1}(u)$).
\end{proof}

It is easy to see that properties (\ref{RGGa})--(\ref{RGGc}) of 
the R--operator of the modular XXZ magnet hold for 
$\SR(u)\equiv \SR^{\SG}_{ss}(u)$ as well. Therefore, using the 
regularity of $\SR(u)$, we can construct a Hamiltonian with nearest 
neighbour interaction of sites by using the same recipe that 
we used to derive~(\ref{HRxxz}). This yields
\begin{equation}\label{Hsg}
\begin{aligned}
 H^{\SG}_{n,n+1} & 
 = \fr{i}{\pi b} \partial_u
 \check{\SR}^{\SG}_{n,n+1}(u) \Bigm|_{u=0} 
 \={Rsg2} (id \otimes \theta) \, H^\XXZ_{n,n+1} 
 + \fr{1}{\gamma} \log(\sk_n \sk_{n+1}) \\
  & = - \frac{1}{\pi} \, \int\limits_{\BR+i0} dt\,
 \frac{\cos (2 b t\, \hat\sS_{n,n+1} )}{\sinh t \, \sinh{b^2 t}  } 
 + \fr{1}{\gamma} \log(\sk_n \sk_{n+1}) \,,
\end{aligned}
\end{equation}
where $\hat\sS =(1 \ot \somega) \,\sS \, (1 \ot \somega)$ 
is the unique positive self--adjoint operator on $\CP_s \ot \CP_s$ 
such that
\begin{equation}\label{ttC}
 4 \cosh^2 \pi b\, \hat\sS = 
	\bigl(\pi_{s} \ot (\theta\circ\pi_{s}) \bigr) \Delta (C) \,. 
\end{equation}

\subsection{Relation with the continuum theory}\label{CL}

To begin with, we may first compute the classical limit, $\hbar\to 0$, 
of (\ref{Hsg}) in the same way as we derived (\ref{Hcl1}). 
Using \rf{thetaEFK}, we obtain
\begin{equation}\label{Hcl2}
\begin{aligned}
 H^{\SG, \rm cl}_{n,n+1} 
 = &\, \fr{1}{\gamma} \log \bigl( \se_n \se_{n+1} + \sff_n \sff_{n+1} 
  + 2 ( \sk_n^{-1} \sk_{n+1}^{-1} + \sk_n \sk_{n+1} ) \\
 & +  2 \cosh(2\pi b s) (\sk_n^{-1} \sk_{n+1} 
  + \sk_n \sk_{n+1}^{-1})  \bigr)  \,.
\end{aligned}
\end{equation}
In order to establish the relation with the sinh--Gordon model it will 
be convenient to change variables as follows: 
\begin{equation}\label{px}
  2 \pi b \, \spp_n =  -\beta \, \Phi_n  \,, \qquad
  4 \pi b \, \sx_n = \beta \, (\fr{1}{2} \Pi_n -\Phi_n) \,, \qquad
  \beta=b \sqrt{8\pi} \,.
\end{equation}
The variables $\Phi_n$  and $\Pi_n$ will then turn out to correspond 
to the (discretized) sinh--Gordon field and its conjugate momentum, 
respectively. The classical field and momentum variables defined 
in~\rf{px} satisfy the Poisson--bracket relations 
$\{\Pi_n,\Phi_m\}=\delta_{nm}$. The Hamiltonian \rf{Hcl2} now looks 
as follows 
\begin{align}
\label{Hsgl}
 H^{\SG, \rm cl}_{n,n+1} 
 &= \fr{1}{\gamma}\log \fr{4}{\mu} \Bigl( 
 \fr{1}{2}\cosh \fr{\beta}{4} \, (\Pi_n + \Pi_{n+1}) 
 + \fr{1+\mu^2}{2} \,
 \cosh \fr{\beta}{2} \, (\Phi_n - \Phi_{n+1}) \\
\nonumber
 &+ \fr{\mu}{2} \, \cosh \beta \bigl(\Phi_n -
	\fr{1}{4}(\Pi_n +\Pi_{n+1})\bigr) 
 + \fr{\mu}{2} \, \cosh \beta \bigl(\Phi_{n+1} -
	\fr{1}{4}(\Pi_n +\Pi_{n+1})\bigr) \\
\nonumber
 & + \mu \, \cosh \fr{\beta}{2} (\Phi_n + \Phi_{n+1}) 
 + \fr{\mu^2}{4} \cosh \beta \bigl( \Phi_n + \Phi_{n+1} - 
 	\fr{1}{4}(\Pi_n +\Pi_{n+1}) \bigr) \Bigr)\,, 
\end{align}
where $\mu=e^{-2\pi bs}$. In order to define the relevant limit 
leading to the {\em continuous} sinh--Gordon model, let us combine 
the limit of vanishing lattice spacing $\SRN\to\infty$, $\Delta\to 0$
($R =\SRN\Delta/2\pi$ is kept fixed) with the limit where the 
representation parameter $s$ goes to 
infinity in such a way that the mass parameter $m$ defined via
\begin{equation}\label{mds}
 \fr{1}{4} m\Delta\,=\,e^{-\pi b s}  \,.
\end{equation}
stays finite. In addition we shall assume the standard
correspondence between lattice and continuous variables:
\begin{equation}\label{cont}
 \Pi_n \to \Pi(x) \, \Delta \,, \quad 
 \Phi_n \to \Phi(x) \,, \quad 
 x = n\Delta \,.
\end{equation}
We then find the following limiting expression for the Hamiltonian 
density:
\begin{equation}\label{limH}
 \sum_n \fr{1}{\Delta}  H^{\SG, \rm cl}_{n,n+1} \to
 {\rm const} + \int\nolimits_0^{2\pi R} dx \,
 \bigl( \fr{1}{2} \Pi^2 + \fr{1}{2} (\partial_x\Phi)^2
 + \fr{m^2}{\beta^2} \cosh \beta\Phi \bigr)
\end{equation}
thus recovering the continuous $\sinh$--Gordon model.

It is also instructive to see what happens to the L--matrix in this limit.
In the classical continuous limit, i.e., when $m$ in (\ref{mds}) 
is kept fixed and $\hbar,\Delta\to 0$, Eqs.~(\ref{hbar}) and
(\ref{cont}) show that L--matrix (\ref{lSG}) becomes
\begin{equation}\label{LU}
 L^{\SG}(\fr{u}{\pi b})  \to 
 \bigl(\begin{smallmatrix} 1 & 0\\ 0 & 1
 	\end{smallmatrix} \bigr)
 + \Delta \, U^{\SG}(u) + O(\Delta^2)\,,
\end{equation}
where $U^{\SG}(u)$ is the well--known $U$--matrix from the Lax 
pair for the classical continuous sinh--Gordon model~\cite{KBI,S1},
\begin{equation}\label{Ucc}
  U^{\SG}(u) =   \left( \begin{array}{cc}  
  \fr{\beta}{4} \, \Pi(x)  & 
  \fr{m}{2i} \, \sinh \bigl(u - \frac{\beta}{2} \Phi(x) \bigr)  \\  
  \fr{m}{2i} \, \sinh \bigl(u + \frac{\beta}{2} \Phi(x) \bigr)  &
  - \fr{\beta}{4} \, \Pi(x)
 \end{array} \right)\,.
\end{equation}

\begin{rem}
The classical lattice Hamiltonian density (\ref{Hcl2}) resembles 
that found in \cite{Ta} for the lattice sine--Gordon model. 
However, relation between the {\em quantum} Hamiltonians is less 
clear because the fundamental R--operator proposed in \cite{Ta}
is represented as a product of \hbox{R--operators} of the
type $r(\sS,\lambda)$ which we mentioned at the end of 
Subsection~\ref{CSM}.
Possibly, recent results on factorization of R--operators \cite{DKK}
will help to clarify the connection between our construction and
that used in~\cite{Ta,FV,V1}.
\end{rem}

\subsection{Failure of the algebraic Bethe ansatz}\label{ABA}

For the sake of clarity it may be worthwhile explaining in some detail
why the algebraic Bethe ansatz is not suitable for the solution of the 
lattice sinh--Gordon model. 

To begin with, let us observe that the L--matrix (\ref{lSG}) has no 
pseudo--vacuum state, i.e., a vector 
$\Psi$ such that $L^{\SG}_{21}(u) \, \Psi =0$.
Indeed, this would require that $\sk_n\Psi=0=\sk_n^{-1}\Psi$ 
for all $n=1,\dots,\SRN$. 
Such a vector
does not exist.

For the sine--Gordon model, one circumvents this difficulty by
considering the composite \hbox{L--matrix},  ${\cal L}(u)$, which 
is product of two L--matrices~\cite{FST,IK}.
For $\gamma=\pi \fr{m}{n}$, $n,m \in {\mathbb N}$, exponential 
operators $e^{i \beta \Phi_n}$, $e^{i \beta \Pi_n}$ admit 
finite--dimensional representations. In this case there exists
a vector $\Psi$ that is annihilated by ${\cal L}_{21}(u)$. 
This makes it possible to apply the algebraic Bethe ansatz technique.
Let us therefore consider the analogous construction for the 
sinh--Gordon model. 
Let ${\cal L}^{\SG}(u) = L^{\SG}_{2}(u) \, 
 L^{\SG}_{1}(u+\varpi)$, where we introduced the shift by
the constant $\varpi\in\BR$ in order to increase the generality
of our consideration. We then have
\begin{equation}\label{CLL}
 {\cal L}^{\SG}_{21}(u) =
 i e^{\pi b u} \bigl( \sk \ot \se + 
	e^{\pi b \varpi} \, \sff \ot \sk  \bigr) -  
 i e^{-\pi b u} \bigl( \sk^{-1} \ot \se + 
	e^{-\pi b \varpi} \, \sff \ot \sk^{-1} \bigr) \,.
\end{equation}
The requirement that a vector $\Psi$ is annihilated by
${\cal L}^{\SG}_{21}(u)$ is equivalent to the two equations
\[
\bigl( \sk^{\pm 1} \ot \se + 
 e^{\pm \pi b \varpi} \, \sff \ot \sk^{\pm 1} \bigr)\Psi=0 \,. 
\]
We claim that there does not exist a reasonable (even in the
distributional sense) state $\Psi$ with such properties. Indeed, note that 
\[	
 \sk \ot \se + e^{\pi b \varpi} \, \sff \ot \sk=(\sk^{i\frac{\varpi}{b}}
 \somega\ot {\sf 1})\cdot(\sk^{-1}\ot\se+\se\ot\sk)\cdot 
 (\sk^{i\frac{\varpi}{b}}\somega\ot {\sf 1})^{-1} \,,
\]
where $\somega$ is the parity operation introduced in Subsection~\ref{FRSG}.
The operator $(\sk^{i\frac{\varpi}{b}} \somega\ot {\sf 1})$ is unitary, 
which allows us to conclude that 
$\sk \ot \se + e^{\pi b \varpi} \, 
	\sff \ot \sk$ and $\sk^{-1}\ot\se+\se\ot\sk$ 
have the same spectrum. However, the latter operator represents 
$(2\sin\gamma)\De(E)$ on $\CP_{s}\ot\CP_s$ (cf. eq.~\rf{De}). 
Unitarity of the Clebsch--Gordan maps (see Appendix~\ref{CGmaps})
implies that this operator has the same 
spectrum as~$\se_s$. The unitary transformation used in the proof of 
Lemma~\ref{Pslem} in Appendix~B maps $\se_s$ to $e^{2\pi b\sx}$. It is 
now clear that all these operators do not have an eigenfunction with 
eigenvalue zero, as would be necessary to construct a Bethe vacuum.

\begin{rem}
Keeping in mind that the sinh--Gordon variables are just linear
combinations of $\spp$ and~$\sx$, cf.~\rf{px}, 
we now see quite clearly that the failure
of the Bethe ansatz is connected with the fact that the 
target space (the space in which the fields take their values) 
is {\em non--compact}. We expect this to be a general lesson.
\end{rem}

\section{Q--operator and Baxter equation}\label{QB} 

As an important first step towards the solution of the 
Auxiliary Spectral Problem we shall now find {\em necessary}
conditions for a function $t(u)$ to be eigenvalue of the operator
$\ST(u)$. In order to do this we are going to construct an operator 
$\SQ(u)$ which satisfies the following properties:
\begin{equation}\label{defQ1} 
\begin{aligned}
 {\rm (i)} \quad & \SQ(u) \;\;{\rm is~a~normal~operator},\;\;
 \SQ(u)\SQ^{*}(v)=\SQ^{*}(v) \SQ(u),\\
 {\rm (ii)} \quad & \SQ(u) \, \SQ(v) = \SQ(v) \, \SQ(u) \,, \\
 {\rm (iii)} \quad & \SQ(u) \, \ST(u) = \ST(u) \, \SQ(u) \,,\\
 {\rm (iv)} \quad &  \SQ(u) \, \ST(u)\,=\,
 \bigl(a(u)\bigr)^{\SRN} \, \SQ(u-ib)
 +\bigl(d(u)\bigl)^{\SRN} \, \SQ(u+ib). 
\end{aligned}
\end{equation}
The first and the second property imply that all operators $\SQ(u)$, 
$u\in\BC$ can be simultaneously diagonalized and their eigenvectors
form a complete system of states in the Hilbert space.
The third and the fourth property imply that $\ST(u)$ will be 
diagonal whenever $\SQ(u)$ is. One may therefore consider the spectral 
problem for $\SQ(u)$ as a refinement of the spectral problem for~$\ST(u)$.

Let us now consider an eigenstate $\Psi_t$ for $\ST(u)$ with 
eigenvalue $t(u)$, $\ST(u)\Psi_t=t(u)\Psi_t$. Thanks to property~(iii) 
above we may assume that it is an eigenstate for $\SQ(u)$ as well,
\begin{equation}\label{Qpq}
 \SQ(u) \, \Psi_t\,=\,q_t(u) \, \Psi_t\,.
\end{equation}
It follows from property (iv) that the eigenvalue $q_t(u)$ must 
satisfy the so--called Baxter equation
\begin{equation}\label{BaxterEV}
\boxed{\quad t(u) \, q_t(u) \,=\, \bigl(a(u)\bigr)^{\SRN}q_t(u-ib)
 + \bigl(d(u)\bigr)^{\SRN}q_t(u+ib) \,.\quad}
\end{equation}
We will construct the operator $\SQ(u)$ explicitly --- see 
Subsection~\ref{SR}. This will allow us to determine the analytic and --
for the lattice sinh--Gordon model with $\SRN$ odd (the SGo--model) 
-- the asymptotic properties that the eigenvalues $q_t(u)$ must 
have, namely
\begin{equation}\label{analQ} 
\left[ \;\;\begin{aligned}
{\rm (i)} \;\; & q_t(u)\;\,\text{is meromorphic in}\;\,\BC, 
 \,\;\text{with poles of maximal order $\SRN$ }
 \text{in}\;\,\Upsilon_{-s}\cup
\bar{\Upsilon}_{s},\\
& \text{where}\;\,
\Upsilon_s = \bigl\{ s + i \bigl( \fr{Q}{2} + nb + mb^{-1} \bigl),
 \quad n,m \in {\mathbb Z}^{\geq 0} \bigl\}\,,\quad
 \bar{\Upsilon}_s\equiv ({\Upsilon}_s)^*\,,\\ 
{\rm (ii)} \;\; & q_t^{\SGo}(u)\;{\sim} \,\left\{
\begin{aligned}
& \exp\big(+\pi i \SRN\big(s+\fr{i}{2}Q\big) u\big)\;\;{\rm for}\;\;
|u|\ra\infty,\;\, |{\rm  arg}(u)|<\fr{\pi}{2}, \\
 & \exp\big(-\pi i \SRN\big(s+\fr{i}{2}Q\big) u\big)\;\;{\rm for}\;\;
|u|\ra\infty,\,\; |{\rm  arg}(u)|>\fr{\pi}{2}.
\end{aligned}\right.
\end{aligned}\;\;\right]
\end{equation}
The derivation of these properties is discussed in Subsection~\ref{AnaQ}.
This means that there is the following {\em necessary} condition for a 
polynomial $t(u)$ to be eigenvalue of the operator $\ST(u)$: $t(u)$ can 
only be an eigenvalue of $\ST(u)$ if there exists a meromorphic function 
$q_t(u)$ 
with singular behavior and asymptotic behavior given in \rf{analQ} 
which is related to $t(u)$ by the Baxter equation~\rf{BaxterEV}.

The problem to classify the solutions to this condition is of course
still rather nontrivial. 
However, previous experience from other integrable
models suggests that the Baxter equation supplemented by the 
analytic and asymptotic properties \rf{analQ} is indeed a 
useful starting point for the determination of the spectrum of
the model, see also our Subsections \ref{Baxtersub} and~\ref{contSG} 
for some further remarks. 
We will discuss in the next section how the 
separation of variables method may allow us to show that
the conditions above are also {\em sufficient} for $t(u)$ to be 
an eigenvalue of~$\ST(u)$.

\noindent{\em Convention:} We will use the superscripts $\SG$ and
$\XXZ$ to distinguish analogous operators within the two 
models we consider. However, we will simply omit these superscripts 
in any equation which holds in the two cases alike.

\subsection{Explicit form of $\SQ(u)$}\label{SR}

Let us now describe explicitly the Q--operators for the 
models that we introduced in Sections~\ref{XXZ} and~\ref{LSG}.
For this purpose we will work in the representation where the 
operators $\sx_r$, $r=1,\ldots,\SRN$ are diagonal. This representation
will be called the {\em Schr{\"o}dinger representation}
for the Hilbert space of a lattice model.
Let ${\bf x}\equiv(x_\1,\ldots,x_{\klein \SRN})$,
${\bf x'} \equiv(x'_\1,\ldots,x'_{\klein \SRN})$.
We will denote the integral kernel of the 
operator $\SQ(u)$ in the Schr{\"o}dinger 
representation by $Q_u({\bf x},{\bf x'})$.
We will also use the following notations
\begin{equation}\label{ssb}
 \sigma \equiv s + \fr{i}{2}Q \,, \qquad
 \bar\sigma \equiv s - \fr{i}{2}Q \,,
\end{equation}
where $s$ stands for the spin of the representation~$\CP_s$.

\begin{defn}
Let the Q--operators $\SQ^\flat_\pm(u)$, $\flat= \XXZ,\, \SG$, be 
defined in the Schr{\"o}dinger representation by the following kernels
\begin{align}\label{Qxxz+}
 Q^{\,\flat}_{+;u} ({\bf x},{\bf x'}) \,= &\\
 =\bigl( D_{-s}(u) \bigr)^\SRN  
 & \prod_{r=1}^\SRN
 D_{\frac{1}{2}(\bar\sigma -u)}(x_{r} - x'_{r}) \,
 D_{\frac{1}{2}(\bar\sigma +u)}(x_{r-1} - \ve_\flat \, x'_{r}) \,
 D_{-s}(x_{r} - \ve_\flat \, x_{r-1}) \,,\nonumber \\
\label{Qxxz-}
 Q^{\,\flat}_{-;u} ({\bf x},{\bf x'}) =  &\prod_{r=1}^\SRN
 D_{\frac{1}{2}(u-\sigma)}(x_{r} - x'_{r}) \,
 D_{-\frac{1}{2}(u+\sigma)}(x_{r} - \ve_\flat \, x'_{r-1}) \,
 D_{s}(x'_{r} - \ve_\flat \, x'_{r-1}) \,,
\end{align}
where $\ve_\XXZ = 1$, $\ve_\SG = -1$, and, in the sinh--Gordon case, 
$s$ is related to the parameters $m$, $\Delta$ as in~(\ref{mds}).
\end{defn}

\begin{thm}\label{QXXZSG}
Let $\ST^\flat(u)$, $\flat= \XXZ,\, \SG$ be the 
transfer--matrices corresponding to the L--matrices
(\ref{Lxxz}) and (\ref{lSG}).
\begin{itemize}
\item[(i)] 
The operators $\SQ^\flat_\pm(u)$ satisfy all relations 
in (\ref{defQ1}). 
\item[(ii)]
The Baxter equation \hbox{(\ref{defQ1}--{\rm iv})} holds for 
$\SQ^\flat_\pm(u)$ with the following coefficients
\begin{equation}\label{baxxz}
\begin{aligned}
 a^\XXZ(u) &= 2 \sinh \pi b (u-\sigma) \,, \quad
 d^\XXZ(u) = 2 \sinh \pi b (u+\sigma) \,, \\
 d^\SG(u) &= a^\SG(-u) =e^{ \pi b (u+i\frac{b}{2})} +
    (\fr{m \Delta}{4})^2 \, e^{ -\pi b (u+i\frac{b}{2})} \,.
\end{aligned}
\end{equation}
\item[(iii)]
The operators $\SQ^\flat_\pm(u)$ satisfy the relation
\begin{equation}\label{QQc}
   \SQ^\flat_+(u) \, \SQ^\flat_-(v) =
   \SQ^\flat_-(v) \, \SQ^\flat_+(u) \,.
\end{equation}
\end{itemize}
\end{thm}
\noindent
The proof of this Theorem is given in Appendix~\ref{PQT}. 
It is worth noting that $\SQ^{\,\flat}_{-}(u)$
and $\SQ^{\,\flat}_{+}(u)$ are related by hermitian conjugation as follows.
\begin{equation}\label{xqpm'}
 \SQ^{\,\flat}_{-}(u) =  \bigl( D_{-s}(u) \bigr)^\SRN \, 
 \bigl( \SQ^{\,\flat}_{+}(\bar{u}) \bigr)^*  \,.
\end{equation}
This allows us to mostly 
focus on $\SQ^{\,\flat}(u)\equiv\SQ^{\,\flat}_{+}(u)$,
but it is nevertheless 
sometimes useful to consider $\SQ^{\,\flat}_{-}(u)$ as well.
The corresponding eigenvalues $q^{\pm}_t (u)$ 
are consequently related as
\begin{equation}\label{qqpm}
 q^{-}_t (u) = \bigl( D_{-s}(u) \bigr)^\SRN \
   \overline{q^{+}_t (\bar{u})} \,.
\end{equation}
Relation \rf{qqpm} will imply that $q^{+}_t$ and $q^{-}_t$
have the same analytic and asymptotic properties \rf{analQ}.

\begin{rem}
It is sometimes useful to observe (see Subsection~\ref{QYZ}) 
that the operator $\SQ^\flat(u)$ can be factorized as follows:
\begin{equation}\label{qyz'}
 \SQ^\flat(u) = \SY^\flat(u) \cdot \SZ \,,
\end{equation}
The operators $\SY^\flat(u)$ and $\SZ$ in \rf{qyz'} 
are represented by the kernels
\begin{align}\label{y1'}
  Y_u^\flat({\bf x},{\bf x'}) &= \prod_{r=1}^\SRN  
 D_{\frac{1}{2}(u-\sigma)}(x_r - \ve_\flat\, x'_{r+1}) \, 
 D_{-\frac{1}{2}(u+\sigma)}(x_r - x'_{r}) \,, \\
 Z({\bf x},{\bf x'}) &=  \bigl(w_b(i\fr{Q}{2} - 2s)\bigr)^\SRN  \, 
	\prod_{r=1}^\SRN \,  D_{\bar{\sigma}}(x_r -x'_r)\,,
\end{align}
respectively, 
where $\ve_\XXZ = 1$, $\ve_\SG = -1$.
\end{rem}

\begin{rem}\label{gaugerem}
The relations \rf{defQ1} do not define the Q--operator $\SQ(u)$ 
uniquely. For instance, for a given $\SQ(u)$, relations \rf{defQ1} are 
also fulfilled for 
$\SQ'(u)= \big(\varphi(u)\big)^\SRN \, \SO \, \SQ(u)$, where 
$\varphi(u)$ is a scalar function and $\SO$ is a unitary operator that 
commutes with $\SQ(u)$ and $\ST(u)$. 
The coefficients in Baxter equation \hbox{(\ref{defQ1}--{\rm iv})} 
for $\SQ(u)$ and $\SQ'(u)$ are related via
\begin{equation}\label{baxxztil}  
 a'(u) =  \frac{\varphi(u)}{\varphi(u-ib)}\, a(u)\,, 
 \qquad d'(u) =  \frac{\varphi(u)}{\varphi(u+ib)} \, d(u) \,.
\end{equation}
Thus, there is no canonical way to fix these coefficients.
However, their combination $a(u)d(u-ib)$ remains invariant; its value is 
related to the quantum determinant if the latter can be defined for 
the L--matrix of the model in question (see Appendix~\ref{qDet}).
\end{rem}

\subsection{Analytic properties of eigenvalues of $\SQ(u)$}\label{AnaQ}

We now turn to the derivation of the analytic properties of 
eigenvalues of $\SQ(u)$. 
More precisely we shall prove the following:
\begin{thm} $\quad$
\begin{itemize}
\item[(i)] 
The 
operators $\SQ^{\XXZ}_\pm(u)$ and $\SQ^{\SG}_\pm(u)$ are meromorphic 
functions of $u$ in $\BC$ with poles of maximal order $\SRN$ contained 
in $\Upsilon_{-s}\cup\bar{\Upsilon}_{s}$. 
\item[(ii)] Denote $\SQ_\pm^\SGo(u)\equiv\SQ^\SG_\pm(u)$ for $\SRN$ odd. 
This operator has the following asymptotic behavior
\begin{equation}\label{qtas}
\SQ_+^\SGo(u)\;{\sim} \,\left\{
\begin{aligned}
& \SQ_{\klein+\infty} \,
\exp\big(+i \pi \SRN \sigma \, u \big)\qquad {\rm for}\;\;
|u|\ra\infty,\;\, |{\rm  arg}(u)|<\fr{\pi}{2} \,, \\
 & \SQ_{\klein-\infty} \,
\exp\big(-i\pi \SRN \sigma \, u\big)\qquad {\rm for}\;\;
|u|\ra\infty,\,\; |{\rm  arg}(u)|>\fr{\pi}{2} \,,
\end{aligned}\right.
\end{equation}
where $\SQ_{\klein\pm\infty}$ are commuting unitary operators 
related to each other as follows
\begin{equation}\label{QQpar}  
  \SQ_{\klein-\infty} = \SOmega \, \SQ_{\klein+\infty} \,.
\end{equation}
Here $\SOmega$ is the parity operator (its action in
the Schr{\"o}dinger representation is given by~\rf{upar}).
\item[(iii)]
$\SQ^\SGo_-(u)$ has asymptotic behavior of the same form \rf{qtas}
with $\SQ_{\klein\pm\infty}$ replaced by $\SQ^*_{\klein\pm\infty}$.
\end{itemize}
\end{thm}

\begin{proof} In order to prove part (i) of the Theorem 
it clearly suffices to
consider the corresponding statements for the eigenvalues 
$q^\pm_t(u)$. Let us first explain why the properties of $q^{+}_t (u)$ 
and~$q^{-}_t (u)$ described in the theorem are the same. 
Given that the theorem holds for $q^{+}_t (u)$, we infer that poles 
of $\overline{q^{+}_t (\bar{u})}$ are contained 
in $\Upsilon_{s}\cup\bar{\Upsilon}_{-s}$. But, since they are
of maximal order $\SRN$, they cancel in \rf{qqpm} against the 
$\SRN$--th order zeroes of $\bigl( D_{-s}(u) \bigr)^\SRN$ 
(see properties of $D_{\alpha}(x)$ in Appendix~\ref{Qdil2}). 
Thus, the only possible poles of $q^{-}_t (u)$ are those
of $\bigl( D_{-s}(u) \bigr)^\SRN$, i.e., they are of
maximal order $\SRN$ and contained in 
$\Upsilon_{-s}\cup\bar{\Upsilon}_{s}$.

The proof of part (i) will be exactly analogous for the cases of the 
XXZ magnet and the sinh--Gordon model. Only the latter case will therefore
be discussed explicitly.
We will study the equation (\ref{Qpq}), which is equivalent to
\begin{equation}\label{EVphipsi}
\int_{\BR^{2\SRN}} d\bx \, d\bx' \; {\Phi}(\bx)\,
 Q_{+;u}^{\SG}({\bf x},\bx')\,\Psi_t(\bx')\,=\,
	q^+_t(u)\,\langle\,\Phi\,|\,\Psi_t\,\rangle,
\end{equation}
for some test--function $\Phi(\bx)\in\CT_s^{\ot \SRN}$,
where $\CT_s$ is the space of test--functions which is 
canonically associated to the representations $\CP_s$ 
as shown in Appendix \ref{posapp}. In order to 
find the analytic properties of $q^+_t(u)$ let us use  
$\SQ_+^\SG(u)=\SY^{\SG}(u)\SZ$ to represent the left hand side 
of \rf{EVphipsi} as $\langle\,\Phi'\,|\,\SZ\Psi_t\,\rangle$, where 
\begin{equation}\label{EVexpl}
 {\Phi}'(\bx')\,\equiv \, \int_{\BR^\SRN} d\bx \; {\Phi}(\bx) \,
 Y_{u}^{\SG}({\bf x},\bx')\,.
\end{equation}
With the help of the Paley--Wiener theorems one easily finds that
the condition $\Phi(\bx)\in\CT_s^{\ot \SRN}$ implies that
${\Phi}(\bx)$ is entire analytic w.r.t. each variable $x_k$ and 
decays exponentially as
\begin{equation}\label{Psiasym}
 |{\Phi}(\bx)|\,\sim\, e^{-\pi Q|x_k|} \qquad
	{\rm for}\;\; |x_k|\ra\infty \,.
\end{equation}
The kernel $Y_{u}^{\SG}({\bf x},\bx')$ has the same asymptotics
w.r.t. its $x_k$ variables, as seen from equation~\rf{y1'} and
relation~\rf{Das1}. Therefore the convergence of the integrals does 
not represent any problem. 
Combined with the observation that the left hand side of \rf{EVexpl} 
is the convolution of two meromorphic functions we conclude that the 
only source of singular behavior is  the possibility that the contours 
of integration in \rf{EVexpl} may become pinched between poles of the 
integrand  approaching the contour from the upper and lower half planes, 
respectively. With the help of \rf{wan} one easily compiles a list of the 
relevant poles of the kernel $Y_{u}^{\SG}({\bf x},\bx')$ 
as given in~\rf{y1'}:
\begin{align*}
 \text{Upper half plane}\;\; {\mathbb H}_- :\quad (1)\;\; 
{} & x_r\in x_r'-\fr{1}{2}(u+{\sigma})+\Upsilon_0,\\
 (2)\;\; & x_r\in -x_{r+1}' + \fr{1}{2}(u-{\sigma})+\Upsilon_0,\\
 \text{Lower half plane}\;\;\ {\mathbb H}_+:\quad (1')\;\; 
{} & x_r\in x_r'+\fr{1}{2}(u+{\sigma})-\Upsilon_0,\\
(2')\;\; & x_r\in -x_{r+1}'-\fr{1}{2}(u-{\sigma})-\Upsilon_0.
\end{align*}
Pinching of the contour between poles from the upper and lower half 
planes would produce the following series of poles:
\begin{equation} \label{poles}
\begin{aligned}
{} &(11')\quad u+s\in +\Upsilon_{0},\\
{} &(22')\quad u-s\in -\Upsilon_0,
\end{aligned}
\quad
\begin{aligned}
{} &(12')\quad x_r'+x_{r+1}'-s\in -\Upsilon_0,\\
{} &(21')\quad x_r'+x_{r+1}'+s\in +\Upsilon_0.
\end{aligned}
\end{equation}
We observe in particular that none of the poles of ${\Phi}'(\bx')$ 
happens to lie on the real axis, which represents the contour of 
integration for each of the integrals over the variables $x_k'$ 
in~\rf{EVphipsi}. Taking into account the exponential decay of 
$Y_{u}^{\SG}({\bf x},\bx')$ for $|x_k'|\ra\infty$, we may conclude that 
the integration over $\bx'$ in \rf{EVphipsi} converges nicely. It follows
that the left hand side of \rf{EVphipsi} defines a meromorphic function  
of $u$ with poles listed on the l.h.s. of~\rf{poles}.

In order to verify part (ii) of the theorem let us note that 
$Y_{u}^{\SG}({\bf x},\bx')$ has the asymptotic behavior
\begin{equation}\label{Yasym}
 Y_{u}^{\SG}({\bf x},\bx')\,\sim e^{\pm\pi i\SRN{\sigma}u}
 \prod_{r=1}^{\SRN} \,e^{\mp 2\pi i x_r^{}(x_{r+1}'+x_{r}')} 
 \quad{\rm for}\;\; |u|\ra\infty,\,\; \left\{
\begin{aligned}
& |{\rm  arg}(u)|<\fr{\pi}{2}, \\
& |{\rm  arg}(u)|>\fr{\pi}{2}.
\end{aligned}\right.
\end{equation}
as follows straightforwardly from \rf{Das2}. In order to check that the 
integral obtained by exchanging the limit for $u\ra\pm\infty$ with the 
integrations in \rf{EVphipsi} is convergent let us note that performing 
the integration over the variables $x_r$ yields the Fourier 
transformation $\tilde{\Phi}(\bk)$ of ${\Phi}(\bx)$, with argument 
$\bk\equiv(x'_\1+x'_\2,\dots,x'_{\SRN}+x'_\1)$.
Note that the change of variables $\bk=\bk(\bx')$ is invertible for 
$\SRN$~odd. We may therefore represent the integration over $\bx'$ by 
an integration over~$\bk$. The nice asymptotic properties of 
$\tilde{\Phi}(\bk)$ which follow from our requirement 
$\Phi\in(\CT_s)^{\ot\SRN}$ ensure the convergence of the resulting 
integrals. Part (ii) of the theorem therefore follows from~\rf{Yasym} 
and~\rf{qyz'}.

The proof of part (iii) of the theorem is immediate, if 
\rf{xqpm'} and \rf{Das1} are taken into account. 
\end{proof}

\begin{rem} Let us comment on the nature of the problems which
prevented us to determine the asymptotics of the Q--operators in the
remaining cases. In both remaining cases one must observe that the 
change of variables $\bk=\bk(\bx')$ is not invertible, which implies 
divergence of the integral over~$\bx'$. 
This is closely related to the fact that the leading 
asymptotics of $\ST(u)$ for $|u|\ra \infty$
introduces a quasi--momentum $\spp_\0$ which has purely continuous 
spectrum (see equations \rf{badinf}, and \rf{eveninf} in 
Appendix \ref{monoapp}). 
It follows that $\ST(u)$ can not have any {\em normalizable} 
eigenstate. Instead one should work with the spectral representation 
for~$\spp_\0$,
$
\CH\,\simeq\,\int_{\BR}dp_\0\;\CH_{p_\0}
$
where 
the elements of $ \CH_{p_\0}$ are represented by 
wave--functions of the form
$
 \Psi_{p_\0}(\bx)=e^{2\pi i p_{\0}x_\0}
	\Psi(x_{\SRN-1}-x_{\0},\dots,x_\1-x_\0).
$
This seems to complicate the analysis considerably. We nevertheless
expect results similar to \hbox{(\ref{analQ}--{\rm ii})} to hold 
for the remaining cases as well.  
\end{rem}

\subsection{Self--duality and quantum Wronskian relation}

The explicit form \rf{Qxxz+}--\rf{Qxxz-} of the Q--operators
along with the properties of $D_\alpha(x)$ listed in 
Appendix~\ref{Qdil2} show that $\SQ^{\flat}_\pm(u)$ are
self--dual with respect to the replacement \hbox{$b\to b^{-1}$}.
Therefore, the Q--operators also satisfy the dual Baxter equations,
\begin{equation}\label{baxdual}
 \widetilde{\ST}^{\flat}(u) \cdot \SQ^{\flat}_\pm(u) =
  \bigl( \tilde{a}^\flat(u) \bigr )^\SRN \, 
  	\SQ^{\flat}_\pm(u - ib^{-1})  +
 \bigl( \tilde{d}^\flat(u) \bigr)^\SRN \, 
 	\SQ^{\flat}_\pm(u + ib^{-1}) \,, 
\end{equation}
where $\widetilde{\ST}^{\flat}(u)$, $\flat = \XXZ,\, \SG$,
denote the transfer--matrices corresponding to the L--matrices
(\ref{Lxxz}) and (\ref{lSG}) with $b$ replaced by $b^{-1}$.
These are the transfer--matrices
of the modular XXZ magnet and lattice sinh--Gordon model 
with ${\cal U}_{\tilde{q}}({\mathfrak sl}(2,\BR))$ symmetry, where
\hbox{$\tilde{q}=e^{i \pi b^{-2}}$}. The coefficients 
$\tilde{a}(u)$, $\tilde{d}(u)$ in \rf{baxdual} are 
similarly obtained from 
those in \rf{baxxz} by the replacement $b\ra b^{-1}$. 
In the sinh--Gordon case 
the mass $m_{b^{-1}}$ is related to the representation parameter 
$s$ via \hbox{$\frac{1}{4} m_{1/b} \Delta = e^{- \pi b^{-1} s} $}.

This self--duality has remarkable consequences, which we shall work 
out explicitly for the case of the sinh--Gordon model with odd~$\SRN$.
We will take advantage of the freedom pointed out in 
Remark~\ref{gaugerem}
to renormalize the operator $\SQ^\SG(u)$ as follows:
\begin{equation}\label{checkSQdef}
 \check\SQ(u) \,\equiv\,\,
\SQ_{+\infty}^*\,e^{-i\fr{\pi }{2} \SRN \left( u^2 +
 \sigma^2 + \de_+^2 \right)} \, 
\SQ_+^{\klein SG}(u)\,,
\end{equation}
where $\de_+=\frac{1}{2}(b+b^{-1})$ and
$\SQ_{+\infty}$ is the unitary operator which appears in the 
asymptotics \rf{qtas}.
The Baxter equation for $\check\SQ(u)$
will then take the following form:
\begin{equation}\label{baa2}
\begin{aligned}
  \ST^{\klein SG}(u) \cdot \check{\SQ}(u) &=
 \bigl( \check a(u) \bigr)^\SRN \, \check{\SQ}(u - ib) +
 \bigl( \check d(u) \bigr)^\SRN \,  
	\check{\SQ}(u + ib) \,,\\[2pt]
 \text{where}\quad & \check d(u) = \check a(-u) = 
 1 + (\fr{m \Delta}{4})^2 \, e^{ -\pi b (2u+ib)} \,.
\end{aligned}
\end{equation}
The normalization of the operator $\check\SQ(u)$
was for later convenience chosen in such a way that
$\check\SQ(u)\,\sim\, e^{\pm\pi i \sigma u -i\frac{\pi }{2} 
 \SRN \left( u^2 +\sigma^2 + \de_+^2 \right)} \cdot {\sf 1}$\; 
for ${\rm Re}(u) \to \pm\infty$.

\begin{thm} The operator $\check\SQ(u)$ fulfills the 
following quantum Wronskian relation:
\begin{equation}\label{q-Wr}
\check\SQ(u+i\de_+)\check\SQ(u-i\de_+)-
	\check\SQ(u+i\de_-)\check\SQ(u-i\de_-)
 \,=\,\, W_\SRN(u) \cdot {\sf 1} \,.
\end{equation}
where $W_\SRN(u)= 
 e^{-i\pi\SRN \left(u^2 +
 \sigma^2\right)}
  \big(D_{\sigma}(u)\big)^{-\SRN}$ 
and $\de_\pm\equiv\frac{1}{2}(b^{-1} \pm b)$.
\end{thm}
\begin{proof}
Let $\SW(u)$ be the left hand side of \rf{q-Wr}.
A straightforward calculation, using the Baxter equation and its 
dual form, shows that $\SW(u)$ satisfies the
following two functional relations:
\begin{equation}\label{q-Wr-diff}
\SW\big(u+\fr{i}{2}b^{\pm 1})\,=\,
\left( e^{2\pi b^{\pm 1} u}  \frac{\cosh\pi b^{\pm 1}(u-\sigma)}%
{\cosh\pi b^{\pm 1}(u+\sigma)}\right)^\SRN
\SW\big(u-\fr{i}{2}b^{\pm 1})\,.\end{equation}
A solution to both functional relations is given by the expression on
the right hand side of \rf{q-Wr}. 
If $b$ is irrational it suffices to notice
that $\SW(u)$ is meromorphic in order to conclude that the solution to the 
system \rf{q-Wr-diff} must be unique up to multiplication by an
operator which does not depend on~$u$. This freedom can be fixed
by comparing the asymptotics of both sides for ${\rm Re}(u)\to\pm\infty$ 
using equations~\rf{qtas},\rf{QQpar} (for considering the asymptotics
as ${\rm Re}(u)\to-\infty$, it is helpful to notice that 
$\SOmega^2={\sf 1}$). In order to cover the case of rational $b$ let us 
notice that both sides 
of the relation \rf{q-Wr} can be analytically continued from irrational 
values of $b$ to the case where $b$ is rational.
\end{proof}

From the proof of this theorem it is clear that the main ingredient is
the self--duality of our representations $\CP_s$. We therefore expect 
that a similar result will hold for the remaining cases as well. 
However, at present we do not control the asymptotics of the Q--operators 
sufficiently well in these cases.

\subsection{Parity and cyclic shift}

Let us consider the 
cyclic shift and parity operators $\SU$ and $\SOmega$ defined
respectively by 
\begin{align}\label{upar}
\SU\, f(x_\1,x_\2,\ldots,x_\SRN) = f(x_\2,\ldots,x_\SRN,x_\1)\,,\qquad
\SOmega \, f(x_\1,\ldots,x_\SRN) = f(-x_\1,\ldots,-x_\SRN)\,.
\end{align}
These operators commute with each other and also with 
$\SQ^{\flat}_\pm(u)$ (as can be easily seen from 
(\ref{Qxxz+})--(\ref{Qxxz-})). Hence they must
commute with $\ST^{\flat}_\pm(u)$, as can also be verified 
directly\footnote{
Indeed, for $\SU$ this is obvious from the definition (\ref{Monotr}) 
and for $\SOmega$ it follows from the observation that
 $\SOmega \, L^\XXZ(u) \, \SOmega = \sigma_1 \, e^{-\pi b u \sigma_3} \, 
	L^\XXZ(u) \, e^{\pi b u \sigma_3} \, \sigma_1$  and
 $\SOmega \, L^\SG(u) \, \SOmega = \sigma_1 \, L^\SG(u) \, \sigma_1$.
}.
It follows that eigenstates $\Psi_t$ may be assumed to be 
simultaneously eigenstates of $\SU$ and $\SOmega$, 
\begin{equation}\label{UWP} 
\SOmega \, \Psi_t = \pm \Psi_t \,, \qquad
 \SU\, \Psi_t = e^{2\pi i \frac{m}{\SRN}} \, \Psi_t \,, \qquad 
 m=1,\ldots,\SRN \,.
\end{equation}
It is therefore useful to observe that $\SU$ and $\SOmega$ can be
recovered from $\SQ^\flat_\pm(u)$ as follows. First
one may notice that the integral kernel (\ref{Qxxz+}) 
simplifies for the special 
values $u= \pm \sigma$ thanks to the relation~(\ref{Ddel}). 
Explicitely, we have
\begin{align}\label{Qsp}
 \SQ_+^{\flat}(\sigma) &= 
    \bigl( w_b(2s{+}\fr{i}{2}Q) \bigr)^{-\SRN} \cdot {\sf 1} \,, \qquad
  \SQ_+^{\flat}(-\sigma) = \bigl( w_b(2s{+}\fr{i}{2}Q)\bigr)^{-\SRN}  
     \cdot \SU^{-1} \cdot \SOmega^\flat \,,
\end{align}
where $\SOmega^\flat$ is defined in~(\ref{Wb}). Combining this
observation with (\ref{UWP}), we conclude that
\begin{equation}\label{qspm} 
 q^+_t(\sigma) = \bigl( w_b(2s{+}\fr{i}{2}Q) \bigr)^{-\SRN} \,, \qquad
 q^+_t(-\sigma) = \pm  e^{2\pi i \frac{m}{\SRN}} \,
 \bigl( w_b(2s{+}\fr{i}{2}Q)\bigr)^{-\SRN} \,,
\end{equation}
where the $-$ sign in the second expression
can occur only in the sinh--Gordon model.

Secondly let us observe that
eqs.~(\ref{qqpm}) and \rf{qspm} imply that $u=\pm \bar{\sigma}$ 
are poles of order~$\SRN$ for~$q^-_t(u)$. Indeed, using (\ref{Qxxz-}), 
(\ref{Ddel}), and (\ref{wres}),  we find
\begin{align}\label{Qsm}
 \SQ_-^{\flat}(\bar{\sigma} + \epsilon)  \sim  
	\bigl( \fr{i}{2\pi\epsilon} \bigr)^\SRN \cdot  {\sf 1} \,, \qquad
 \SQ_-^{\flat}(-\bar{\sigma} + \epsilon)  \sim  
	\bigl( \fr{1}{2\pi i \epsilon} \bigr)^\SRN \cdot 
	\SOmega^\flat \cdot \SU \,,
\end{align}
as~$\epsilon\to 0$. This implies 
\begin{equation}\label{qmspm} 
 q^-_t(\bar{\sigma} + \epsilon) \sim 
   \bigl( \fr{i}{2\pi\epsilon} \bigr)^\SRN \,, \qquad
 q^-_t(-\bar{\sigma} + \epsilon) \sim 
  \pm  e^{2\pi i \frac{m}{\SRN}} \, 
  \bigl( \fr{1}{2\pi i \epsilon} \bigr)^\SRN \,,
\end{equation}
where the $-$ sign in the second expression 
can occur again only in the sinh--Gordon model.

\section{Separation of variables}\label{SOV} 

In the previous section we have identified necessary conditions  
for a function $t(u)$ to be an eigenvalue of $\ST(u)$. If we were 
able to show that these conditions are also {\em sufficient},
we would have arrived at a useful reformulation 
of the Auxiliary Spectral Problem.

A promising approach to this problem is offered by the 
separation of variables method pioneered by Sklyanin \cite{Sk2,Sk3}.
The basic idea is to introduce a representation for the 
Hilbert space $\CH$ of the model in which the off--diagonal
element of the monodromy matrix $\SM(u)$, the operator
$\SB(u)$, is diagonal. 

For simplicity of exposition let us temporarily restrict attention 
to the case of the sinh--Gordon model with $\SRN$ odd. The operator 
$\SB(u)$ has the following form:
\begin{equation}\label{SBexp}
\SB(u)\,=\,- i^{\SRN} \, e^{\SRN \pi b (u-s)} \sum_{m=0}^{\SRN} (-)^m
	e^{-2m\pi b u} \, \SB_{m}\,.
\end{equation}
By Lemma~\ref{poslem}, the operators $\SB_{m}$, $m=0,\dots,\SRN$
are positive self--adjoint. Basic for the separation of variables 
method is the validity of the following conjecture.

\begin{conj}\label{simplicity}
The joint spectrum of the family of operators
$\{\SB_{m};m=0,\dots,\SRN\}$ is simple. This means that 
eigenstates of $\SB(u)$ are  uniquely parameterized by the
corresponding eigenvalue $b(u)$.
\end{conj}

This conjecture can be supported by counting the degrees of freedom.
However, it is not easy to provide
a rigorous proof (see also Remark~\ref{Bspecrem} below).

The function $e^{-\pi b \SRN u}b(u)$ is a polynomial in the
variable $\la=e^{-2\pi b u}$. It can conveniently be represented
in the following form
\begin{equation}\label{bfactor}
b(u) \,= \,-(2i)^{\SRN}  \, e^{-\pi b \SRN s}\, 
\prod_{k=1}^{\SRN} \sinh\pi b(u-y_k)\,.
\end{equation}
The variables $y_k$, $k=1,\dots,\SRN$ are uniquely defined
up to permutations once we adopt the convention that ${\rm Im}(y_k)\in
(-\fr{1}{2b},\fr{1}{2b}].$ This means that the representation 
for the Hilbert space $\CH$ in which $\SB(u)$ is diagonal may be
described by wave--functions $\Psi(\by)$, $\by=(y_1^{},\dots,y_{\SRN}^{})$.
This representation for the vectors 
in $\CH$ will subsequently be referred to as the SOV representation.

We will then show that the Auxiliary Spectral Problem, 
$\ST(u)\Psi_t=t(u)\Psi_t$, gets transformed into the system of Baxter 
equations 
\begin{equation}\label{baxtereqn}
\begin{aligned}
{}& t(y_k) \, \Psi(\by) = \Big[\,\big(a(y_k)\big)^\SRN\,
 \ST_k^{-}+\big(d(y_k)\big)^\SRN\,\ST_k^{+}\,\Big]\,\Psi(\by)
 \,,\quad k =1,\dots,\SRN \,, 
\end{aligned}
\end{equation} 
where the operators $\ST_k$ are shift operators defined as
\begin{equation}\label{shiftdef}
\ST_k^\pm\Psi(\by)\,\equiv\,\Psi(y_1,\dots, y_k\pm ib,\dots,
 y_{\SRN}).
\end{equation}
The coefficients in front of the shift operators in (\ref{baxtereqn})
depend only on a single variable $y_k$, which 
is the crucial simplification that is gained by working
in the representation where $\SB(u)$ is diagonal. 

The key observation to be made at this point is that 
the {\em same} finite difference equation \rf{baxtereqn} was found 
in the previous Section~\ref{QB} in connection with the 
necessary conditions for a function $t(u)$ to represent a point
in the spectrum. It now remains to observe that {\em any} function
$q_t(u)$ that fulfills the necessary conditions \rf{BaxterEV},\rf{analQ}
can be used to construct
\begin{equation}\label{factorrep}
 \Psi_t(\by)\,=\, \prod_{k=1}^{\SRN}\,q_t(y_k) \,.
\end{equation}
The fact that \rf{factorrep} defines an 
eigenstate of $\ST(u)$ is verified by comparing \rf{baxtereqn}
with \rf{BaxterEV}. 
The main point that needs to be verified is whether the 
function $\Psi_t(\by)$ actually represents an element of
$\CH$, i.e. whether it has finite norm. The scalar product of vectors 
in $\CH$ can be represented in the form
\begin{equation}\label{SOVscprod}
\langle\,\Psi_\2\,|\,\Psi_\1\,\rangle\,=\,
\int_{\mathbb Y} d\mu(\by)\;
\langle\,\Psi_\2\,|\,\by\,\rangle\langle\,\by\,|\,\Psi_\1\,\rangle\,
\end{equation}
We clearly need to 
know both the range ${\mathbb Y}$ of values $\by$ that we need
to integrate over, as well as the measure $d\mu(\by)$ of 
integration to be used. We will be able to determine the
measure $d\mu(\by)$ provided that the following conjecture is true.

\begin{conj}\label{Bspec}
The functions $b(u)$ of the product form \rf{bfactor} that describe 
the spectrum of $\SB(u)$ have only real roots, i.e.,
$y_k\in\BR$\, for $k=1,\dots,\SRN$.
\end{conj}

We will discuss the status of this conjecture after
having explained its consequences.
Conjecture~\ref{Bspec} directly implies that ${\mathbb Y}=\BR^{\SRN}$ 
in~\rf{SOVscprod}. Assuming the validity of Conjecture~\ref{Bspec}, 
we will show in Proposition~\ref{Sklmeasure} that the measure 
$d\mu(\by)$ can be represented in the following explicit form
\begin{equation}\label{mexx'}
  d\mu(\by)\,=\, \prod_{k=1}^{\SRN}\,dy_k
 \prod_{l<k}\,4 \sinh\pi b(y_k-y_l)\, \sinh\pi b^{-1}(y_k-y_l)\,.
\end{equation}

Knowing explicitly how to represent the scalar product of $\CH$
in the SOV representation finally allows us to check that 
{\em any solution of the necessary conditions \rf{BaxterEV},\rf{analQ}
defines an eigenvector $|\,\Psi_t\,\ket$ of $\ST(u)$ via 
\rf{factorrep}.}
In other words: The conditions \rf{BaxterEV}, \rf{analQ} are not only
necessary but also {\em sufficient} for $t(u)$ to be an eigenvalue of 
a vector $|\,\Psi_t\,\ket\in\CH$.

\begin{rem}\label{Bspecrem}
Our claim that the conditions \rf{BaxterEV}, \rf{analQ} 
are also sufficient for a function $t(u)$ to represent a point
in the spectrum of $\ST(u)$ does not seem to depend very strongly on
the validity of the Conjecture~\ref{Bspec}. In this sense 
the conjecture mainly serves us to simplify the exposition. 

In any case it is a problem of fundamental importance for the
separation of variable method to determine the spectrum of 
$\SB(u)$ precisely. Even in simpler models which have been studied
along similar lines like the Toda chain \cite{KL} or the 
XXX spin chains \cite{DKM} there does not seem to exist a 
rigorous proof of the analogous statements. 
The explicit construction of the eigenfunctions of $\SB(u)$, 
which may proceed along similar lines as followed for the Toda 
chain in \cite{KL} or for the XXX chain in \cite{DKM}, should provide 
us with the basis for a future proof of Conjecture~\ref{Bspec} or 
some modification thereof.
\end{rem}

\begin{rem}
Within the Separation of Variables method the Auxiliary Spectral 
Problem gets transformed into the separated Baxter 
equations~\rf{baxtereqn}. However, these finite difference equations 
will generically have many solutions that do {\em not} 
correspond to eigenstates of~$\ST(u)$. 

In order to draw a useful analogy let us compare 
the situation with the spectral problem for a differential 
operator like $\sh=-\partial_y^2+V(y)$. One generically has two linearly 
independent solutions to the second order differential equation like 
$(-\partial_y^2+V(y))\psi={\cal E}\psi$ for {\em any} choice of~${\cal E}$.
The spectrum of $\sh$ is determined by restricting attention to the subset 
of square--integrable solutions within the set of all solutions to the 
eigenvalue equation. 

From this point of view we may identify the conditions \rf{analQ}  
on analyticity and asymptotics of the function $q_t(u)$ as the 
quantization conditions which single out the subset which constitutes 
the spectrum of $\ST(u)$ among the set of {\em all} 
solutions of~\rf{baxtereqn}.
\end{rem}

\subsection{Operator zeros of $\SB(u)$}

The adaption of Sklyanin's observation to the case at hand is based
on the following observations. First, by Lemma~\ref{poslem},
the operators $\SB_{m}$ introduced in \rf{SBexp} 
are {positive} self--adjoint. Taking into 
account the mutual commutativity (which follows from~\rf{ABD}),
\begin{equation}
{[}\,\SB_{m}\,,\,\SB_{n}\,{]}\,=\,0,
\end{equation}
leads us to conclude that the family of operators 
$\{\SB_m;m=1,\dots \SRN\}$ can be simultaneously
diagonalized\footnote{By Lemma~\ref{leadterm}, 
 we have $\SB_0=(\SB_\SRN)^{-1}$.}.

Conjecture~\ref{simplicity} implies that the spectral representation
for the family $\{\SB_m;m=1,\dots \SRN\}$ can be 
written in the form 
\begin{equation}\label{bspec}
 |\Psi\rangle\,=\, \int_{\BR_+^\SRN} 
   d\nu(\bb)\,|\,\bb\,\rangle\,\langle\,\bb\,|\,\Psi\,\rangle\,
\end{equation}
where $|\,\bb\,\rangle$ is a (generalized) eigenvector
of $\SB_m$ with eigenvalue $b_m$, and we have
assembled the eigenvalues into the vector $\bb=(b_1^{},\dots,b_{\SRN}^{})$.

It now turns out to be particularly useful to parameterize the
polynomial of eigenvalues $b(u)$ in terms of its roots. This 
representation may always be written as follows
\begin{equation}\label{b(u)exp}
b(u) \equiv b(u|\by)\equiv
	- (2i)^{\SRN} \,
	e^{- \pi b \SRN  s } \, 
	\prod\limits_{k=1}^{\SRN } \sinh\pi b(u-y_k) \,,
\end{equation}
where $\by=(y_1^{},\dots,y_\SRN^{})$. 
The variables $y_k$ are either real or they come in pairs
related by complex conjugation. The variables $y_k$ are
uniquely defined up to permutations
if one requires that ${\rm Im}(y_k)\in (-\frac{1}{2b},\frac{1}{2b}]$. 
We will assume that $y_k\in\BR$ according to Conjecture~\ref{Bspec}.
It then follows that the spectral representation 
\rf{bspec} can be rewritten as
\begin{equation}\label{bspec2}
|\Psi\rangle\,=\,
\int_{\BR^\SRN} d\mu(\by)\,|\,\by\,\rangle\,
	\langle\,\by\,|\,\Psi\,\rangle\,.
\end{equation}
However, points $\by$, $\by'$ in $\BR^\SRN$ which are obtained
from each other by the permutation $y_k\leftrightarrow y_l$ will
correspond to the same eigenstate of $\SB(u)$. This means that
the spectral representation for $\SB(u)$ can be used to 
define an isomorphism
\begin{equation}\label{CHiso}
\CH\,\simeq\,L^2(\BR^\SRN;d\mu)^{\rm Symm}\,,
\end{equation}
where $L^2(\BR^\SRN;d\mu)^{\rm Symm}$ is the subspace 
within $L^2(\BR^\SRN;d\mu)$ which consists of totally symmetric 
wave--functions.

Despite the fact that $\CH$ is isomorphic only to a subspace in 
$L^2(\BR^\SRN;d\mu)$ it will turn out to be useful to extend the 
definition of the operators $\SA(u)$, $\SB(u)$, $\SC(u)$, $\SD(u)$
{}from $L^2(\BR^\SRN;d\mu)^{\rm Symm}$, where it is canonically 
defined via \rf{CHiso} to $L^2(\BR^\SRN;d\mu)$. 
As a first step let us introduce the operators $\sy_k$ which act
as $\sy_k\,|\,\by\,\rangle=y_k\,|\,\by\,\rangle$.
Substituting $y_k\ra \sy_k$ in \rf{b(u)exp}
leads to a representation of the operator $\SB(u)$ in terms of its 
{\em operators zeros}~$\sy_k$.

\subsection{Operators $\SA(u)$ and $\SD(u)$}

Monodromy matrices of the modular XXZ magnet and the lattice
sinh--Gordon model satisfy the exchange relations (\ref{rLL}),
where the R--matrix is given by (\ref{R12}) and (\ref{Rlsg}), 
respectively. 
Among these relations we have, in particular,
the following
\begin{align}
\label{ABD}
{}& [\SB(u) , \SB(v)] = 0 \,, \qquad
 [\SA(u) , \SA(v) ]= 0 \,,\qquad
 [\SD(u) , \SD(v) ]= 0 \,, \\
\label{AD}
& \sinh \pi b (u-v+ib) \, \SB(u) \, \SA(v)  \\
\nonumber
& \qquad\qquad\quad = \sinh \pi b (u-v) \, \SA(v) \, \SB(u)
 + R^{\,\flat}_{23}(u-v) \, \SB(v) \, \SA(u)\,,\\
\label{BD}
& \sinh \pi b (u-v+ib) \, \SB(v)\SD(u) \\
\nonumber
& \qquad\qquad\quad =
 \sinh \pi b (u-v) \, \SD(u) \, \SB(v)
 + R^{\,\flat}_{32}(u-v) \, \SB(u) \, \SD(v) \,,\\[1ex]
\label{R23}
& R^\XXZ_{23}(u) = R^\XXZ_{32}(-u) = 
	i e^{\pi b u} \, \sin\gamma \,, \qquad
 R^\SG_{23}(u) = R^\SG_{32}(u) = 
	i \sin\gamma \,.
\end{align}

We are now going to show that there is an essentially unique
representation of the commutation relations \rf{ABD}--\rf{BD} 
on wave--functions $\Psi(\by)$ 
which is such that $\SB(u)$ is represented as operator of 
multiplication by $b(u)\equiv b(u|\by)$, cf. \rf{b(u)exp}.

To this aim let 
us consider for $k\geq 1$ the following distributions:
\[
\langle\by|\,\SA(y_k)\,\equiv\,
	\lim_{u\ra y_k}\langle\by|\,\SA(u)\,,
\qquad
\langle\by|\,\SD(y_k)\,\equiv\,
	\lim_{u\ra y_k}\langle\by|\,\SD(u)\,.
\]
These distributions, which will be defined
on suitable dense subspaces of $L^2(\BR^{\SRN},d\mu)$, 
can be regarded as 
the result of action on $\langle\by|$ by operators $\SA(\sy_k)$ 
and $\SD(\sy_k)$ with operator arguments substituted into 
(\ref{aexp}), (\ref{dexp}), (\ref{aaexp}), (\ref{ddexp}) from the left.
The commutation relations (\ref{AD})--(\ref{BD}) imply that 
$\langle\by|\,\SA(y_k)$ and $\langle\by|\,\SD(y_k)$
are eigenstates of $\SB(u)$ with eigenvalues $b(u|\by')$,
with $y_k'= y_k \mp ib$, respectively, $y'_l=y_l$ otherwise.
This, along with relations (\ref{ABD}), leads to the conclusion that 
the action of the operators $\SA(\sy_k)$ and $\SD(\sy_k)$ on 
wave--functions $\Psi(\by)=\langle\by|\Psi\rangle$ can be represented 
in the form
\begin{equation}\label{SADshift}
\SA(\sy_k)\Psi(\by)\,=\,a_{\SRN,k}(y_k)\,\ST_k^{-}\Psi(\by)\,,\qquad
\SD(\sy_k)\Psi(\by)\,=\,d_{\SRN,k}(y_k)\,\ST_k^{+}\Psi(\by)\,,
\end{equation}
where $\ST_k^\pm$ are the shift operators defined in equation 
 \rf{shiftdef}. The functions $a_{\SRN,k}(y_k)$ and $d_{\SRN,k}(y_k)$ 
are further restricted by the following identities:
\begin{align}
 \det\nolimits_q \SM^\SG(u) &\equiv
 \SA^\SG(u) \, \SD^\SG(u-ib) -  \SB^\SG(u) \, \SC^\SG(u-ib) \\
{} &= \bigl( 4 e^{-2\pi b s}\, \cosh \pi b( s + u - i \fr{b}{2}) \,
 \cosh \pi b( s - u + i \fr{b}{2})  \bigr)^\SRN \,.
\end{align}
These identities are proven in Appendix~\ref{qDet}. It follows
then from \rf{SADshift} that
$\det\nolimits_q \SM(y_k) =a_{\SRN,k}(y_k)d_{\SRN,k}(y_k-ib)$.
Not having specified the measure $\mu(\by)$ yet leaves us the
freedom to multiply all wave--functions $\Psi(\by)$ by 
functions of the form $\prod_{k} f_k(y_k)$. 
This allows us to choose 
\begin{align}\label{acoeff}
 & a_{\SRN,k}(y_k)= \bigl(a(y_k)\bigr)^\SRN \,, \qquad\qquad
 d_{\SRN,k}(y_k) = \bigl( d(y_k) \bigr)^\SRN \,,\\[1ex]
& \begin{aligned}
\text{where}\;\;a^\SG(u)=
 d^\SG(-u)&=e^{-\pi b s}\,2\cosh\pi b (u-s-i\fr{b}{2})\\
&= e^{-\pi b(u-i\frac{b}{2})}+
	\big(\fr{m\Delta}{4}\big)^2 e^{\pi  b(u-i\frac{b}{2})}\,.
\label{adcoeffs}\end{aligned}\end{align}
We have used that the sinh--Gordon parameters $m$, $\Delta$ are related 
to $s$ as in~(\ref{mds}). 

In the special case that $\Psi(\by)$ is 
an eigenfunction of the transfer--matrix $\ST(u)$ with eigenvalue 
$t(u)$ we get the Baxter equations \rf{baxtereqn} from
$\ST(\sy_k)=\SA(\sy_k)+\SD(\sy_k)$ and equations \rf{SADshift} and
\rf{acoeff}, as advertised.

It will be useful for us to have explicit formulae for $\SA(u)$
and $\SD(u)$ in terms of the operators $\sy_k$ and~$\ST_k^{\pm}$.
In the case of $\SRN$ odd we may use the following
formulae:
\begin{align}
\SA^\SG(u)&= 
\sum_{k=1}^{\SRN}\,\prod_{l\neq k}\,
\frac{\sinh\pi b(u-\sy_l)}{\sinh\pi b (\sy_k-\sy_l)}\,
 \bigl(a^\SG(\sy_k)\bigr)^\SRN\,\ST_k^{-}\, , \\ 
\SD^\SG(u)&= 
\sum_{k=1}^{\SRN}\,\prod_{l\neq k}\,
\frac{\sinh\pi b(u-\sy_l)}{\sinh\pi b (\sy_k-\sy_l)}\,
 \bigl(d^\SG(\sy_k)\bigr)^\SRN\,\ST_k^{+}\,.
\end{align}
These formulae are easily verified by noting
that the number of variables $\sy_k$ coincides
with the number of coefficients in the expansion (\ref{aaexp})
and (\ref{ddexp}). It follows that the polynomials 
$\SA^\SG(u)$ and $\SD^\SG(u)$ are uniquely determined by their values
$\SA^\SG(\sy_k)$ and $\SD^\SG(\sy_k)$, $k=1,\dots,\SRN$.

\subsection{Sklyanin measure}

We furthermore know that the 
operators $\SA_m$  and $\SD_m$ which are defined by the expansion
\begin{align}
\SA^\SG(u)&= -ie^{-\pi bu} i^{ \SRN} \,
	e^{ \pi b \SRN(u- s)}
	\sum_{m=0}^{\SRN -1} (-)^m 
	e^{-2m\pi b u} \, \SA_{m}\,,\\
\SD^\SG(u)&= -ie^{-\pi bu} i^{ \SRN} \,
	e^{ \pi b \SRN(u- s)}
	\sum_{m=0}^{\SRN -1} (-)^m 
	e^{-2m\pi b u} \, \SD_{m}\,
\end{align}
are positive (Lemma \ref{poslem} in Appendix~\ref{monoapp}).
\begin{propn}\label{Sklmeasure}
There exists a unique measure $d\mu(\by)$ such that the operators
$\SA_m$ and $\SD_m$ on
$L^2(\BR^\SRN;d\mu)$ are positive. This measure $d\mu$ can be
represented explicitly as
\begin{equation}\label{mexx}
 d\mu(\by)\,=\, \prod_{k=1}^{\SRN}\,dy_k \,
 \prod_{l<k}\,4 \sinh\pi b(y_k-y_l) \, \sinh\pi b^{-1}(y_k-y_l)\,.
\end{equation}
\end{propn}
\begin{proof}
The similarity transformation 
$\Psi(\by)= \chi_{\scriptscriptstyle\SA}(\by) \, \Phi(\by)$, where
\begin{equation}\label{similarA}
 \chi_{\scriptscriptstyle\SA}(\by) \,=\,\prod_{k=1}^{\SRN}\,
 \big(e^{\pi i y_ks} w_b(y_k-s)\big)^{-\SRN}
\prod_{l<k} \bigl(w_b(y_k-y_l+ \fr{i}{2}Q)\bigr)^{-1} \,,
\end{equation}
maps to a representation in which the operator $\SA^\SG(u)$
is represented as 
\begin{align}
\SA^{\SG}(u)\,=\,\sum_{k=1}^{\SRN}\,\prod_{l\neq k}\,
{\sinh\pi b(u-\sy_l)}\,
\ST_k^{-}\, .
\end{align}
Expanding in powers of $e^{\pi b u}$ yields a representation for the 
coefficients $\SA^{\SG}_{m}$ that appear in the 
expansion \rf{aexp} which takes the form
\begin{equation}\label{SDexp}
\SA^{\SG}_{m} \,=\,
	\sum_{k=1}^{\SRN}p_{mk}(\by)\ST_k^{-}\,,
\end{equation}
The coefficients $p_{mk}(\by)$ in \rf{SDexp} are positive for all 
$\by\in\BR^{\SRN}$, and $p_{mk}(\by)$ does not depend on~$y_k$.
We are next going to show that the positivity of $\SA^{\SG}_{m}$ 
implies that $\ST_k^{-}$ must be a positive operator in 
$L^2(\BR^\SRN;d\mu)$. Let us keep in mind that $\ST_k^{-}$ satisfies 
the commutation relations
\begin{equation}\label{Cweyl}
e^{-it\sy_l}\,\ST_k^{-}\,e^{it\sy_l}\,=\,e^{bt\de_{kl}}\,\ST_k^{-} \,.
\end{equation}
If there was any negative contribution to the expectation value
$\langle \Phi\,|\,\sum_{k=1}^{\SRN}p_{mk}(\by)
\ST_k^{-}\,|\,\Phi\rangle$ we could make it
arbitrarily large by means of the unitary transformation 
$|\Phi\rangle\ra e^{it\sy_k}|\Phi\rangle$. It follows that
$\left\langle \Phi\,|\,p_{mk}(\by)
\ST_k^{-}\,|\,\Phi\right\rangle>0$ for any $k=1,\dots,\SRN$.

It remains to notice that, since the $p_{mk}(\by)$ are non--vanishing, 
vectors of the form $\sqrt{p_{mk}(\by)}\,|\,\Phi\,\ket$ form a dense 
subset in $L^2(\BR^\SRN;d\mu)$. 
This finally allows us to conclude that $\ST_k^-$ must be 
a positive operator.
But this furthermore implies that $(\ST^{-}_k)^{il}$ is a unitary 
operator which satisfies the commutation relations 
\[ 
(\ST^{-}_k)^{il}\,e^{it\sy_m}\,=\,
\exp(iblt\de_{km})\,e^{it\sy_m}\,(\ST^{-}_k)^{il}\,.
\]
It is well--known that the representation of these commutation
relations by unitary operators is essentially unique. The measure which 
defines the corresponding Hilbert space is just 
$d\nu(\by)=\prod_{k=1}^{\SRN}dy_k$.

It remains to return in (\ref{bspec2}) to the original representation 
via~\rf{similarA}, that is to compute 
\hbox{$d\mu(\by)=|\chi_{\scriptscriptstyle\SA}(\by)|^2 \, d\nu(\by)$}.
Using \rf{wdual}--\rf{wcc} to simplify the resulting expression yields 
the formula for $d\mu(\by)$ stated in Proposition~\ref{Sklmeasure}.

For the operator $D^\SG(u)$, a completely analogous consideration 
applies with the transformation 
$\Psi(\by)= \chi_{\scriptscriptstyle\SD}(\by) \, \Phi(\by)$, where
\begin{equation}\label{similarD}
 \chi_{\scriptscriptstyle\SD}(\by) \,=\,\prod_{k=1}^{\SRN}\,
 \big(e^{-\pi i y_ks} w_b(y_k+s)\big)^{\SRN}
\prod_{l<k} \bigl(w_b(y_k-y_l+ \fr{i}{2}Q)\bigr)^{-1} \,.
\end{equation}
Thanks to the relation (\ref{wcc}) and the Conjecture~\ref{Bspec},
we have $|\chi_{\scriptscriptstyle\SA}(\by)|^2=
 |\chi_{\scriptscriptstyle\SD}(\by)|^2$. This leads to
the same measure $d\mu(\by)$ given by~\rf{mexx}.
\end{proof}

\subsection{Remaining cases}

To end this section let us briefly discuss the necessary modification 
in the cases of the modular XXZ magnet and the lattice sinh--Gordon 
model with even~$\SRN$. The main new feature that arises in these cases 
is the existence of a quasi--momentum $y_\0$ which first appears
in the expansions
\begin{align}
\label{opzero2a}
&b^\XXZ(u) = 2^{\SRN-1} \, i \, e^{\pi b (u + y_\0)}\, 
\prod_{k=1}^{\SRN-1} \sinh\pi b(u-y_k) \,,\\
\label{opzero2b}
&b^\SG(u) =        - (2i)^{\SRN -1} \,
	e^{\pi b ( y_\0 - \SRN  s) } \, 
	\prod\limits_{k=1}^{\SRN -1} \sinh\pi b(u-y_k) \,,
	\;\;\text{$\SRN$ -- even} \,,
\end{align}
The variable $y_\0$ requires a slightly different treatment compared 
to the $y_k$, $k\geq 1$.
Considering the states (where $\kappa=0$ for the modular 
magnet and \hbox{$\kappa=i/2+bs$} for the sinh--Gordon 
model with even~$\SRN$)
\begin{align*}
{}& \langle\by|\,\SA_{\0} \equiv
 \lim_{u\to +\infty} e^{\pi \SRN (\kappa -bu)} \, 
	\langle\by|\,\SA(u)\,,
 \\
{}&  \langle\by|\,\SD_{\0} \equiv 
  \lim_{u\to +\infty} e^{\pi \SRN (\kappa - bu)} \, 
	\langle\by|\,\SD(u) \,,\\
{}& \langle\by|\,\SA_{\SRN} \equiv 
 \lim_{u\to -\infty} (-)^\SRN e^{\pi \SRN (\kappa +bu)} \, 
	\langle\by|\,\SA(u)\,,\\
{}& \langle\by|\,\SD_{\SRN} \equiv 
  \lim_{u\to -\infty} (-)^\SRN e^{\pi \SRN (\kappa + bu)} \, 
	\langle\by|\,\SD(u) \,,	
\end{align*}
and taking into account the asymptotic behaviour of $\SB(u)$ and of
the coefficients (\ref{R23}) at \hbox{$u \to \pm \infty$}, we infer 
{}from the relations (\ref{AD})--(\ref{BD}) that
\begin{align*}
 \SA_{\0}\Psi(\by)\,=\,
	a_{\0}(y_\0)\,\ST_\0^{+}\Psi(\by)\,,\qquad
 \SD_{\0}\Psi(\by)\,=\,
	d_{\0}(y_\0)\,\ST_\0^{-}\Psi(\by)\,,\\
 \SA_{\SRN}\Psi(\by)\,=\,
	a_{\SRN}(y_\0)\,\ST_\0^{-}\Psi(\by)\,,\qquad
 \SD_{\SRN}\Psi(\by)\,=\,
	d_{\SRN}(y_\0)\,\ST_\0^{+}\Psi(\by)\,,
\end{align*}
where the shift operators $\ST_\0^\pm$ are defined analogously 
to \rf{shiftdef} for the variable~$y_\0$. It follows from 
(\ref{badinf}) and (\ref{eveninf}) that we have
$a_{\0}(y) = d_{\SRN}(y)$, $d_{\0}(y) = a_{\SRN}(y)$,
 $a_{\0}(y) a_{\SRN}(y+ib) = 
	d_{\0}(y) d_{\SRN}(y-ib) =1$. Noting that
\begin{align}
 \Psi(\by) &= \lim_{u\to \pm\infty}  {\bigl[} 
	e^{ \pi b \SRN (2\kappa \pm(i b - 2u))} \,
	{\rm det}_q  \SM(u){\bigr]}\,\Psi(\by)= \\
\nonumber
 &= \begin{cases}
 \SA_{\0}\SD_{\0} \Psi(\by)=a_{\0}(y_\0-ib)
 d_{\0}(y_\0)\Psi(\by), & u \to +\infty; \\ 
  \SA_{\SRN}\SD_{\SRN} \Psi(\by) = a_{\SRN}(y_\0+ib) 
 d_{\SRN}(y_\0)\Psi(\by), & u \to -\infty \,,
 \end{cases}
\end{align}
allows us to choose
\begin{align}
{}& \qquad\qquad  a_{\0} = d_{\0} = a_{\SRN} =
	d_{\SRN} = 1 \,.
\end{align}
The resulting equation for the $y_\0$--dependence may therefore 
be written as
\begin{equation}\label{y0baxter}
 \lim_{u\to \pm\infty}  {\bigl[} (\pm)^\SRN
	e^{ \pi \SRN (\kappa \mp bu)} \, t(u) {\bigr]}
 \,\Psi(\by) = \bigl( \ST_\0^{+}+\ST_\0^{-} \bigr) \, \Psi(\by) \,.
 \end{equation}
This relation supplements the Baxter equations \rf{baxtereqn} in 
the cases of the modular XXZ magnet and the sinh--Gordon model with 
even~$\SRN$.

In the case of the modular XXZ magnet we furthermore find a small 
modification in the form of the coefficient functions $a(u)$ 
and $d(u)$ which appear in the Baxter equations. These follow from
the following formula for the q--determinant (see Appendix~\ref{qDet}):
\begin{align}
 \det\nolimits_q \SM^\XXZ(u) &\equiv 
 A(u) \, D(u-ib) - q^{-1} \, B(u) \, C(u-ib) \\
{} &= \bigl( 4 \, \cosh \pi b( s + u - i \fr{b}{2}) \,
 \cosh \pi b( s - u + i \fr{b}{2})  \bigr)^\SRN \,, 
\end{align}
The resulting expressions for $a^{\XXZ}(u)$ and $d^{\XXZ}(u)$ will be
\begin{equation} a^{\XXZ}(u) \,=\,d^{\XXZ}(-u)\,=\,
-2i\cosh\pi b (u -s-i\fr{b}{2})\,.
\end{equation}

The existence of the ``zero mode'' $y_\0$ also leads to modifications 
in the formulae for $\SA(u)$ and $\SD(u)$.
For the modular magnet and the sinh--Gordon model with even $\SRN$, 
the number of coefficients in the expansion (\ref{aaexp}) and 
(\ref{ddexp}) exceeds by two the number of the operators~$\sy_k$.
However, in these cases we know the asymptotics of $\SA(u)$ and 
$\SD(u)$ and therefore we will need the following 
interpolation formula. 
\begin{lem}\label{interpol}
Let $e^{\SRN \pi b u}P(u)$ be a polynomial in $e^{2\pi b u}$
such that 
\[ 
 P(u)\sim  
	\begin{cases} e^{+\pi b (\SRN u +p_\0)}, & 
	\text{for $u\to +\infty$}\,,\\
	(-)^\SRN e^{-\pi b (\SRN u + p_\0)}, & 
 	\text{for $u\to -\infty$} \,.
	\end{cases}
\]
For an arbitrary set of variables $y_1,\dots,y_{\SRN-1}$ such that
$y_k\neq y_l$ for all $k\neq l$ we may then write 
\begin{equation*}
P(u)=
\sinh\pi b(u+p_\0+\rho_{\klein\SRN})\prod_{k=1}^{\SRN-1}
\sinh\pi b (u-y_k)+
\sum_{k=1}^{\SRN-1}\,\prod_{l\neq k}\,
\frac{\sinh\pi b(u-y_l)}{\sinh\pi b (y_k-y_l)}\,
P(y_k)\, ,
\end{equation*}
where $\rho_{\klein\SRN} \equiv\sum_{k=1}^{\SRN-1}y_k$.
\end{lem}

Thus, for the modular magnet and the sinh--Gordon model with even $\SRN$,
we have the following formulae for the operators $\SA(u)$ and 
$\SD(u)$ (recall that $\kappa=0$ for the modular magnet and 
\hbox{$\kappa=i/2+bs$} for the sinh--Gordon model with even~$\SRN$)
\begin{align}
\SA(u)&= 2^\SRN e^{-\pi \kappa \SRN} \, 
	\sinh\pi b(u+\spp_{\klein\SRN} + \rho_{\klein\SRN})
 \prod_{k=1}^{\SRN-1} \sinh\pi b (u-\sy_k)\label{Apsi}\\& \qquad +
\sum_{k=1}^{\SRN-1}\,\prod_{l\neq k}\,
\frac{\sinh\pi b(u-\sy_l)}{\sinh\pi b (\sy_k-\sy_l)}\,
\bigl(a(\sy_k)\bigr)^\SRN\,\ST_k^{-}\, ,
\nn \\ 
\SD(u)&= 2^\SRN e^{-\pi \kappa \SRN} \, 
	\sinh\pi b(u+\spp_{\klein\SRN}+\rho_{\klein\SRN})
 \prod_{k=1}^{\SRN-1} \sinh\pi b (u-\sy_k)\label{Dpsi}\\&\qquad+
\sum_{k=1}^{\SRN-1}\,\prod_{l\neq k}\,
\frac{\sinh\pi b(u-\sy_l)}{\sinh\pi b (\sy_k-\sy_l)}\,
\bigl(d(\sy_k)\bigr)^\SRN\,\ST_k^{+}\, ,
\nn 
\end{align}
where now ${\sf \rho}_{\klein\SRN}\equiv\sum_{k=1}^{\SRN-1} \sy_k$;
notice that $[\spp_{\klein\SRN},{\sf \rho}_{\klein\SRN}]=0$ and 
$[\spp_{\klein\SRN},\sy_\0]=\fr{i}{\pi b}$.
Furthermore, comparision with (\ref{badinf}) 
and (\ref{eveninf}) shows that
$\spp_{\klein\SRN}=-\sum_{k=1}^{\SRN} \spp_k$ for the modular magnet and 
$\spp_\SRN= \sum_{k=1}^{\SRN} (-)^k \spp_k$ for the sinh--Gordon model.


\section{Concluding remarks --- outlook}  

\subsection{On the Baxter equations}\label{Baxtersub}

Summarizing the results of Sections~\ref{QB} and~\ref{SOV}, we arrive 
at the main result of the present article. We will formulate it only 
for the case of the sinh--Gordon model with $\SRN$ odd for which
our analysis is most complete, but from our previous discussions
and remarks it seems clear that very similar results should
hold in the other cases as well.\\[1ex]
\noindent{\bf Main result:} {\em A function $t(u)$ is an eigenvalue 
of the transfer--matrix $\ST^{\SGo}(u)$ if and only if there exists 
a function $q_t(u)$ which satisfies the following conditions
\begin{equation}\label{summq} 
 \left[ \;\;
\begin{aligned}
{\rm (i)} \;\; & q_t(u)\;\,\text{is meromorphic in}\;\,\BC, 
 \,\;\text{with poles of maximal order $\SRN$ }
 \text{in}\;\,\Upsilon_{-s}\cup
\bar{\Upsilon}_{s},\\
{\rm (ii)} \;\; & q_t(u)\;{\sim} 
\left\{
\begin{aligned}
& \exp\big(+i \pi \SRN\, \sigma\, u - 
	i\fr{\pi }{2} \SRN u^2 \big)\;\;\;{\rm for}\;\;
|u|\ra\infty,\;\,\; |{\rm  arg}(u)|<\fr{\pi}{2} \,, \\
 & \exp\big(-i \pi \SRN\, \sigma\, u -
   i\fr{\pi }{2} \SRN u^2 \big)\;\;\;{\rm for}\;\;
|u|\ra\infty,\;\,\; |{\rm  arg}(u)|>\fr{\pi}{2} \,.
\end{aligned}\right.\\
{\rm (iii)} \;\; & t(u) \, q_t(u)\,=\,
	\bigl(\check{a}(u)\bigr)^{\SRN}q_t(u-ib)
	+\bigl(\check{d}(u)\bigr)^{\SRN}q_t(u+ib) \,,\\
 & \text{where}\;\; \check d(u) = \check a(-u) = 
 1 + (\fr{m \Delta}{4})^2 \, e^{ -\pi b (2u+ib)} \,, \\
{\rm (iv)} \;\; & q_t(u)\;\,\text{satisfies the following 
quantum Wronskian relation}\\
 & \qquad q_t(u+i\de_+) \, q_t(u-i\de_+) 
  -q_t(u+i\de_-) \, q_t(u-i\de_-)
\,=\,W_\SRN(u)\,,\\
& \text{where}\;\;
 W_\SRN(u)=  \,
 e^{-i\pi\SRN \left(u^2 +\sigma^2\right)}
 \bigl(D_{\sigma}(u)\bigr)^{-\SRN}\,.
\end{aligned}
\;\;\right]
\end{equation}
The corresponding eigenstate $\Psi_t$ in the SOV representation 
defined in Section~\ref{SOV} is represented as in~\rf{factorrep}.}\\[1ex]
We have therefore succeeded in reformulating the spectral problem for 
$\ST(u)$ as the problem to determine the set $\mathfrak S$ of solutions to 
the Baxter equation which possess the properties (i)-(iv) above.
 
It should be observed that conditions (i)-(iii) already constrain the 
possible functions $q_t(u)$ rather strongly.  
Let us consider 
\begin{equation}
Q_t(u)\,=\,
\Big(
\Gamma_b\big(\fr{Q}{2}-i(u+s)\big)
\Gamma_b\big(\fr{Q}{2}-i(u-s)\big)
\Big)^{-\SRN}
q_t(u)\,,
\end{equation}
where $\Gamma_b(x)\equiv \Gamma_2(x|b^{-1},b)$, with $\Gamma_2$ being 
the Barnes Double Gamma function defined in Appendix~\ref{DGF}.
The function $Q_t(u)$ will then have the properties
\begin{equation}\label{summQ} \left[ \;\;
\begin{aligned}
{\rm (a)} \;\; & Q_t(u)\;\,\text{is entire analytic of order 2 in}\;\,\BC,\\
{\rm (b)} \;\; & t(u) \, Q_t(u)\,=\,
	\bigl(A(u)\bigr)^{\SRN}Q_t(u-ib)
	+\bigl(D(u)\bigr)^{\SRN}Q_t(u+ib) \,,
\end{aligned}
\;\;\right]
\end{equation}
The explicit form of the
coefficients $A(u)$ and $D(u)$ can easily be figured out with the help 
of the formula \rf{baxxztil} and the functional relations~\rf{Ga2funrel}.

Property $\rm (a)$ combined with the Hadamard factorization 
theorem (see e.g. \cite{Ti}) imply that $Q_t(u)$ can be represented 
by a product representation of the form
\begin{equation}
Q_t(u)\,=\,e^{r(u)}\prod_{k=1}^{\infty} {'}  
\left(1-\frac{u}{u_k}\right)\,,
\end{equation}
where the prime indicates the canonical Weierstrass regularization of the
infinite product. The function $r(u)$ in the prefactor is a second order 
polynomial which can be worked out explicitly. The Baxter 
equation \rf{BaxterEV} then implies that the zeros $u_k$ must satisfy
an infinite set of equations,
\begin{equation}
 -1\,=\,\frac{(A(u_k))^{\SRN}}{(D(u_k))^{\SRN}} \,
\frac{Q_t(u_k-ib)}{Q_t(u_k+ib)},
\qquad k\in\BN \,,
\end{equation}
which may be regarded as a generalization of the Bethe ansatz equations.
However, as it stands it is not quite clear if these equations represent
an efficient starting point for the investigation of the spectrum 
of our models.

The quantum Wronskian relation (iv) encodes remarkable additional 
information which can not easily be extracted from the conditions
(i)-(iii) above. We plan to discuss its implications elsewhere.

\subsection{Continuum limit}\label{contSG}

It is certainly interesting to discuss the consistency of 
our results with existing results and conjectures on the 
sinh--Gordon model in continuous space--time. Let us therefore
now show that our findings are consistent with Lukyanov's 
remarkable conjecture \cite{Lu} on the ground state wave--function 
for the sinh--Gordon model in the SOV 
representation.

Recall from the Subsection~\ref{CL} that we are interested in the
limit $N\to\infty$, $\Delta\ra 0$, $s\ra\infty$ such that
$m=\fr{4}{\De} e^{-\pi b s}$ and $R=\SRN\De/2\pi$ are kept finite in 
the limit. We are interested in the limiting behavior of the 
Baxter equation and of its solutions. 
Let us first note that the poles of 
${q}_t(u)$ move out to infinity when $s\ra\infty$.
Also note that according to property 
\hbox{(\ref{summq}--{\rm ii})} rapid decay
is found only within the strip 
$\CS=\{u\in\BC;|{\rm Im}(u)|< Q/2\}$.
By noting that $m\De=\CO(1/N)$ in the limit under consideration
one sees that the coefficients $\check{a}(u)$, $\check{d}(u)$
in the Baxter equation \hbox{(\ref{summq}--{\rm iii})} become
unity when $\SRN\to\infty$.  
Most importantly, let us 
finally observe that the right hand side of the Wronskian 
relation \hbox{(\ref{summq}--{\rm iv})} approaches a
constant for $\SRN\to\infty$.

Our results therefore strongly suggest the following conjecture
on the conditions which characterize the 
spectrum of the continuum sinh--Gordon model in the SOV representation:
\begin{equation}\label{summqlim} 
\left[ \;\;
\begin{aligned}
{\rm (i)} \;\; & q_t(u)\;\,\text{is entire analytic},\\
{\rm (ii)} \;\; & q_t(u)\;\,\text{decays rapidly for}\;\,
|{\rm Re}(u)|\ra\infty,\;\,u\in\CS,\\
{\rm (iii)} \;\; & q_t(u)\;\,\text{satisfies a difference equation 
of the form}\\
 & \qquad t(u) \, q_t(u)\,=\,
	q_t(u-ib) +q_t(u+ib), \\
& \text{where}\;\,t(u)\;\,\text{is periodic under}\;\,u\ra u+ib^{-1},\\
{\rm (iv)} \;\; & q_t(u)\;\,\text{satisfies the following 
quantum Wronskian relation}\\
 & \;\; q_t(u+i\de_+) \, q_t(u-i\de_+)-q_t(u+i\de_-) \, q_t(u-i\de_-)
\,=\,1\,.
\end{aligned}\;\;\right]
\end{equation}

Our next aim will be to show that, by adding one supplementary
condition, one gets a complete characterization of the function 
$q_0^{}(u)$ which was proposed in \cite{Lu} to describe the 
ground state for the continuum sinh--Gordon model in the SOV 
representation:
\begin{equation}\label{summqlim2} {\rm (v)} \qquad
q_0(u)\;\,\text{is nonvanishing within}\;\, \CS\,.
\end{equation}
We claim that the solution to conditions (i)--(v) is essentially
unique and given by the formula
\begin{equation}\label{Yq}
\log q_0(u)=-\frac{mR}{2}\frac{\cosh\frac{\pi}{Q}u}{\sin\frac{\pi}{Q}b}+
\int\limits_{\BR}\frac{dv}{2Q}\;
\frac{\log(1+Y(v))}{\cosh\frac{\pi}{Q}(u-v)},
\end{equation}
which expresses $q_0(u)$ 
in terms of the solution $Y(u)$ to the nonlinear integral equation 
\begin{equation}\label{TBA}
\log Y(u)=-mR\cosh\fr{\pi}{Q}u+
\int\limits_{\BR}\frac{dv}{2Q}\, S(u-v) \,{\log(1+Y(v))} ,
\end{equation}
where the kernel $S(u-v)$ is explicitly given as follows:
\begin{equation}
S(u)=\frac{2 \, \sin\frac{\pi}{Q}b \, \cosh\frac{\pi}{Q}u}
{\sinh\frac{\pi}{Q}(u+ib)\sinh\frac{\pi}{Q}(u-ib)}\,.
\end{equation}
These equations form the basis for the calculation of the ground state 
energy \cite{Za} and other local conserved quantities of the continuum 
sinh--Gordon model \cite{Lu} within the thermodynamic 
Bethe ansatz framework.

For the reader's convenience we will present the 
outline of an argument\footnote{This argument is inspired by the 
considerations in \cite{Za} and a
suggestion of F.~Smirnov (private communication, see also \cite{Sm2}).
However, the key point of our argument, namely the origin of the
quantum Wronskian relation (iv) seems to be new.}  
which establishes the equivalence between (i)-(v) 
and \rf{Yq}, \rf{TBA}.  
Let us define an auxilliary function $Y(u)$ by the 
formula
\begin{equation}\label{yqq}
1+Y(u)\,=\,q_0^{}(u+i\de_+) \, q_0^{}(u-i\de_+)\,.
\end{equation}
Assuming that $q_0(u)$ satisfies the properties (i), (ii), and (v),
one can take the logarithm of \rf{yqq} and then solve the
resulting difference equation by Fourier 
transform, which leads to the representation~\rf{Yq}.  
Re--inserting this representation into the Wronskian relation
(iv) shows that $Y(u)$ must satisfy~\rf{TBA}.

A proof of Lukyanov's conjecture \cite{Lu} 
therefore amounts to showing that $q_0(u)$ must satisfy the
properties (i)--(v). We find it very encouraging that our study
of the lattice sinh--Gordon model gave us strong support for
the necessity of properties (i)--(iv).

\subsection{Connection with lattice Liouville model}

Relations between the compact XXZ chain, lattice sine--Gordon model,
and the (imaginary field) Liouville model was investigated 
in~\cite{FT} from the view point of the QISM. Let us show that similar 
connections exist between the modular XXZ magent, lattice sinh--Gordon 
model, and the (real field) Liouville model. 
Following~\cite{FT}, we introduce the L--matrix
\begin{equation}\label{LLi}
 L^{\zeta} (u) = 
  e^{-\frac{1}{2}\pi b \zeta \sigma_3} \, L^{\SG} (u+\zeta)
 \, e^{\frac{1}{2} \pi b \zeta \sigma_3} \,,
\end{equation}
where $\zeta$ is related to the representation parameter $s$ and 
the lattice spacing $\Delta$ as 
$e^{\pi b \zeta} = \Delta \, e^{\pi b s}$, and the operators $\spp$ 
and $\sx$ in (\ref{lSG}) are related to the discretized field and its 
conjugate momentum as follows
\begin{equation}\label{pxz}
  2 \pi b \, \spp =  2\pi b \, \zeta -\beta \, \Phi  \,, \qquad
  4 \pi b \, \sx = 2\pi b \, \zeta + 
	\beta \, (\fr{1}{2} \Pi -\Phi) \,, \qquad
  \beta=b \sqrt{8\pi} \,.
\end{equation}
Comparison with \rf{px} and \rf{mds} shows that the new variables
defined by \rf{pxz} are related to these of the sinh--Gordon model 
via a canonical transformation,
\begin{equation}\label{pppp}
  \Pi =  \Pi^{\SG}  \,, \qquad
  \beta \, \Phi =  \beta \, \Phi^{\SG} + 2\pi b \zeta \,, \qquad
  e^{-\pi b \zeta} =\fr{m}{4} \,.
\end{equation}
In the $\zeta\to+\infty$ limit  L--matrix (\ref{LLi}) turns into 
\begin{equation}\label{LL}
 L^{\rm\klein L}(u)  \equiv \lim_{\zeta\to+\infty} L^{\zeta} (u) =
 \left( \begin{array}{cc}  
  e^{\frac{\beta}{8}\, \Pi_n} \,
	\bigl( 1 +\Delta^2  \, e^{-\beta \Phi_n} \bigr) \,
	e^{\frac{\beta}{8}\, \Pi_n} & 
  -i \Delta\, e^{\pi b u - \frac{\beta}{2} \Phi_n }  \\  
  -2i \Delta\, \sinh \bigl(\pi b u + \frac{\beta}{2} \Phi_n \bigr) &
   e^{- \frac{\beta}{4} \, \Pi_n} 
 \end{array} \right) \,.
\end{equation}
It is natural to expect that this L--matrix describes some massless 
limit of the sinh--Gordon model. The  corresponding \hbox{U--matrix} 
obtained according to formula (\ref{LU}),
\begin{equation}\label{UccL}
  U^{\rm\klein L}(u) =   \left( \begin{array}{cc}  
  \fr{\beta}{4} \, \Pi(x)  & 
  -i \, e^{u - \frac{\beta}{2} \Phi(x) }  \\  
  -2i \, \sinh \bigl(u + \frac{\beta}{2} \Phi(x) \bigr)  &
  - \fr{\beta}{4} \, \Pi(x)
 \end{array} \right)\,,
\end{equation}
reproduces the Liouville equations of motion via the zero curvature 
equation (see \cite{FT} for details in the case of sine--Gordon model 
and imaginary Liouville field). This observation suggests that (\ref{LL}) 
is a suitable L--matrix  for describing the quantum lattice Liouville 
model in the QISM framework.

Although the limiting procedure in (\ref{LL}) has not been 
mathematically rigorously developed yet (in particular, there is 
a subtle question of interchangibility of $\zeta\to +\infty$ limit with 
the classical limit), the results of the present article provide further 
support for the proposed connection between the sinh--Gordon and Liouville
models. First, observe that the twist by 
$e^{\frac{1}{2}\pi b \zeta \sigma_3}$ and the shift of the spectral
parameter in \rf{LLi} do not change the corresponding auxiliary
R--matrix (\ref{Rlsg}) and the fundamental R--matrix given in
Proposition~\ref{Rfsg}. Therefore, the local lattice density of the 
classical Hamiltonian corresponding to \rf{LLi} can be obtained by 
substituting (\ref{pppp}) into~(\ref{Hcl2}). Taking then the limit 
$\zeta\to +\infty$, we obtain the following lattice Hamiltonian 
density (up to an additive constant)
\begin{align}\label{Cclcont2}
 H^{\rm {\klein L}, cl}_{n,n+1} &\equiv
 \lim_{\zeta\rightarrow\infty}  H^{\rm \SG, cl}_{n,n+1} 
 = \fr{1}{\gamma}\log \Bigl( 
 \fr{1}{2}\cosh \fr{\beta}{4} \, (\Pi_n + \Pi_{n+1}) 
 + \fr{1}{2} \,
 \cosh \fr{\beta}{2} \, (\Phi_n - \Phi_{n+1}) \\
\nonumber
 &+ \fr{\Delta^2}{2} \, e^{-\frac{\beta}{2}(\Phi_n + \Phi_{n+1})} \,
 \bigl( 1 + e^{\frac{\beta}{4}(\Pi_n + \Pi_{n+1})} \,
 \cosh \fr{\beta}{2} ( \Phi_n - \Phi_{n+1} ) \bigr) \Bigr)\,,
\end{align}
which in the continuum limit (\ref{cont}) yields the 
Liouville Hamiltonian:
\begin{equation}
 \sum_n \fr{1}{\Delta} 
 H^{\rm \scriptscriptstyle L, cl}_{n,n+1}
 \rightarrow  {\rm const} + \int_0^{2\pi R} dx \, 
 \bigl( \fr{1}{2} \, \Pi^2 + \fr{1}{2} \,
 (\partial_x\Phi)^2
 + \fr{1}{\gamma} \, e^{-\beta \,\Phi} \bigr)  \,.
\end{equation}

The second observation that we can make to support the
proposed relationship between the lattice sinh--Gordon and
Liouville models is the following. The transfer--matrix
corresponding to \rf{LLi} is given by 
$\ST^\zeta(u)=\ST^\SG(u +\zeta)$. Therefore, 
$\SQ_\pm^\zeta(u) \equiv \check\SQ_\pm^\SG(u+\zeta)$ 
satisfy (cf.~\rf{baa2}) the Baxter equation
\begin{equation}\label{baz}
\begin{aligned}
{}&  \ST^\zeta(u) \cdot \SQ_\pm^\zeta(u) =
 \bigl( a_\zeta(u) \bigr)^\SRN \, \SQ_\pm^\zeta(u - ib) +
 \bigl( d_\zeta(u) \bigr)^\SRN \,  
  \SQ_\pm^\zeta(u + ib) \,,\\[2pt]
{}& \text{where}\quad  a_\zeta(u) = 
 1 + \Delta^2 \, e^{ \pi b (2u-ib)} \,, \quad d_\zeta(u) =  
 1 + \Delta^2 \, e^{ -\pi b (2u+4\zeta+ib)} \,.
\end{aligned}
\end{equation}
Hence for 
$\ST^{\rm\klein L}(u) \equiv \lim_{\zeta\to +\infty} \ST^\zeta(u)$ and
$\SQ_\pm^{\rm\klein L}(u) \equiv \lim_{\zeta\to +\infty} \SQ_\pm^\zeta(u)$
(the limits are meaningful if $\ST^\zeta(u)$ and $\SQ_\pm^\zeta(u)$ 
are expressed in terms of $\Phi$ and $\Pi$), we obtain
the Baxter equation
\begin{equation}\label{baL}
  \ST^{\rm\klein L}(u) \cdot \SQ_\pm^{\rm\klein L}(u) =
 \bigl( 1 + \Delta^2 \, e^{ \pi b (2u-ib)} \bigr)^\SRN \, 
 \SQ_\pm^{\rm\klein L}(u - ib) + \SQ_\pm^{\rm\klein L}(u + ib) \,,
\end{equation}
which coincides\footnote{modulo notations, in particular,
 $\Delta=1$ in \cite{K1,FKV}} with the Baxter equation derived 
for the lattice Liouville model by a different method in~\cite{K1,FKV}.

We finish by noting that in the continuum limit, $N\to \infty$,
$\Delta = \CO(1/N)$, the coefficient 
$\bigl( 1 + \Delta^2 \, e^{ \pi b (2u-ib)} \bigr)^\SRN$ becomes unity. This 
suggests that the Baxter equation for the eigenvalue $q^{\rm\klein L}_t(u)$ 
of the Q--operator for the continuum Liouville model is
\begin{equation}\label{baLc}
 t^{\rm\klein L}(u) \, q^{\rm\klein L}_t(u)\,=\, q^{\rm\klein L}_t(u-ib) +
	q^{\rm\klein L}_t(u+ib) \,, 
 	\qquad t^{\rm\klein L}(u+ib^{-1}) = t^{\rm\klein L}(u) \,,
\end{equation}
which coincides with that for the continuum sinh--Gordon 
model~(\ref{summqlim}--iii). 

However, it seems to be crucial to observe  that  the asymptotic
 properties of the function 
$q^{\rm\klein L}_t(u)$ will certainly differ from those found in the
case of the sinh--Gordon model. 
Indeed, for any model, the asymptotic properties of 
$q_t(u)$ are related to these of~$t(u)$. Comparing the structure
of the L--matrices \rf{lSG} and \rf{LL}, we see that the transfer--matrix
$\ST^{\rm\klein L}(u)$ corresponding to the latter L--matrix has
asymmetric asymptotics for ${\rm Re}(u)\ra\pm\infty$.
This seems to be related to the fact that the sinh--Gordon potential
$\cosh \beta \Phi(x)$ is spatially symmetric while the Liouville
potential $e^{-\beta \Phi(x)}$ is asymmetric.
As a consequence, we expect that the set of 
solutions to the Baxter equation \rf{baLc} describing the spectrum of the 
continuum Liouville model will be quite different from the set of 
solutions to the same Baxter equation which
characterizes the spectrum of the
continuum sinh--Gordon model. 

These observations seem to offer a key to the understanding of the
relation between massive and massless theories from the point of view
of their integrable structure.


\appendix

\renewcommand{\theequation}{\Alph{section}.\arabic{equation}}
\newcommand{\slabel}[1]{\label{#1}\setcounter{equation}{0}}

\section{Special functions}\slabel{Qdil}

\subsection{Double Gamma function}\label{DGF}

All the special functions that we have to deal with can be
obtained from the Barnes Double Gamma function 
$\Gamma_2(x|\omega_\1,\omega_\2)$ \cite{Ba},
which may be defined by
\begin{equation}\label{dgamma}
 \log \Gamma_2 (x|\omega_\1,\omega_\2) = 
 \Biggl( \frac{\partial}{\partial t} \sum_{n_\1,n_\2=0}^\infty
 (x + n_\1 \, \omega_\1 + n_\2 \, \omega_\2)^{-t} \Biggr)_{t=0} \,.
\end{equation}
The infinite sum in \rf{dgamma} is defined by analytic continuation
from its domain of convergence (${\rm Re}(t)>2$) to the point of 
interest ($t=0$). One may alternatively use the integral representation
\begin{equation}
\log\Gamma_{2}(x|\omega_{\1},\omega_{\2})=\frac{C}{2}
 B_{2,2}(x|\omega_{\1},\omega_{\2})
 +\frac1{2\pi i}\int_{\mathcal{C}}\frac{e^{-xt}\log(-t)}{\left(
1-e^{-\omega_{\1}t}\right)  \left(  1-e^{-\omega_{\2}t}\right)  }\frac
{dt}t\label{cint}%
\end{equation}
where the contour $\mathcal{C}$ goes from $+\infty$ to $+\infty$ encircling
$0$ counterclockwise, $C$ is the Euler's constant and
\begin{equation}
B_{2,2}(x|\omega_{\1},\omega_{\2})=\frac{(2x-\omega_{\1}-\omega_{\2})^{2}}%
{4\omega_{\1}\omega_{\2}}-\frac{\omega_{\1}^{2}+\omega_{\2}^{2}}
{12\omega_{\1}\omega_{\2}}\label{B2}%
\end{equation}
The integral is well defined if ${\rm Re}(\omega_{\1})>0,$
${\rm Re}(\omega_{\2})>0$, and ${\rm Re}(x)>0$.
It satisfies the basic functional relations
\begin{equation}\label{Ga2funrel}
\frac{\Gamma_2 (x+\omega_\1|\omega_\1,\omega_\2)}%
{\Gamma_2 (x|\omega_\1,\omega_\2)}
=\sqrt{2\pi}
\frac{\omega_\2^{\frac{1}{2}-\frac{x}{\omega_\2}}}{\Gamma(x/\omega_\2)},
\qquad 
\frac{\Gamma_2 (x+\omega_\2|\omega_\1,\omega_\2)}%
{\Gamma_2 (x|\omega_\1,\omega_\2)}
=\sqrt{2\pi}
\frac{\omega_\1^{\frac{1}{2}-\frac{x}{\omega_\1}}}{\Gamma(x/\omega_\1)}\,.
\end{equation}
$\big(\Gamma_2 (x|\omega_\1,\omega_\2))^{-1}$ is an 
entire analytic function of order 2 w.r.t. its variable $x$
with  simple zeros at $x=-m\omega_{\1}-n\omega_{\2}$, where $m$ and
$n$ are non--negative integers.

\subsection{Function $w_b(x)$}\label{Qdil1}

In what follows we will be dealing with 
\begin{equation}\label{Gb}
 w_b(x) \equiv 
 \frac{\Gamma_2(\fr{Q}{2} - ix|b^{-1},b)}%
 {\Gamma_2(\fr{Q}{2} +i x|b^{-1},b)} \,.
\end{equation}
In the strip $|{\rm Im}(x)| < \fr{Q}{2}$, function $w_b(x)$ has the 
following integral representation
\begin{equation}\label{wint}
 w_b(x)= 
 \exp \Biggl\{ \frac{i \pi}{2} x^2 \+ 
      \frac{i \pi}{24}(b^2 \+ b^{-2}) -
 \int\limits_{\BR+i0} \frac{dt}{4\, t} \,
 \frac{ e^{-2i t x}}{\sinh b t \, \sinh{\frac{t}{b}} } \Biggr\} \,,
\end{equation}
where the integration contour goes around the pole $t=0$ in the 
upper half--plane. 
This function is closely related (cf.~Eq.~(\ref{gw})) to the 
remarkable special 
function introduced under the name of {\em quantum dilogarithm}
in \cite{FK2} and studied in the context of quantum groups
and integrable models in \cite{F2,Ru,Wo,PT2,K1,K2,BT,T2,V2}.

Analytic continuation of $w_b(x)$ to the
entire complex plane is a meromorphic function with the
following properties 
\begin{align}
\text{self--duality} \label{wdual}  \quad& 
 w_b(x)=w_{b^{-1}}(x) \,,  \\[0.5mm]
\text{functional equation}  \label{wfunrel} \quad&
 \frac{w_b(x + \fr{i}{2}b^{\pm 1})}{w_b(x - \fr{i}{2}b^{\pm 1})} = 
 2 \, \cosh (\pi b^{\pm 1} x) \,, \\[0.5mm]
\text{reflection property} \label{wrefl} \quad&
 w_b(x) \; w_b(-x) = 1  \,, \\[0.5mm]
  \label{wcc} \text{complex conjugation} \quad&
  \overline{w_b(x)} = w_b(-\bar{x}) \,,\\[0.5mm]
 \label{wan} 
 \text{zeros\,/\,poles} \quad& 
 (w_b(x))^{\ppm 1} = 0 \ \Leftrightarrow  
  \pm x \in  \big\{ i\fr{Q}{2}{+}nb{+}mb^{-1};n,m\in\BZ^{\geq 0}\big\}   
	\,, \\[0.5mm]
 \label{wres}
 \text{residue} \quad& 
  \Res_{x=-i\frac{Q}{2}} w_b(x)=\frac{i}{2\pi}
 \,, \\[0.5mm]
 \text{asymptotics} \quad& w_b(x) \sim 
\left\{
\begin{aligned}
& e^{- \frac{i \pi}{2}(x^2 + \frac{1}{12}(b^2+b^{-2}))}\;\;{\rm for}\;\,
|x|\ra\infty,\;\, |{\rm  arg}(x)|<\fr{\pi}{2} \,, \\
 & e^{+ \frac{i \pi}{2}(x^2 + \frac{1}{12}(b^2+b^{-2}))}\;\;{\rm for}\;\,
|x|\ra\infty,\;\, |{\rm  arg}(x)|>\fr{\pi}{2} \,.
\end{aligned}\right.
\end{align}
Notice that $|w_b(x)|=1$ if $x\in\BR$. Therefore, $w_b(\SO)$ is
unitary if $\SO$ is a self--adjoint operator.

The function $w_b(x)$ allows us to define a whole class of 
new special functions. In Appendix~\ref{proofSH} we will use in 
particular the following b--analogues of the hypergeometric 
functions defined by
\begin{equation}\label{Phidef}
\Phi^{}_r(U_1\dots U_r;V_1 \dots V_r;x)\;\equiv\;
\frac{1}{i}
\int\limits_{i\BR-0}d\tau\;e^{\pi\tau x}\,\prod_{k=1}^{r}
\frac{S_b(U_k+\tau)}{S_b(V_k+\tau)},
\end{equation}
where the special function $S_b(x)$ is defined by
\begin{equation}\label{Sbx}
 S_b(x) = w_b(ix -\fr{i}{2}Q) 
\end{equation}
and has the properties
\begin{align}
\label{Sb1}
\text{self--duality} \quad&  
 S_b(x) = S_{b^{-1}}(x) \,, \\
\label{Sb2}
\text{functional equation} \quad&
 S_b(x \+ b^{\ppm 1}) = 2 \, \sin (\pi b^{\ppm 1} x) \, S_b(x) \,,\\
\label{Sb3}
\text{reflection property} \quad&
  S_b(x) \, S_b(Q-x) = 1 \,.
\end{align}

\subsection{Function $D_\alpha(x)$}\label{Qdil2}

Let us also introduce another useful function
\begin{equation}\label{Ds}
  D_\alpha(x) = \frac{ w_b(x+\alpha)}{w_b(x-\alpha)} \,.
\end{equation}
Combining (\ref{wint}) with (\ref{Dpar}), we derive the integral 
representation
\begin{equation}\label{Dint}
 D_\alpha(x)= 
 \exp \Biggl\{ i 
 \int\limits_{\BR+i0} \frac{dt}{2t} \,
 \frac{ \cos(2 t x) \, \sin(2 \alpha t)}%
 {\sinh b t \, \sinh{\frac{t}{b}} } \Biggr\} \,.
\end{equation}
$D_{\al}(x)$ is a meromorphic function with zeros at 
$\ppm x \in \Upsilon_{-\alpha}$ and poles at 
$\ppm x \in \Upsilon_{\alpha}$, where the set $\Upsilon_\alpha$ 
is defined in~(\ref{ups}).
The function $D_\alpha(x)$ is self--dual in $b$ (but we will 
omit this index) and has the following properties
\begin{align}
\text{functional equation}  \label{Dfunrel} \quad&
 \frac{D_\alpha(x + \fr{i}{2}b^{\pm 1})}%
	{D_\alpha(x - \fr{i}{2}b^{\pm 1})} = 
 \frac{\cosh \pi b^{\pm 1} (x+\alpha)}{\cosh \pi b^{\pm 1} (x-\alpha)} 
	\,, \\ 
\text{$x$--parity} \label{Dpar} \quad&
 D_\alpha(x) = D_\alpha(-x)  \,, \\[0.5mm]
\text{reflection property} \label{Drefl} \quad&
 D_\alpha(x) \; D_{-\alpha}(x) = 1  \,, \\ 
  \label{Dcc} \text{complex conjugation} \quad&
  \overline{D_\alpha(x)} = D_{-\bar\alpha}(\bar{x}) \,,\\[0.5mm]
\label{Das1}
 \text{$x$--asymptotics} \quad D_\alpha(x) \sim & 
 \left\{
\begin{aligned} e^{- 2 \pi i \alpha x} \;\;{\rm for}\;\,
|x|\ra\infty,\;\, |{\rm  arg}(x)|<\fr{\pi}{2} \,, \\
e^{+ 2 \pi i \alpha x} \;\;{\rm for}\;\,
|x|\ra\infty,\;\, |{\rm  arg}(x)|>\fr{\pi}{2} \,,
\end{aligned}\right. \\ 
\label{Das2}
 \text{$\alpha$--asymptotics} \quad D_\alpha(x) \sim & 
 \left\{
\begin{aligned} & e^{- i \pi(x^2 + \alpha^2 + \frac{1}{12}(b^2+b^{-2}))} 
\;\,{\rm if}\;
|\al|\ra\infty,\;\, |{\rm  arg}(\al)|<\fr{\pi}{2} \,, \\
& e^{+ i \pi(x^2 + \alpha^2 + \frac{1}{12}(b^2+b^{-2}))} 
\;\,{\rm if}\;
|\al|\ra\infty,\;\, |{\rm  arg}(\al)|>\fr{\pi}{2} \,.
\end{aligned}\right.
\end{align}
Also, the following identity is obvious from the definition~(\ref{Ds})
\begin{equation}\label{Dsp2}
 D_\alpha(x) \, D_\beta(y) = 
 D_{\frac{\alpha+\beta+x-y}{2}}\bigl(\fr{x+y+\alpha-\beta}{2}\bigr) \, 
 D_{\frac{\alpha+\beta-x+y}{2}}\bigl(\fr{x+y-\alpha+\beta}{2}\bigr) \,.
\end{equation}
Notice that $|D_\alpha(x)|=1$ if $\alpha\in\BR$ and $x\in \BR$
or $x\in i\BR$. Therefore, 
$D_\alpha(\SO)$ is unitary if $\alpha\in\BR$ and 
$\SO$ is a self--adjoint or anti--self--adjoint operator.

\subsection{Integral identities for $D_\alpha(x)$}\label{Qdil3}

Here we will give some integral identities involving
products of D--functions. These identities can be regarded as
summation formulae for the b--hypergeometric 
functions~$\Phi^{}_r$ introduced in~(\ref{Phidef}).

Let us denote $\alpha^\star\equiv -\fr{i}{2}Q -\alpha$ and 
introduce the function
\begin{equation}\label{Adef}
A(\alpha_1,\alpha_2,\ldots) = w_b(\alpha_1 {-} \alpha^\star_1) \,
 w_b(\alpha_2 {-} \alpha^\star_2) \,  \ldots \ .
\end{equation}
Notice that $(\alpha^\star)^\star =\alpha$ and hence
$A(\alpha^\star_1,\alpha^\star_2,,\ldots) \,
 A(\alpha_1,\alpha_2,,\ldots) = 1$.

Lemma~15 in \cite{PT2} and Eqs.~(26)--(27) in \cite{FKV} can be 
rewritten as the following property of the function 
$D_\alpha(x)$ under the Fourier transform:
\begin{equation}\label{D1}
 \int_\BR dx\, e^{2\pi i x y} \, D_\alpha(x) = 
  A(\alpha) \, D_{\alpha^\star}(y) \,.
\end{equation}
Taking into account that $\lim_{\alpha\to 0} D_{\alpha}(x) =1$,
we obtain from (\ref{D1}) as a special case
\begin{equation}\label{Ddel}
 \lim_{\alpha\to -\frac{i}{2}Q} 
 \,A(\alpha^\star) \, D_{\alpha}(x) = \delta(x) \,.
\end{equation}
Here $\delta(x)$ on the r.h.s.\ is the Dirac delta--function 
and this relation should be understood in the sense of distributions.
Indeed $\lim_{\alpha\to -\frac{i}{2}Q}  \,A(\alpha^\star)
 = w_b(\frac{i}{2}Q)=0$ and the l.h.s.\ of (\ref{Ddel})
vanishes almost everywhere. On the other hand, the
only double pole of $D_{-\frac{i}{2}Q}(x)$ is at $x=0$.

Using (\ref{D1}), it is easy to derive the following relation
\begin{equation}\label{D2}\begin{aligned}
 \int_\BR dx\, e^{2\pi i z x} & \, D_\alpha(x-u) \, D_\beta(x-v) = \\
 &=  A(\alpha,\beta) \, e^{\pi i z (u+v)} \,
 \int_\BR dy\, e^{2\pi i y(u-v)} \, 
  D_{\alpha^\star}(y+\fr{z}{2}) \, D_{\beta^\star}(y-\fr{z}{2}) \,.
\end{aligned}\end{equation}
Choosing $z={\alpha^\star}+{\beta^\star}$, we can use (\ref{Dsp2})
in order to rewrite the product of $D$'s in the integrand on the 
r.h.s.\ as a single function, 
$D_{\alpha^\star + \beta^\star}(y+\frac{\alpha^\star - \beta^\star}{2})$,
and then apply~(\ref{D1}). This yields
\begin{equation}\label{D21}\begin{aligned}
 \int_\BR dx\, & e^{2\pi i (\alpha^\star + \beta^\star) x} \, 
 D_\alpha(x-u) \, D_\beta(x-v) = \\
&=  A(\alpha,\beta,\alpha^\star \+ \beta^\star) \, 
  e^{2\pi i (v \alpha^\star + u \beta^\star)} \,
  D_{\alpha+\beta+\frac{i}{2}Q}(u-v)  \,.
\end{aligned}\end{equation}
In the case $\alpha^\star = -\beta^\star$, Eq.~(\ref{Ddel}) 
can be used and we conclude that
\begin{equation}\label{D2del}
 \int_\BR dx\, D_\alpha(x-u) \, D_\beta(x-v) = 
  A(\alpha,\beta) \, \delta(u-v) 
\end{equation}
holds in the sense of distributions provided that $\alpha+\beta=-iQ$.  

\begin{lem}
 The identities 
\begin{align}
\label{D3}
 &\int_{\BR} dx \,  D_\alpha(x {-}u) \, D_\beta(x {-} v) \,
 D_\gamma(x {-} w)  \\ \nonumber &
\hspace{4cm}= A(\alpha,\beta,\gamma) \,
 D_{\alpha^\star}(w {-} v) \, D_{\beta^\star}(u {-} w) \, 
 D_{\gamma^\star}(v {-} u)  \,, \\
\label{D4}
 &\int_{\BR} dx \,  D_\alpha(x {-}u) \, D_\beta(x {-} v) \,
 D_\gamma(x {-} w) \, D_\omega(x {-}z) \\
 \nonumber
 &= A(\alpha,\beta,\gamma,\omega) \, 
 \frac{ D_{\alpha+\beta+\frac{i}{2}Q}(u {-} v) }%
	{ D_{\alpha+\beta+\frac{i}{2}Q}(w {-} z) } \,
 \int_{\BR} dx\,  D_{\alpha^\star}(x {-}v) \, D_{\beta^\star}(x {-} u) \,
 D_{\gamma^\star}(x {-} z) \, D_{\omega^\star}(x {-} w)
\end{align}
are valid provided that $\alpha+\beta+\gamma= -iQ$ 
in (\ref{D3}),
and $\alpha+\beta+\gamma +\omega= -iQ$ in~(\ref{D4}).
\end{lem}
\begin{proof}
Relation (\ref{D3}) follows straightforwardly 
from Eq.~(19) in~\cite{K2}, where function $\varphi_b(x)$ is 
our $g_b(e^{2\pi b x})$ (cf.~Eq.~(\ref{gw})). Also, in other
notations, relation (\ref{D3}) is Eq.~(11) in~\cite{V2}.

Eq.~(\ref{D3}) provides two expressions for a function which
we denote as $I(u,v,w;\alpha,\beta,\gamma)$. In order to prove 
(\ref{D4}) we multiply two copies of (\ref{D3}) and compute the 
following integral
\begin{equation}\label{D3D3}
 \int_\BR dt\, I(t,u,v;\nu,\alpha,\beta) \,
 I(t,w,z;\mu,\gamma,\omega) \,,
\end{equation}
where $\nu+\alpha+\beta=\mu+\gamma+\omega=-iQ$ and we
impose an additional condition $\nu+\mu=-iQ$.
Then the l.h.s.\ of (\ref{D4}) is obtained if we substitute
for the $I$'s the expressions on the l.h.s.\ of (\ref{D3}) 
and use relation (\ref{D2del}). The r.h.s.\ of (\ref{D4})
is obtained directly from (\ref{D3D3}) if we substitute
for the $I$'s the expressions on the r.h.s.\ of (\ref{D3})
and use also that $\mu^\star = - \nu^\star$. 
\end{proof}

\section{Positivity versus self--duality of the 
 representations $\CP_s$}\slabel{posapp}

The representations $\CP_s$ are distinguished by the property 
that the operators $\pi_s(u)$, $u\in\{E,F,K\}$ are positive 
self--adjoint. We are now going to show that this property
is closely related to the remarkable self--duality
of these representations under $b\ra b^{-1}$ which has
such profound consequences for the physics of our models.

To begin with, let us remark that there exists a linear basis 
$\CB_q(\fsl(2,\BR))$ for $\Uq$ such that all elements $u$ of 
$\CB_q(\fsl(2,\BR))$ are realized by positive 
operators $\pi_s(u)$. Such a basis is, e.g., given by the monomials
\[
\begin{aligned}
{}& q^{+\frac{mn}{2}}\,C^{l}\,E^m \,K^n\;\,
	{\rm represented ~by}\;\,\SC_s^{l}\,
(\SK_s)^\frac{n}{2} \, (\SE_s)^m \, (\SK_s)^{\frac{n}{2}},\\
{}& q^{-\frac{mn}{2}}\,C^{l}\,F^m \,K^n \;\,{\rm represented ~by}\;\,
\SC_s^{l}\,(\SK_s)^{\frac{n}{2}}\,(\SF_s)^m\,(\SK_s)^{\frac{n}{2}},
\end{aligned}
\quad l,m,n\in\BZ,\;\;l,m\geq 0.
\]

The elements of $\Uq$ are clearly realized by unbounded 
operators on $L^2(\BR)$. It is therefore useful to consider suitable
subspaces $\CT_s\subset L^2(\BR)$ of test--functions on which
all operators $\pi_s(u)$,  $u\in\Uq$ are well--defined. In order 
to describe a canonical choice for $\CT_s$ let us
represent the elements of $\CT_s$ by functions 
$f(k)$ such that $\spp$ acts as $(\spp f)(k)=k f(k)$. 
\begin{defn}
Let $\CT_s$ be the space of functions ${f} (k)$ 
which satisfy $e^{a|k|}f\in L^2(\BR)$ for all $a>0$, and 
which have an analytic continuation to  
$\BC\setminus (\Upsilon_s\cup\bar{\Upsilon}_{-s})$, where 
\begin{equation}\label{ups}
\begin{aligned}
 \Upsilon_s &= \Bigl\{ s + i \bigl( \fr{Q}{2} + nb + mb^{-1} \bigl),
 \quad n,m \in {\mathbb Z}^{\geq 0} \Bigl\} \,,\\
 \bar{\Upsilon}_s &= 
 \Bigl\{ s - i \bigl( \fr{Q}{2} + nb + mb^{-1} \bigl),
 \quad n,m \in {\mathbb Z}^{\geq 0} \Bigl\}  \,.
\end{aligned}
\end{equation}
\end{defn}

On the spaces $\CT_s$ the action of $\Uq$ is given by
\begin{equation}
\label{efjm} 
\begin{aligned}
  \SE_s \, {f}(k) &= 
 \bigl[\fr{Q}{2} + i s - i k \bigr]_b \,
  {f}(k + ib) \,, \\
 \SF_s \, {f}(k) &= 
  \bigl[\fr{Q}{2} + i s + i k \bigr]_b \, 
     {f}(k - ib) \,, 
\end{aligned}
\quad 
  \SK_s \, {f}(k) = e^{-\pi bk} \, {f}(k) \,,
\end{equation}
where $[x]_b \equiv \fr{\sin \pi b x}{\sin \pi b^2}$\,.

The distinguished role of the space $\CT_s$ is explained by the 
following result, which shows that the space $\CT_s$ is canonically 
associated to the representation $\pi_s$:
\begin{lem}\label{Pslem}
$\CT_s$ is the largest space on which all $\pi_s(u)$, $u\in\Uq$ are 
well--defined, i.e.,
\begin{equation}\label{intersect}
\CT_s\,=\,\bigcap_{u\in\CB_q(\fsl(2,\BR))} \CD_u\,,
\end{equation}
where $\CD_u$ is the domain of the unbounded operator 
$\pi_s(u)$, $u\in\CB_q(\fsl(2,\BR))$.
The spaces $\CT_s$, $s\in\BR$ are Fr\'echet spaces with topology 
defined by the family of seminorms 
\begin{equation}\label{seminorms}
\lVert \,f\,\rVert_{u}^{}\,\equiv\,\sup_{k\in\BR}\big| \,
 (\pi_s(u)f)(k)\,\big|,\quad u\in\CB_q(\fsl(2,\BR)).
\end{equation}
\end{lem}
\begin{proof}
It is easy to check that $\pi_s(u)f\in\CT_s$ for all $f\in\CT_s$. 
In order to show that the conditions in the definition of $\CT_s$ are 
all necessary, let us first observe that $e^{a|k|}f\in L^2(\BR)$ is 
clearly necessary for $\SK_s^{n}f$ to be well--defined for all~$n\in\BZ$. 
In order to determine the conditions on $f$ for $\SE_s^nf $ to be 
well--defined, let us consider the unitary operator 
$\SU_s\equiv w_b(\spp-s)$, where the special function $w_b(x)$ 
and its properties are described in Appendix~\ref{Qdil1}. We then have 
\[
\SU_s^{}\cdot\SE_s^n\cdot\SU_s^{-1} =e^{2\pi nb\sx} .
\]
The intersection of the domains of $e^{2\pi nb\sx}$ for all $n\in\BN$ 
consists of functions $g(k)$ that are analytic in the upper half plane 
${\mathbb H}_+$, see, e.g.,~\cite[Lemma 1]{S}. The corresponding 
functions $f(k)=(\SU_s^{-1}g)(k)= (w_b(k-s))^{-1}g(k)$ may have poles 
in~$\Upsilon_s$. Similar arguments applied to $\SF_s^n$ allow us to 
complete the proof of the first statement in Lemma~\ref{Pslem}. 

In order to verify the second statement, we mainly have to show that 
the space $\CT_s$ is complete w.r.t. the topology defined by the 
seminorms~\rf{seminorms}. This follows from \rf{intersect} together with 
the observation that the self--adjoint operators 
$\pi_s(u)$, $u\in\CB_q(\fsl(2,\BR))$ are  {\em closed} on~$\CD_u$.
\end{proof}

We regard Lemma \ref{Pslem} as the key to the mathematical 
understanding of the duality $b\ra b^{-1}$ of our representations
$\CP_s$. Indeed, let us introduce
the operators $\tilde{\SE}_s$, 
$\tilde{\SF}_s$, $\tilde{\SK}_s$, obtained by replacing $b\to b^{-1}$
in (\ref{EFK1}). These operators generate
a representation $\tilde{\CP}_s$ of $\CU_{\tilde{q}}(\fsl(2,\BR))$, 
$\tilde{q}=e^{i\pi b^{-2}}$ on the {\em same} space~$\CT_s$. The
space $\CT_s$ is associated to the representation $\tilde{\CP}_s$
as canonically
as it is associated to $\CP_s$. Moreover,
it is easy to see that the representation $\tilde{\CP}_s$ 
commutes\footnote{Commutativity of $\CP_s$ and 
$\tilde{\CP}_s$ only holds on the dense domain $\CT_s$ but not
in the usual sense of commutativity of spectral projections!} 
with $\CP_s$ on $\CT_s$. 
It is therefore natural to regard $\CT_s$ as the 
natural space on which a representation of the 
{\em modular double} $\Uq\ot \CU_{\tilde{q}}(\fsl(2,\BR))$~\cite{F3}
is realized.

Another way to make 
the self--duality of the representations $\CP_s$ transparent
uses the rescaled generators introduced in \rf{efk}.
These generators and their counterparts
$\tilde{\se}_s$, $\tilde{\sff}_s$,
$\tilde{\sk}_s$, obtained by replacing $b$ with $b^{-1}$
are related as~\cite{BT} 
\begin{equation}\label{bdual}
 (\se_s)^{\frac{1}{b}} =  (\tilde{\se}_s)^b \,, \quad 
 (\sff_s)^{\frac{1}{b}} =  (\tilde{\sff}_s)^b  \,, \quad
 (\sk_s)^{\frac{1}{b}} =  (\tilde{\sk}_s)^b \,.
\end{equation}
These observations express quite clearly that 
the representations of the two halves of the
modular double, $\CU_{q}(\fsl(2,\BR))$ and
$\CU_{\tilde{q}}(\fsl(2,\BR))$,  are related to 
each other like the two sides of the same coin.

\section{Structure of the monodromy matrix}\slabel{monoapp}

This appendix is devoted to the derivation of some simple, but important
structural properties of the monodromy matrix~$\SM(u)$,
\begin{equation}\label{Mono'}
 \SM(u) \,\equiv\, \left(\begin{matrix} \SA_\SRN(u) & \SB_\SRN(u) \\
 \SC_\SRN(u) & \SD_\SRN(u) \end{matrix} \right)\,\equiv\,
 L_\SRN(u) \cdot \ldots \cdot L_2(u) \cdot L_1(u) \,.
\end{equation}

\subsection{Expansions in the spectral parameter}

Introduce the following notations for $n \in {\mathbb N}$,
\begin{equation}\label{npm}
 \lceil n \rceil \equiv 
	\begin{cases} 
		n-1 & \text{if $n$ -- odd;} \\
		n & \text{if $n$ -- even;}
	\end{cases} \,, \quad
\lfloor n \rfloor \equiv 
	\begin{cases} 
		n & \text{if $n$ -- odd;} \\
		n-1 & \text{if $n$ -- even;}
	\end{cases}
\,.
\end{equation}

\begin{lem}\label{poslem}
The elements $\SA_{\SRN}(u)$, $\SB_{\SRN}(u)$, and 
$\SD_{\SRN}(u)$ of $\SM(u)$ have the following form:
\begin{align}
\label{aexp}
\SA^\XXZ_{\SRN}(u)&= e^{\SRN \pi b u}\sum_{m=0}^{\SRN} (-)^m 
	e^{-2m\pi b u} \, \SA^\XXZ_{\SRN,m}\,,\\
\label{bexp}
\SB^\XXZ_{\SRN}(u)&= i \, e^{\SRN \pi b u} \sum_{m=0}^{\SRN-1} (-)^m
	e^{-2m\pi b u} \, \SB^\XXZ_{\SRN,m}\,,\\
\label{dexp}
\SD^\XXZ_{\SRN}(u)&=  e^{\SRN \pi b u}\sum_{m=0}^{\SRN} (-)^m 
	e^{-2m\pi b u} \, \SD^\XXZ_{\SRN,m}\,, 
\end{align}
\begin{align}	 
\label{aaexp}
\SA^\SG_{\SRN}(u)&= i^{\lceil \SRN \rceil} \,
	e^{ \pi b (\lceil \SRN \rceil u- \SRN s)}
	\sum_{m=0}^{\lceil \SRN \rceil} (-)^m 
	e^{-2m\pi b u} \, \SA^\SG_{\SRN,m}\,,\\
\label{bbexp}
\SB^\SG_{\SRN}(u)&= - i^{\lfloor \SRN \rfloor} \,
	e^{\pi b (\lfloor \SRN \rfloor u- \SRN s) }
	\sum_{m=0}^{\lfloor \SRN \rfloor} (-)^m 
	e^{-2m\pi b u} \, \SB^\SG_{\SRN,m}\,,\\
\label{ddexp}
\SD^\SG_{\SRN}(u)&= i^{\lceil \SRN \rceil} \,
	e^{ \pi b (\lceil \SRN \rceil u- \SRN s)}
	\sum_{m=0}^{\lceil \SRN \rceil} (-)^m 
	e^{-2m\pi b u} \, \SD^\SG_{\SRN,m}\,,
\end{align}
where $\SA_{\SRN,m}$, $\SB_{\SRN,m}$, and $\SD_{\SRN,m}$ are positive
self--adjoint operators. 
\end{lem}

\begin{proof} 
Let us consider the case of the XXZ chain, the other case being
very similar. The definition of the monodromy matrix $\SM_\SRN(u)$ 
yields the following recursion relations
\begin{align} \label{Arecrel}
{}&\SA^\XXZ_{\SRN}(u)=(e^{\pi b u}\sk_{\SRN}- e^{-\pi b u}\sk_{\SRN}^{-1})
 \SA^\XXZ_{\SRN-1}(u)+ie^{\pi bu}\sff_{\SRN} \SC^\XXZ_{\SRN-1}(u),\\
{}&\SB^\XXZ_{\SRN}(u)=(e^{\pi b u}\sk_{\SRN}-e^{-\pi b u}\sk_{\SRN}^{-1})
 \SB^\XXZ_{\SRN-1}(u)+ie^{\pi bu}\sff_{\SRN} \SD^\XXZ_{\SRN-1}(u),\\
{}&\SC^\XXZ_{\SRN}(u)=(e^{\pi b u}\sk_{\SRN}^{-1}-e^{-\pi b u}\sk_{\SRN})
 \SC^\XXZ_{\SRN-1}(u) +ie^{-\pi bu}\se_{\SRN} \SA^\XXZ_{\SRN-1}(u),\\
{}&\SD^\XXZ_{\SRN}(u)=(e^{\pi b u}\sk_{\SRN}^{-1}-e^{-\pi b u}\sk_{\SRN})
 \SD^\XXZ_{\SRN-1}(u) +ie^{-\pi bu}\se_{\SRN}\SB^\XXZ_{\SRN-1}(u),
\label{Drecrel}\end{align}
where $\sk_{\SRN}\equiv \sk_s\ot1\ot\dots 1$ etc. Using these recursion 
relations one may inductively show that the operators $\SA_{\SRN,m}$,
$\SB_{\SRN,m}$ and $\SD_{\SRN,m}$ are linear combinations of monomials
of the form
\[
\su_s^{(\SRN)}\ot\dots\ot\su_s^{(1)},\qquad \su_s^{k}\in
\big\{\se_s^{},\sff_s^{},\sk_s^{},\sk_s^{-1}\big\}
\]
with {\em positive} integer coefficients. It remains to note that
an operator which is the sum of positive self--adjoint operators
will be self--adjoint on the intersection of the domains of the
individual summands. These obserations 
reduce our claim to the self--adjointness 
and positivity of $\se_s$, $\sff_s$,~$\sk_s$.
\end{proof}

\begin{lem}\label{leadterm}
 The leading terms of $\SA_{\SRN}(u)$, $\SB_{\SRN}(u)$, and 
$\SD_{\SRN}(u)$ at $e^{\pi b u} \to \pm \infty$ are given by
\begin{equation}
\begin{aligned}
\label{badinf}
{}& \SB^\XXZ_{\SRN,\0} = \Delta^{(\SRN-1)} \, \sff \,, 
{}&& \SA^\XXZ_{\SRN,\0} = \SD^\XXZ_{\SRN,\SRN}= 
	\Delta^{(\SRN-1)} \, \sk \,, \\
{}& \SB^\XXZ_{\SRN,\SRN-1} = \bar\Delta^{(\SRN-1)} \, \sff \,, \qquad
{}&& \SA^\XXZ_{\SRN,\SRN} = \SD^\XXZ_{\SRN,\0} =
	\Delta^{(\SRN-1)} \, \sk^{-1} \,, 
\end{aligned}
\end{equation}
where $\Delta^{(n)}$ is the $n$--fold co--product defined via
$\Delta^{(n+1)}=(\Delta^{(n)} \otimes id) \circ \Delta$,
with $\Delta^{(0)} \equiv id$ and $\Delta^{(1)} \equiv \Delta$,
and $\bar\Delta^{(n)}$ is defined analogously for the opposite 
co--product $\bar\Delta^{(1)} \equiv \Delta'$.
\begin{align}
\label{oddinf}
\text{$\SRN$ -- odd:\qquad } &
	\begin{cases}
	\SB^\SG_{\SRN,\0} =
	\theta_{\rm odd} \, \bigl( \Delta^{(\SRN-1)} \, 
	\sk^{-1} \bigr) \,, & 
	\SB^\SG_{\SRN,\SRN} = 
	\theta_{\rm odd} \, \bigl( \Delta^{(\SRN-1)} \, 
	\sk \bigr) \,,  \\
	\SA^\SG_{\SRN,\0} =
	\theta_{\rm odd} \, \bigl( \Delta^{(\SRN-1)} \, 
	\sff \bigr) \,, & 
	\SA^\SG_{\SRN,\SRN -1} = 
	\theta_{\rm odd} \, \bigl( \bar\Delta^{(\SRN-1)} \, 
	\sff \bigr) \,,  \\
	\SD^\SG_{\SRN,\0} =
	\theta_{\rm odd} \, \bigl( \bar\Delta^{(\SRN-1)} \, 
	\se \bigr) \,, & 
	\SD^\SG_{\SRN,\SRN -1} = 
	\theta_{\rm odd} \, \bigl( \Delta^{(\SRN-1)} \, 
	\se \bigr) \,,  
   \end{cases}
 \\
\label{eveninf}
\text{$\SRN$ -- even:\qquad } &
	\begin{cases}
	\SB^\SG_{\SRN,\0} =
	\theta_{\rm even} \, \bigl( \Delta^{(\SRN-1)} \, 
	\sff \bigr) \,, & 
	\SB^\SG_{\SRN,\SRN -1} = 
	\theta_{\rm even} \, \bigl( \bar\Delta^{(\SRN-1)} \, 
	\sff \bigr) \,,  \\
	\SA^\SG_{\SRN,\0} =
	\theta_{\rm even} \, \bigl( \Delta^{(\SRN-1)} \, 
	\sk \bigr) \,, & 
	\SA^\SG_{\SRN,\SRN } = 
	\theta_{\rm even} \, \bigl( \Delta^{(\SRN-1)} \, 
	\sk^{-1} \bigr) \,,  \\
	\SD^\SG_{\SRN,\0} =
	\theta_{\rm even} \, \bigl( \Delta^{(\SRN-1)} \, 
	\sk^{-1} \bigr) \,, & 
	\SD^\SG_{\SRN,\SRN } = 
	\theta_{\rm even} \, \bigl( \Delta^{(\SRN-1)} \, 
	\sk \bigr) \,,  
   \end{cases}
\end{align}
where $\theta_{\rm odd} \equiv 
 \theta_{\SRN-1} \circ \ldots \theta_{3} \circ \theta_{1}$ and
$\theta_{\rm even} \equiv 
 \theta_{\SRN} \circ \ldots \theta_{4} \circ \theta_{2}$
are compositions of the automorphism (\ref{thetaEFK}) at odd/even sites.
\end{lem}
\begin{proof}
Eqs.~\rf{badinf} follow easily from the decomposition 
\rf{Lpm} of the corresponding \hbox{L--matrix}. Eqs.~\rf{eveninf} 
are obtained by analogous consideration if $L^{\SG}(u)$ is replaced
with $\sigma_1\,{L}'(u)$ (see \rf{Lsg2}) and formula \rf{Lt}
is used. In order to apply this approach in the $\SRN$ odd case,
one has to multiply the monodromy matrix with an extra $\sigma_1$
{}from the right (which leads to the interchange 
$\SA \leftrightarrow \SB$ and $\SC \leftrightarrow \SD$).
\end{proof}

\subsection{Quantum determinant}\label{qDet}

Let us discuss connection between the so called {\em quantum determinant}
and coefficients $a(u)$ and $d(u)$ which arise in the Baxter
equation~(\ref{baxxz}). Since for the modular magnet we use R--matrix 
(\ref{R12}) which is not symmetric, the corresponding quantum determinant 
will differ from the ``standard'' formula applicable, e.g., for the 
sinh--Gordon model.  Therefore we commence by deriving the 
required expression.

\begin{lem}\label{GQD}
Let $L(u)$ be an L--matrix, satisfying the relation (\ref{rLL})
with the auxiliary R--matrix of the form
\begin{equation}\label{Rxi}
  R(u;\xi) = 
 \left( \begin{array}{cccc}
  \sinh \pi b(u + i b) & & & \\ [-1mm]
   &  \sinh \pi b u  & 
    i\sin\pi b^2\, e^{ \pi b \xi u}  & \\ 
   &  i\sin\pi b^2 \, e^{- \pi b \xi u} 
   &  \sinh \pi b u & \\ [-1mm]
   & & & \sinh\pi b(u + i b)
 \end{array} \right) \,.
\end{equation}
and let $\SM(u)$ be the corresponding monodromy matrix defined 
by~(\ref{Mono'}). The following element (quantum determinant)
\begin{align}\label{qdM0}
 \det\nolimits_q \SM(u) &=  
 \SA(u) \, \SD(u-ib) - q^{-\xi} \, \SB(u) \, \SC(u-ib) 
\end{align}
is central, i.e., $[\SM(v),\det\nolimits_q \SM(u)]=0$,
and can be written as follows
\begin{equation}\label{qdML}
 \det\nolimits_q \SM(u) = \bigl( \det\nolimits_q L(u) \bigr)^\SRN \,,
\end{equation}
where the quantum determinant of $L(u)$ is defined by the same
formula (\ref{qdM0}) (with $\SM(u)$ replaced by~$L(u)$).
\end{lem}

\begin{proof}
Existence of the quantum determinant is due to the degeneration of the 
auxiliary R--matrix $\check{R}(u;\xi)\equiv \SP \, R(u;\xi)$ 
at $u=-ib$,
\begin{equation}\label{Pxi}
  \check{R}(-ib;\xi) = i\sin\pi b^2 \,
 \left( \begin{array}{cccc}
  0 & & & \\ [-1mm]
   &  q^{\xi}  & -1 & \\ 
   & -1 &  q^{-\xi} & \\ [-1mm]
   & & & 0
 \end{array} \right) \,.
\end{equation}
It is interesting to notice that this matrix is proportional (in the 
standard basis) to the one--dimensional projector $P^-_\xi$ onto 
the spin~$0$ representation in the tensor square of spin~$\fr{1}{2}$
representations of ${\cal U}_{q^\xi}(\mathfrak{su}(2))$.

For $R(u;0)$ the statement of the Lemma is well--known (see, e.g.,
\cite{KBI}). In the generic case $\xi\neq 0$, one can observe that 
the gauge transformation
\begin{equation}\label{LMg}
  \tilde{L}(u) = g_u^{-1} \, L(u) \, g_u \,, \qquad
  \tilde{\SM}(u) = g_u^{-1} \,\SM(u) \, g_u \,, \qquad
  g_u = e^{\frac{1}{2} \pi b \xi u \sigma_3}
\end{equation}
yields L--matrix and monodromy matrix which satisfy the relation 
(\ref{rLL}) with the auxiliary R--matrix $R(u;0)$. Therefore
\begin{equation}\label{qdMt}
\det\nolimits_q \tilde{\SM}(u) =  
 \tilde{\SA}(u) \, \tilde{\SD}(u-ib) - 
 	\tilde{\SB}(u) \, \tilde{\SC}(u-ib)
\end{equation}
commutes with entries of $\tilde{\SM}(u)$ and hence with entries 
of~$\SM(u)$. Using (\ref{LMg}) in order to rewrite (\ref{qdMt}) 
in terms of entries of $\SM(u)$, we obtain~(\ref{qdM0}).

In order to prove (\ref{qdML}) it suffices to observe that
\begin{equation}\label{detMP}
 \det\nolimits_q \SM(u) \cdot \check{R}_{12}(-ib;\xi) =
 \SM_1(u) \, \SM_2 (u-ib) \, \check{R}_{12}(-ib;\xi) \,,
\end{equation}
which yields also three different expressions equivalent to (\ref{qdM0})
if we take into account the relation 
$\check{R}_{12}(-ib;\xi) \, \SM_1(u-ib) \, \SM_2 (u) =
 \SM_1(u) \, \SM_2 (u-ib) \, \check{R}_{12}(-ib;\xi)$
which is a particular case of~(\ref{rLL}). 
Now if $\SM'(u)$, $\SM''(u)$  satisfy (\ref{rLL}) with
the same R--matrix and their entries commute, then we have
\begin{align}
\nonumber
{}& \det\nolimits_q \bigl(\SM'(u) \, \SM''(u) \bigr)
 \cdot \check{R}_{12}(-ib;\xi) = 
 \SM'_1(u) \, \SM''_1 (u) \, 
 \SM'_2(u-ib) \, \SM''_2 (u-ib) \, \check{R}_{12}(-ib;\xi) \\
\label{detMM}
{}& = \SM'_1(u) \, \SM'_2 (u-ib) \, 
 \SM''_1(u) \, \SM''_2 (u-ib) \, \check{R}_{12}(-ib;\xi) =
 \det\nolimits_q \SM'(u) \cdot \det\nolimits_q \SM''(u) \,.
\end{align}
Whence (\ref{qdML}) follows immediately.
\end{proof}

For the models that we consider, Lemma~\ref{GQD} yields
\begin{align}
\label{qdxxz1}
 \det\nolimits_q \SM^\XXZ(u) &=  
 \SA(u) \, \SD(u-ib) - q^{-1} \, \SB(u) \, \SC(u-ib) \\
\label{qdxxz2}
{} &= \bigl( 4 \, \cosh \pi b( s + u - i \fr{b}{2}) \,
 \cosh \pi b( s - u + i \fr{b}{2})  \bigr)^\SRN \,, \\
\label{qdsg1}
 \det\nolimits_q \SM^\SG(u) &=  
 \SA(u) \, \SD(u-ib) -  \SB(u) \, \SC(u-ib) \\
\label{qdsg2}
{} &= \bigl( 4 e^{-2\pi b s}\, \cosh \pi b( s + u - i \fr{b}{2}) \,
 \cosh \pi b( s - u + i \fr{b}{2})  \bigr)^\SRN \,.
\end{align}

Proving the Proposition~\ref{Yxxz} in Appendix~E, we will deal with
monodromy matrices $\tilde{\SM}(u)$ obtained from $\SM(u)$ by
a gauge transformation with a matrix~$G(u)$. 
Notice that such $\tilde{\SM}(u)$ satisfies the exchange relation 
(\ref{rLL}) with an R--matrix 
\begin{equation}\label{Ruv}
\tilde{R}_{12}(u,v;\xi) = G_1(u) \, G_2(v) \,R(u-v;\xi) \,
 G_1^{-1}(u) \, G_2^{-1}(v)
\end{equation}
which {\em is not} of the form~(\ref{Rxi}). Therefore, for
general $G(u)$, Lemma~\ref{GQD} does not apply to~$\tilde{\SM}(u)$.
Nevertheless, there exists a class of gauge transformations which
preserve the quantum determinant in the following sense.

\begin{lem}\label{QD2}
Let $\SM(u)$ satisfy (\ref{rLL}) with the auxiliary 
R--matrix $R(u;\xi)$ of the form~(\ref{Rxi}). Let $\tilde{\SM}(u)=
\Bigl(\begin{smallmatrix} \tilde{A}(u) & \tilde{B}(u)\\ 
 \tilde{C}(u) & \tilde{D}(u) \end{smallmatrix} \Bigr)$ be its gauge 
transform defined by
\begin{equation}\label{qdM3}
 \tilde{\SM}(u) = 
 G(u) \cdot \SM(u) \cdot G^{-1}(u) \,, \qquad
 G(u) = \bigl(\begin{smallmatrix} 1 & 0\\ \rho & 1
 \end{smallmatrix} \bigr) \,, \qquad
 \rho = e^{-\pi b \xi u} \, \rho_0 \,,
\end{equation}
where $\rho_0$ is a c--number, which does not depend on~$u$. Then
\begin{equation}\label{qdMMt}
 \det\nolimits_q \tilde{\SM}(u) \equiv  
 \tilde{\SA}(u) \, \tilde{\SD}(u-ib) - 
 q^{-\xi} \, \tilde{\SB}(u) \, \tilde{\SC}(u-ib) =
 \det\nolimits_q \SM(u) \,,
\end{equation}
where the r.h.s. is defined according to Lemma~\ref{GQD}.
\end{lem}

Since entries of $\tilde{\SM}(u)$ are linear combinations of 
entries of $\SM(u)$, it follows that $\det\nolimits_q \tilde{\SM}(u)$
is central, i.e., $[\tilde{\SM}(v),\det\nolimits_q \tilde{\SM}(u)]=0$.
Thus, the l.h.s. defines the quantum determinant corresponding
to the R--matrix~(\ref{Ruv}).

\begin{proof}
Although $R(u-v;\xi)$ and $\tilde{R}(u,v;\xi)$ are in general
not equal, they coincide at $u-v=-ib$. Indeed, using the explicit
form of $G(u)$, it is easy to check that
\begin{equation}\label{cRGG}
  G_1(u) \, G_2(u-ib) \, \check{R}(-ib;\xi) =
 \check{R}(-ib;\xi) \,.
\end{equation}
Observing also that (\ref{detMP}) holds for 
$\det\nolimits_q \tilde{\SM}(u)$ as well, we derive
\begin{align*}
{}& \det\nolimits_q \tilde{\SM}(u) \cdot \check{R}_{12}(-ib;\xi) =
 \tilde{\SM}_1(u) \, \tilde{\SM}_2 (u-ib) \, \check{R}_{12}(-ib;\xi) \\
{}& = G_1(u) \, \SM_1(u) \, G_1^{-1}(u) \, 
  G_2(u-ib) \, \SM_2(u-ib) \, G_2^{-1}(u-ib) \, 
 \check{R}_{12}(-ib;\xi)  \\
{}& = G_1(u) \, G_2(u-ib) \, \SM_1(u) \,  
   \SM_2(u-ib) \,  G_2^{-1}(u-ib) \, G_1^{-1}(u) \,
 \check{R}_{12}(-ib;\xi) \\
{}& = G_1(u) \, G_2(u-ib) \, \SM_1(u) \,  
   \SM_2(u-ib) \, \check{R}_{12}(-ib;\xi) \\
{}&= G_1(u) \, G_2(u-ib) \, \check{R}_{12}(-ib;\xi) \cdot
 \det\nolimits_q \SM(u) = \check{R}_{12}(-ib;\xi) \cdot
 \det\nolimits_q \SM(u) \,,
\end{align*}
which proves the assertion of the Lemma.
\end{proof}

It is important that the gauge transformations used in the proof
of the Proposition~\ref{Yxxz} belong to the class of 
gauge transformations described in Lemma~\ref{QD2}; they
correspond to $\xi=1$ and $\xi=0$, respectively. This fact allows 
us to relate the quantum determinants of the models in question 
and the coefficients of the corresponding Baxter equations. 
A quick inspection of the proof of Proposition~\ref{Yxxz} shows that
\begin{equation}\label{adqdet}
 \bigl( a(u)\, d(u-ib) \bigr)^\SRN = \det\nolimits_q \SM(u) \,,
\end{equation}
where the r.h.s. is given by (\ref{qdxxz2}) and (\ref{qdsg2}),
respectively.

\section{Construction of the fundamental R--operator $\SR(u)$} 
\slabel{proofSH}

For the proof of Theorem \ref{Rthm} we will need 
the following material from \cite{PT2,BT}.

\subsection{Clebsch--Gordan maps}\label{CGmaps}

Let the space $\CM$ be defined by the direct integral
\begin{equation} 
\CM\equiv\int_{\BR^+}^{\oplus}ds\; \CP_{s}.
\end{equation}
Realizing elements of $\CT_s$ as functions $f(k)$ leads us to
represent the elements of $\CM$ by families of functions
$f\equiv(\,f_s\,;\,s\in\BR^+)$, where $f_s\equiv f_s(k)\in\CT_s$ for 
all $s\in\BR^+$. We shall define the multiplication operator
$\sS$ by 
\begin{equation}\label{canact0}
\sS f\,=\,\big(\,sf_s\,;
\,s\in\BR^+\,\big).
\end{equation}
To any family $(\,\SO_s\,;\,s\in\BR^+)$ of operators on $\CT_s$ 
we may then associate an operator $\SO_{\sS}$ on $\CM$ 
in the obvious manner. We have the corresponding
canonical action of  $\Uq$ on $\CM$ via 
\begin{equation}\label{canact}
\hat{\pi}_{{\sS}}(\xx) f\,=\,\big(\,{\pi}_{{s}}(\xx) \, f_s\,;
\,s\in\BR^+\,\big),\quad\forall\, \xx\in\Uq.
\end{equation}

The Clebsch--Gordan maps $\SC_{s_\2 s_\1}$ were defined 
in \cite{PT1,PT2} as a family of operators
\begin{align} \SC_{s_\2 s_\1}:
\CP_{s_\2}\ot\CP_{s_\1}&
\longrightarrow \CM \,.
\end{align}
The Clebsch--Gordan maps $\SC_{s_\2 s_\1}$
intertwine the action of $\Uq$ on 
$\CP_{s_\2}\ot\CP_{s_\1}$ with the canonical action on
$\CM$ in the sense that
\begin{equation}
\SC_{s_\2 s_\1} \cdot (\pi_{s_\2} \ot \pi_{s_\1}) \Delta (\xx)=
\hat{\pi}_{{\sS}}(\xx)\cdot\SC_{s_\2 s_\1}
\quad\forall\, \xx\in\Uq.
\end{equation}

\subsection{R--operator and braiding}

Let us introduce
\begin{equation}\label{Rdef}
\SR_{s_\2s_\1} = q^{\SH_{s_\2}\ot \SH_{s_\1}} \,
 g_b\bigl(\se_{s_\2}\ot \sff_{s_\1}\bigr)\, q^{\SH_{s_\2}\ot \SH_{s_\1}} 
\,.\end{equation}
Here the anti--self--adjoint operator $\SH_s$ is defined by 
$\sk_s=q^{\SH_s}$, and $g_b(x)$ is the {\em non--compact quantum 
dilogarithm} related to the special function $w_b(x)$,
defined in Appendix~\ref{Qdil}, via 
\begin{equation}\label{gw}
 g_b(e^{2 \pi b x})=  e^{\frac{\pi i}{24}(b^2 +b^{-2}) + 
	\frac{\pi i}{2} x^2} \, w_b(-x) \,.
\end{equation}
The R--operator $\SR_{s_\2s_\1}$ satisfies the following 
relations~\cite{BT}
\begin{align}
\label{Rrel1}
& \SR_{s_\2 s_\1} \cdot (\pi_{s_\2} \ot \pi_{s_\1}) \De(\xx) = 
 (\pi_{s_\2} \ot \pi_{s_\1})\De^\prime(\xx) \cdot \SR_{s_\2 s_\1} \,, \\
&\label{Rrel2}
 \SR_{s_\2 s_\1} \, (\se_{s_\2}^{} \otimes \sk^{-1}_{s_\1}) = 
(\se_{s_\2}^{} \otimes \sk_{s_\1}^{}) \, \SR_{s_\2 s_\1} \,,\\
& 
 \SR_{s_\2 s_\1} \, (\sk_{s_\2}^{} \ot \sff_{s_\1}^{} ) = 
 (\sk^{-1}_{s_\2} \ot \sff_{s_\1}^{}) \, \SR_{s_\2 s_\1}  \,,
\end{align}
where $\xx \in \Uq$ and
$\Delta^\prime$ stands for the opposite co--product.

The braiding operator 
$\SB : \CP_{s_\2} \otimes \CP_{s_\1} \to \CP_{s_\1} \otimes \CP_{s_\2}$
is defined by $\SB_{s_\2 s_\1} = \SP \, \SR_{s_\2 s_\1}$,
where $\SP$ is the operator that permutes the two tensor
factors. In what follows we will need the following statement.

\begin{propn}[Theorem~6 in \cite{BT}]\label{TBCC}
The braiding operator is diagonalized by the Clebsch--Gordan maps
in the following sense:
\begin{equation}\label{BCC}
 \SC_{s_\1 s_\2} \cdot \SB_{s_\2 s_\1} = 
 \Omega^{\;\sS}_{s_\2 s_\1} \cdot \SC_{s_\2 s_\1} \,,
\end{equation}
where $\Omega^{\;\sS}_{s_\2 s_\1}$ is the operator 
on $\CM$ associated via~\rf{canact0} to the scalar function
\begin{equation}\label{Om}
 \Omega^{\;s}_{s_\2 s_\1} = 
 e^{\pi i (s_\1^2 +s_\2^2 - s^2 +\frac{Q^2}{4})} \,.
\end{equation}
\end{propn}

In the particular case, \hbox{$s_\2 = s_\1$}, one can regard 
the permutation $\SP$ as an endomorphism of 
$\CP_{s_\1} \otimes \CP_{s_\1}$ and then
Proposition~\ref{TBCC} allows to relate it to the R--operator:
\begin{equation}\label{PRO}
 \SP = \SR_{s_\1 s_\1} \, 
 e^{-\pi i (\frac{Q^2}{4} + 2 s_\1^2  - \sS^2) } \,.
\end{equation}

\subsection{Proof of  Theorem~\ref{Rthm}}

We will consider the more general R--operator
defined on $\CP_{s_\2}\ot\CP_{s_\1}$ by the formula
\begin{equation}\label{Rxxz}
 \SR_{s_\2 s_\1}(u) 
 = \SR_{s_\2 s_\1} \, e^{+\pi i  (\sS^2_{\2\1} 
 -s_\1^2 - s_\2^2 - \frac{Q^2}{4})} \,
 D_u(\sS_{\2\1}) 
\end{equation}
where
the special function $D_\alpha(x)$ is defined by Eq.~(\ref{Ds})
in Appendix~\ref{Qdil}, 
and $\sS_{\2 \1}$ is the unique positive 
self--adjoint operator such that
\begin{equation}\label{tC}
 4 \cosh^2 \pi b \sS_{\2\1} = 
	(\pi_{s_\2} \ot \pi_{s_\1}) \Delta (C) \,. 
\end{equation}
It follows easily from
relation (\ref{PRO}) that $\SR(u)\equiv\SR_{ss}(u)$ 
coincides with the fundamental R--operator defined earlier in 
\rf{Rsimpledef}:
\begin{equation}\label{RGG}
 {\SR}(u) = \SP \, w_b(u + \sS) \, w_b(u - \sS) =\SP \, D_{u}(\sS) \,.
\end{equation}


To begin the proof of  Theorem \ref{Rthm}, which is somewhat more
involved than the proofs of the analogous results in the
case of highest weight representations \cite{Ji,F1,BD},
let us observe that Eq.~(\ref{RLL}) is equivalent to
the following set of equations
\begin{align}
\label{Rdd}
{}& \SR_{s_\2 s_\1}(u) \,  \De(L^{\ppm}) = 
   \De^\prime(L^{\ppm}) \, \SR_{s_\2 s_\1}(u) \,,\\ 
\label{Rabcd}
{}&\SR_{s_\2 s_\1}(u) \, \ell_{ij}(u) = 
 \tau\bigl(\ell_{ij}(-u)\bigr) \, \SR_{s_\2 s_\1}(u) \,,
 \quad i,j=1,2 \,,
\end{align}
where $L^\pm$ were introduced in (\ref{dLL})--(\ref{Lpm}),
$\tau$  is the flip operation:
$\tau(a_{s_\2} {\ot} b_{s_\1})= b_{s_\2}{\ot} a_{s_\1}$, 
and the operators $\ell_{ij}(u)$ are given by
\begin{align}
\label{l11}
 \ell_{\1\1}(u) &= e^{\pi b u} \, \se_{s_\2} \ot \sff_{s_\1}
  + e^{\pi b u} \, \sk^{-1}_{s_\2} \ot \sk_{s_\1} +
  e^{-\pi b u} \, \sk_{s_\2} \ot \sk^{-1}_{s_\1} \,, \\
\label{l12}
  \ell_{\1\2}(u) &= e^{\pi b u} \, \sk_{s_\2} \ot \sff_{s_\1}
  + e^{-\pi b u} \, \sff_{s_\2} \ot \sk^{-1}_{s_\1} \,, \\
\label{l21}
  \ell_{\2\1}(u) &= e^{\pi b u} \, \se_{s_\2} \ot \sk^{-1}_{s_\1}
  + e^{-\pi b u} \, \sk_{s_\2} \ot \se_{s_\1} \,, \\
\label{l22}  
  \ell_{\2\2}(u) &= e^{-\pi b u} \, \sff_{s_\2} \ot \se_{s_\1}
  + e^{-\pi b u} \, \sk^{-1}_{s_\2} \ot \sk_{s_\1} +
  e^{\pi b u} \, \sk_{s_\2} \ot \sk^{-1}_{s_\1} \,.
\end{align}

The verification of the relations \rf{Rdd} is easy. Introducing
\begin{equation}\label{rdef}
 \sr_{s_\2 s_\1}(u) = \SR_{s_\2 s_\1}^{-1} \cdot
 \SR_{s_\2 s_\1}^{}(u) \,, 
\end{equation}
it follows from \rf{Rrel1} that
\rf{Rdd} is equivalent to the system of equations
\begin{equation}
\label{rdd}
 \sr_{s_\2 s_\1}(u) \,  \De(\SX) =
    \De(\SX) \, \sr_{s_\2 s_\1}(u)\, .
\end{equation}
Validity of the relation \rf{rdd} follows from
$[\sS,\De(\SX) ]=0$.

The verification of the relations \rf{Rabcd} is somewhat harder. To begin 
with, let us observe that it suffices to verify relation \rf{l12}, say.

\begin{lem} Validity of \rf{Rabcd} for $(i,j)=(1,2)$ implies that
$\SR_{s_\2 s_\1}(u)$ satisfies the
 three remaining relations in \rf{Rabcd} as well.
\end{lem}

\begin{proof}
Equivalence of the $(1,2)$ and $(2,1)$ equations in \rf{Rabcd}
can be established by invoking the automorphism $\theta$,
defined in~\rf{thetaEFK}. Introduce 
$\tau{\!\vphantom |}_\theta \equiv (\theta \ot \theta) \circ \tau$.
It is easy to see that 
$\tau{\!\vphantom |}_\theta \bigl(\SR_{s_\2 s_\1}\bigr)=\SR_{s_\2 s_\1}$
and $\tau{\!\vphantom |}_\theta \bigl( \Delta(C) \bigr) = \Delta(C)$.
Therefore, 
$\tau{\!\vphantom |}_\theta \bigl(\SR_{s_\2 s_\1}(u)\bigr)
 =\SR_{s_\2 s_\1}(u)$.
The claimed equivalence of the $(1,2)$ and $(2,1)$ equations 
in \rf{Rabcd} follows now by observing that 
$\tau{\!\vphantom |}_\theta \bigl( \ell_{\1\2}(u) \bigr) =
\ell_{\2\1}(u)$.

In order to prove that 
relation \rf{Rabcd} for $(i,j)=(1,1)$ 
follows from the validity of relations \rf{Rabcd} for 
$(i,j)=(1,2)$, $(i,j)=(2,1)$ and \rf{Rdd},
let us consider the following object:
\begin{align}
\nn
\SX(u) &= 
 q \bigl( \sk^2_{s_\2} \ot \sk^2_{s_\1}  -
 \sk^{-2}_{s_\2} \ot \sk^{-2}_{s_\1} \bigr) \, \ell_{\1\1}(u) 
 + \sk_{s_\2} \ot \sk_{s_\1} 
 \bigl( e^{-\pi b u} \SC_{s_\2} \ot {\sf 1} 
 + e^{\pi b u} {\sf 1} \ot \SC_{s_\1} \bigr) \\ 
\label{X1} 
 & + \sk^{-1}_{s_\2} \ot \sk^{-1}_{s_\1} 
 \bigl( e^{\pi b u} \SC_{s_\2} \ot {\sf 1} + 
 e^{-\pi b u} {\sf 1} \ot \SC_{s_\1} \bigr) \,.
\end{align}
The operator $( \sk^2_{s_\2} \ot \sk^2_{s_\1}  -
 \sk^{-2}_{s_\2} \ot \sk^{-2}_{s_\1})$ does not have a normalizable
zero mode.
Validity of relation \rf{Rabcd} for $(i,j)=(1,1)$ 
therefore follows from
\begin{equation}
\label{RX}
\SR_{s_\2 s_\1}(u) \, \SX(u) =
 \tau \bigl(\SX(-u)\bigr) \, \SR_{s_\2 s_\1}(u) \,.
\end{equation}
Validity of this relation follows from the observation
that $\SX(u)$ can be represented in the form
\begin{align}
\label{X2}
 \SX(u) = 
 (\sk_{s_\2} \ot \sk_{s_\1}) \, \ell_{\1\2}(u) \, 
 (\pi_{s_\2} {\ot} \pi_{s_\1}) \Delta(e) -
 (\sk^{-1}_{s_\2} \ot \sk^{-1}_{s_\1}) \ell_{\2\1}(u) \, 
 (\pi_{s_\2} {\ot} \pi_{s_\1}) \Delta(f) \,.
\end{align}

Exchanging $\sk \leftrightarrow \sk^{-1}$ in \rf{X2},
we can derive relation \rf{Rabcd} for $(i,j)=(2,2)$  in  a
completely analogous way.
\end{proof}

It remains to verify 
relation \rf{Rabcd} for $(i,j)=(1,2)$. 
This relation may be rewritten in terms of the 
operator $\sr_{s_\2 s_\1}(u)$ defined in \rf{rdef} 
by using the relation \rf{Rrel2}. 
We conclude that validity of \rf{Rabcd} follows from 
the validity of 
\begin{align}
\label{rek}
 \sr_{s_\2 s_\1}(u) \, &\bigl( e^{2 \pi b u} \SO_1 + \SO_2 \bigr) 
 =  \bigl( e^{2 \pi b u} \, \SO_1 +  \SO_3
  \bigr) \,  \sr_{s_\2 s_\1}(u) \,,
\end{align}
where we have introduced the convenient abbreviations
\begin{equation}\label{EB}
\begin{aligned}
\SO_\1  \equiv &\  \se_{s_\2} \otimes \sk_{s_\1}^{-1}\,,\\
 \SO_\2 \equiv &\  \sk_{s_\2} \otimes \se_{s_\1} \,,
\end{aligned}\qquad
 \SO_\3 \equiv   \SR_{s_\2 s_\1}^{-1} \, (\sk_{s_\2}^{-1} 
 \otimes \se_{s_\1}) \, \SR_{s_\2 s_\1}^{} \,.
\end{equation}


In order to prove that $\sr_{s_\2 s_\1}(u)$ satisfies
\rf{rek} it will be convenient to use the Clebsch--Gordan maps 
$\SC_{s_\2 s_\1}$ introduced in Subsection~\ref{CGmaps}.
Let us observe that (\ref{rdd}) implies that 
the r--matrix is diagonalized by the Clebsch--Gordan maps:
\begin{equation}
\label{rCG}
 \SC_{s_\2 s_\1} \cdot \sr_{s_\2 s_\1}(u) =
 {r}^{\,\sS}_{s_\2 s_\1}(u) \cdot \SC_{s_\2 s_\1} \,,
\end{equation}
where $r^{\,\sS}_{s_\2 s_\1}(u)$ is the operator 
on $\CM$ associated to the scalar function
$r^{\,s}_{s_\2 s_\1}(u)$ via~\rf{canact0}.

In order to further evaluate equation \rf{rek} we will need to 
describe the images $\tilde{\SO}_{\ell}$, $\ell{=}1,2,3$  of the 
operators $\SO_\ell$ under the maps $\SC_{s_\2 s_\1}$
which are defined by \hbox{$\SC_{s_\2 s_\1}\cdot
\SO_\ell\equiv \tilde{\SO}_{\ell}\cdot\SC_{s_\2 s_\1}$}.
\begin{propn}\label{SH}
The operators $\tilde{\SO}_{\ell}$ can be represented as follows:
If $g\equiv(g_s;s\in\BR^+)\in\CM$ then
$\tilde{\SO}_{\ell}
g\equiv \big(\,(\tilde{\SO}_{\ell} g)_s^{}\,;\,s\in\BR^+\,\big)$, 
where
\begin{equation}\label{Ekjn}
 (\tilde{\SO}_{\ell} g)_s^{}(k)  \equiv
  \! \sum_{\nu=-1}^{1} \, A_{\ell;s}^{\nu;s_\2s_\1}(k)\,\ST_{s}^{\nu}
  g_{s}^{}(k \+ib) \,, 
\end{equation}
where $\ST_{s}^{\nu}g_s=g_{s \+ i \nu b}$.
The coefficients $A_{\ell;s}^{\nu;s_\2s_\1}(k)$ are symmetric under 
exchange of $s_\2$ and $s_\1$,
\begin{equation}
\label{AA}
A_{\ell;s}^{\nu;s_\2s_\1}(k) = A_{\ell;s}^{\nu;s_\1s_\2}(k) 
 \end{equation}
and are otherwise related to each other by
\begin{equation}\label{al23'}
\begin{aligned}
 A_{2;s}^{\nu;s_\2s_\1}(k)
 = &\   e^{+2\pi b\nu s + i \pi b^2 \nu^2} 
 A_{1;s}^{\nu;s_\1s_\2}(k)\,, \\
 A_{3;s}^{\nu;s_\2s_\1}(k)
 = &\  e^{-2\pi b\nu s - i \pi b^2 \nu^2} 
 A_{1;s}^{\nu;s_\1s_\2}(k)\,.
\end{aligned}
\end{equation}
\end{propn}
The proof of Proposition~\ref{SH}, which is somewhat technical, 
is given in Subsection~\rf{proppf}.

Proposition~\ref{SH} together with equation \rf{rCG} allows us to
rewrite the defining relation (\ref{rek}) as a commutation relation
satisfied by the corresponding operators on $\CM$. 
Applying (\ref{rek}) to a function $g_s$ and  matching the 
coefficients in front of $g_{s \+ i\nu}(k \+ib)$ in the
resulting equation, we derive functional equations
on $r^{\,s}_{s_\2 s_\1}(u)$,
\begin{equation}\label{reqs}
\begin{aligned}
 r^{s\+ i b\nu }_{s_\2 s_\1}(u) \,&\
 \big(e^{2\pi b u } \, A_{1;s}^{\nu;s_\2s_\1}(k)
 + A_{2;s}^{\nu;s_\2s_\1}(k) \big) \\
 &\ = \big(e^{2\pi b u } \, A_{1;s}^{\nu;s_\2s_\1}(k)
 + A_{3;s}^{\nu;s_\2s_\1}(k) \big)\;
 r^{\,s}_{s_\2 s_\1}(u) \,.
\end{aligned}
\end{equation}
The case $\nu=0$ holds trivially. Taking
into account (\ref{AA}) and \rf{al23'}, Eqs.~(\ref{reqs})
for $\nu=\pm 1$ are equivalent to a single
functional equation,
\begin{equation}\label{req}
 r^{s + ib}_{s_\2 s_\1}(u) \, 
 \bigl(e^{2\pi b u} + \zeta  \bigr) = 
 \bigl(e^{2\pi b u} + \zeta^{-1}  \bigr) \, 
 r^{\,s}_{s_\2 s_\1}(u) \,, 
\end{equation}
where $ \zeta =  e^{i \pi b^2 + 2\pi b s}$. In terms of
$\tilde{r}^{\,s}_{s_\2 s_\1}(u)=
 e^{-i\pi s^2 } r^{\,s}_{s_\2 s_\1}(u)$ 
one may rewrite this functional relation as follows
\begin{equation}\label{req2}
 \tilde{r}^{s+ i \frac{b}{2}}_{s_\2 s_\1}(u) 
 \, \cosh \pi b (s + u) = 
 \tilde{r}^{s- i \frac{b}{2}}_{s_\2 s_\1}(u) \,
 \cosh \pi b (s - u)  \,.
\end{equation}
Recalling that the special function $w_b(x)$ satisfies 
(\ref{wfunrel}), we conclude that equation (\ref{req}) is solved 
by the expression for $r^{\,s}_{s_\2 s_\1}$ which
follows from eqn.~\rf{Rxxz} via \rf{rdef} and \rf{rCG}, namely
\begin{equation}\label{rl}
 r^{\,s}_{s_\2 s_\1} (u) = 
 e^{\pi i  (s^2 -s_\1^2 - s_\2^2 - \frac{Q^2}{4})} \,
 \frac{w_b(u + s)}{w_b(s - u)} =
 e^{\pi i  (s^2 -s_\1^2 - s_\2^2 - \frac{Q^2}{4})} \,
 D_{u}(s) \,,
\end{equation}
where $D_{\alpha}(x)$ is defined in (\ref{Ds}). 

Property (\ref{Dcc}) implies that $|r^{\, s}_{s_\2 s_\1}(u)|=1$ if 
$u, s \in \BR$. Since $\sS_{\2 \1}$ is self--adjoint,
we infer that $r^{\,\sS}_{s_\2 s_\1}(u)$ is a unitary operator
for $u\in\BR$. On the other hand property (\ref{wcc}) implies 
that $g_b(x)$ given by (\ref{gw}) satisfies $|g_b(x)|=1$ for 
$x\in\BR^+$. Since $\se_\2\otimes\sff_\1$ in (\ref{Rdef})
is positive self--adjoint, we infer that $\SR_{s_\2 s_\1}$ 
is unitary. The unitarity of $\SR_{s_\2 s_\1}(u)$ for 
$u\in\BR$ follows.\hfill $\square$\\[1ex]
Let us finally remark that relation (\ref{Das2}) leads to the 
following asymptotics of $\SR_{s_\2 s_\1}(u)$ 
\begin{equation}\label{Ras}
\begin{aligned}
 \SR_{s_\2 s_\1}(u) &\sim e^{-i\pi (u^2 + \ldots)} \, \SR_{s_\2 s_\1}
 \quad \text{for } {\rm Re}(u) \to +\infty \\
 \SR_{s_\2 s_\1}(u) &\sim e^{+i\pi (u^2 + \ldots)} \, \bar\SR_{s_\2 s_\1}
 \quad \text{for } {\rm Re}(u) \to -\infty \,,
\end{aligned}
\end{equation}
where $\bar\SR_{s_\2 s_\1}^{}\equiv
\SP\cdot\SR_{s_\2 s_\1}^{-1}\cdot
\SP$.

\subsection{Clebsch--Gordan and Racah--Wigner coefficients for $\CP_s$}

In order to prove Proposition \ref{SH} we will need to describe the 
Clebsch--Gordan maps $\SC_{s_\2 s_\1}$ more explicitly.
For the following it will be convenient
to use the
variables $\al_r=\frac{Q}{2}+is_r$ in order parameterize the 
representations $\CP_{s_r}$, $r=0,1,2\dots$.
The Clebsch--Gordan maps $\SC_{s_\2 s_\1}$ can then be represented
explicitly as an integral transformation \cite{PT2,BT}:
\begin{equation}
\tilde{f}_s(k)\,\equiv\,
\int dk_2dk_1 \;\big[\begin{smallmatrix}
\al_3 \\
k_3\end{smallmatrix}\big|
\begin{smallmatrix}\al_2 & \al_1 \\ k_2 & k_1\end{smallmatrix}
\big]\,f(k_2,k_1),
\end{equation}
The kernel which appears on the right hand side, the so--called 
b--Clebsch--Gordan kernel, was calculated 
in~\cite{PT2,BT}. It is of the general form
\begin{equation}\label{CGexpl}
\big[\begin{smallmatrix}
\al_3 \\
k_3\end{smallmatrix}\big|
\begin{smallmatrix}\al_2 & \al_1 \\ k_2 & k_1\end{smallmatrix}
\big]\,=\,\de(k_3-k_2-k_1)\,
\big[\begin{smallmatrix}
\al_3 \end{smallmatrix}\big|
\begin{smallmatrix}\al_2 & \al_1 \\ k_2 & k_1\end{smallmatrix}
\big]
\end{equation}
where the function $\big[\begin{smallmatrix}
\al_3 
\end{smallmatrix}\big|
\begin{smallmatrix}\al_2 & \al_1 \\ k_2 & k_1\end{smallmatrix}
\big]$ is given as
\begin{equation}
\big[\begin{smallmatrix}
\al_3 \end{smallmatrix}\big|
\begin{smallmatrix}\al_2 & \al_1 \\ k_2 & k_1\end{smallmatrix}
\big]=
\frac{
e^{-\frac{\pi i}{2}\al_{123}^{(1)}\al_{123}^{(0)}}
e^{\pi (k_1\al_2-k_2\al_1)}}
{S_b(\al_1+\al_2+\al_3-Q)}
\,\Phi^{}_3(R_1,R_2,R_3;S_1,S_2,S_3;-\al_{123}^{(0)}).
\end{equation}
In equation  \rf{CGexpl} we have used the special function
$\Phi^{}_3(R_1,R_2,R_3;S_1,S_2,S_3;x)$ defined in \rf{Phidef}
whose arguments have been chosen as follows:
\[
\begin{aligned}
R_1&=\al_1+ik_1\\
R_2&=\al_2-ik_2\\
R_3&=\al_{123}^{(0)}
\end{aligned}
\qquad
\begin{aligned}
S_1&=\al_{123}^{(1)}+\al_1+ik_1\\
S_2&=\al_{123}^{(2)}+\al_2-ik_2\\
S_3&=Q.
\end{aligned}
\]
Here and below we are using the following notations:
\[ 
\begin{aligned}
{} &\al_{ijk}^{(1)}=\al_j+\al_k-\al_i,\\
{} &\al_{ijk}^{(2)}=\al_k+\al_i-\al_j,\\
{} &\al_{ijk}^{(3)}=\al_i+\al_j-\al_k,
\end{aligned}\qquad
\al_{ijk}^{(0)}=\al_i+\al_j+\al_k-Q.
\]
The precise relation to the b--Clebsch--Gordan coefficients from
\cite{BT} is as follows:
\begin{equation}
\big[\begin{smallmatrix}
\al_3 \\
k_3\end{smallmatrix}\big|
\begin{smallmatrix}\al_2 & \al_1 \\ k_2 & k_1\end{smallmatrix}
\big]_{\rm here}
\equiv \nu^{\al_3}_{\al_2\al_1}
\big[\begin{smallmatrix}
\al_3 \\
k_3\end{smallmatrix}\big|
\begin{smallmatrix}\al_2 & \al_1 \\ k_2 & k_1\end{smallmatrix}
\big]_{\rm BT}
\end{equation}
with phases $\nu_{\al_2\al_1}^{\al_3}$ chosen as
\begin{equation}
\big(\nu_{\al_2\al_1}^{\al_3}\big)^{-1}\,=\,
S_b(\al_3+\al_1-\al_2)S_b(\al_3+\al_2-\al_1).
\end{equation}

The b--Racah--Wigner coefficients 
are then defined by the 
relation
\begin{equation}\label{RWdef} \begin{aligned}
\big[\begin{smallmatrix}
\al_s\\
k_s\end{smallmatrix}\big|
\begin{smallmatrix}
\al_1 & \al_0\\
k_1 & k_0
\end{smallmatrix}\big]& 
\big[\begin{smallmatrix}
\al_3 \\
k_3\end{smallmatrix}\big|
\begin{smallmatrix}
\al_2 & \al_s\\
k_2 & k_s
\end{smallmatrix}
\big]\\
&=\frac{1}{i}\int\limits_{\BS} d\al_t\;
\big\{\begin{smallmatrix}
\al_0 & \al_1\\
\al_2 & \al_3
\end{smallmatrix}\big|\begin{smallmatrix}\al_s\\\al_t
\end{smallmatrix}
\big\}\;
\big[\begin{smallmatrix}
\al_3 \\
k_3\end{smallmatrix}\big|
\begin{smallmatrix}\al_t & \al_0 \\ k_t & k_0\end{smallmatrix}
\big]
\big[\begin{smallmatrix}\al_t\\k_t\end{smallmatrix}
\big|
\begin{smallmatrix}\al_2 & \al_1\\ k_2 & k_1
\end{smallmatrix}\big],
\end{aligned}
\end{equation}
where $\BS\equiv \frac{Q}{2}+i\BR_+$.
The coefficients $\{\dots\}$ can be represented explicitly by the 
following formula
\begin{equation}
\begin{aligned}
\big\{\begin{smallmatrix}
\al_0 & \al_1\\
\al_2 & \al_3
\end{smallmatrix}\big|\begin{smallmatrix}\al_s\\\al_t
\end{smallmatrix}
\big\}=&\;
\frac{S_b(\al_t+\al_0+\al_3-Q)S_b(\al_3+\al_0-\al_t)S_b(\al_3+\al_t-\al_0)}
   {S_b(\al_s+\al_2+\al_3-Q)S_b(\al_3+\al_2-\al_s)S_b(\al_3+\al_s-\al_2)}\\
&\;\times\frac{S_b(\al_t+\al_2-\al_1)}
{S_b(\al_s+\al_0-\al_1)}\Phi^{}_4(U_1,U_2,U_3,U_4;V_1,V_2,V_3,V_4;0)
\end{aligned}
\end{equation}
where
\[
\begin{aligned}
U_1&=\al_s+\al_0-\al_1\\
U_2&=\al_s+Q-\al_0-\al_1\\
U_3&=\al_s+\al_2+\al_3-Q\\
U_4&=\al_s+\al_3-\al_2
\end{aligned}
\qquad
\begin{aligned}
V_1&=Q-\al_t+\al_s+\al_3-\al_1\\
V_2&=\al_t+\al_s+\al_3-\al_1\\
V_3&=2\al_s\\
V_4&=Q.
\end{aligned}
\]
For completeness let us note that the  
b--Racah--Wigner coefficients $\{\cdots\}$ are related
to the corresponding objects from \cite{PT2} via
\begin{align}
&\big\{\begin{smallmatrix}
\al_0 & \al_1\\
\al_2 & \al_3
\end{smallmatrix}\big|\begin{smallmatrix}\al_s\\\al_t
\end{smallmatrix}
\big\}_{\rm here}
\,=\,\frac{\nu_{\al_1\al_0}^{\al_s}\nu_{\al_2\al_s}^{\al_3}}
          {\nu_{\al_2\al_1}^{\al_t}\nu_{\al_t\al_0}^{\al_3}}
\big\{\begin{smallmatrix}
\al_0 & \al_1\\
\al_2 & \al_3
\end{smallmatrix}\big|\begin{smallmatrix}\al_s\\\al_t
\end{smallmatrix}
\big\}_{\rm PT2}\,.
\end{align}

The b--Clebsch--Gordan and b--Racah--Wigner coefficients 
are meromorphic functions of all of their arguments. The
complete set of poles may be described as follows 
\cite{PT2,BT}\footnote{We take the opportunity to correct 
some typos in these references.}\\[1ex]
\underline{$\big\{\begin{smallmatrix}
\al_0 & \al_1\\
\al_2 & \al_3
\end{smallmatrix}\big|\begin{smallmatrix}\al_s\\\al_t
\end{smallmatrix}
\big\}$ has poles at}
\begin{equation}\label{RWpoles}
\begin{aligned}
{} & Q-\al_{32s}^{(\iota)}=-nb-mb^{-1},\\
{} &  Q-\al_{s10}^{(\iota)}=-nb-mb^{-1},
\end{aligned}\qquad
\begin{aligned}
{} & \al_{3t0}^{(\iota)}=-nb-mb^{-1},\\ 
{} & \al_{t21}^{(\iota)}=-nb-mb^{-1},
\end{aligned}\qquad \iota=0,1,2,3,
\end{equation}
where $n,m\in\BZ^{\geq 0}$.\\[1ex]
\underline{$\big[\begin{smallmatrix}
\al_3 \\
k_3\end{smallmatrix}\big|
\begin{smallmatrix}\al_2 & \al_1 \\ k_2 & k_1\end{smallmatrix}
\big]$ has poles at}
\begin{equation}\label{CGpoles}
\begin{aligned}
{} &Q-\al_{321}^{(\iota)}=-nb-mb^{-1},\qquad \iota=0,1,2,3,\\
{} & \pm ik_i=\al_i+nb+mb^{-1},\qquad i=1,2,\\
{} & \pm ik_3=Q-\al_3+nb+mb^{-1},
\end{aligned}
\end{equation}
where again  $n,m\in\BZ^{\geq 0}$.

\subsection{Proof of Proposition~\ref{SH}} \label{proppf}

Our proof of Proposition~\ref{SH} will be based on the following
nontrivial identity satisfied by the 
\hbox{b--Clebsch}--Gordan kernel.
\begin{lem}\label{idlem}
\begin{equation}\label{CGid} \begin{aligned}
\big[\!\big[\begin{smallmatrix}
\al_1 \end{smallmatrix}\big|
\begin{smallmatrix}
\al_1 & -b\\
k_1 & \;\, ib
\end{smallmatrix}
\big]\!\big]& 
\big[\begin{smallmatrix}
\al_3\end{smallmatrix}\big|
\begin{smallmatrix}
\al_2 & \al_1\\
k_2 & k_1+ib
\end{smallmatrix}\big]\\
&=\sum_{\tau=-1}^1
F_{\tau}\big[\begin{smallmatrix}
\al_2 & \al_1\\
\al_3 & -b
\end{smallmatrix}
\big]\;
\big[\!\big[\begin{smallmatrix}
\al_3 \end{smallmatrix}\big|
\begin{smallmatrix}\al_3+\tau b & -b\\ k_3 & ib\end{smallmatrix}
\big]\!\big]\,\ST_{\al_3}^{\tau}
\big[\begin{smallmatrix}\al_3\end{smallmatrix}
\big|
\begin{smallmatrix}\al_2 & \al_1\\ k_2 & k_1
\end{smallmatrix}\big],
\end{aligned}
\end{equation}
where $\ST_{\al_3}^{\tau}f(\al_3)=f(\al_3+\tau b)$, $k_3=k_2+k_1$ 
and we 
have furthermore used the notation
\begin{align*}
& \big[\!\big[\begin{smallmatrix}
\al_3 \end{smallmatrix}\big|
\begin{smallmatrix}
\al_2 & \al_1\\
k_2 & -i\al_1
\end{smallmatrix}
\big]\!\big]=
2\pi i\!
\Res_{k_1=  -i\al_1}  \big[\begin{smallmatrix}
\al_3\end{smallmatrix}\big|
\begin{smallmatrix}
\al_2 & \al_1\\
k_2 & k_1
\end{smallmatrix}\big]\,,\\
& F_{\tau}\big[\begin{smallmatrix}
\al_2 & \;\,\al_1\\
\al_3 & -b
\end{smallmatrix}
\big]=2\pi i\Res_{\al_\tau=\al_3+b\tau}
\big\{\begin{smallmatrix}
-b & \al_1\\
\;\al_2 & \al_3
\end{smallmatrix}\big|\begin{smallmatrix}\al_1\\
\al_\tau
\end{smallmatrix}
\big\}\,,
\end{align*}
We have the explicit formulae
\begin{align}\label{CGres}
\big[\!\big[\begin{smallmatrix}
\al_3 \end{smallmatrix}\big|
\begin{smallmatrix}
\al_2 & \al_1\\
k_2 &  -i\al_1
\end{smallmatrix}
\big]\!\big]&=\frac{
e^{\frac{\pi i}{2}(\De_{\al_3}-\De_{\al_2}-\De_{\al_1})}}
{S_b(\al_3+\al_2-\al_1)}
e^{-\pi k_2\al_1}\frac{S_b(\al_2-ik_2)}{S_b(\al_1+\al_3-ik_2)}\,,\\
F_{-}^{}\big[\begin{smallmatrix}
\al_2 & \;\,\al_1\\
\al_3 & -b
\end{smallmatrix}
\big]
&=\frac{S_b(2\al_3-2b-Q)}{S_b(2\al_1+b)}\sin\pi b(\al_2+\al_1-\al_3)
\label{RWres1}\,,\\
F_{+}^{}\big[\begin{smallmatrix}
\al_2 & \;\,\al_1\\
\al_3 & -b
\end{smallmatrix}
\big]
&= 
\frac{S_b(2\al_3-Q)}{S_b(2\al_1+b)}\sin\pi b(\al_3+\al_2+\al_1-Q)
\label{RWres2}\\
&\quad\ti\sin\pi b(\al_3+\al_1-\al_2)\sin\pi b(\al_3+\al_2-\al_1)\,,\nn
\end{align}
where $\De_{\al}=\al(Q-\al)$.
\end{lem}
Given that Lemma~\ref{idlem} holds, it becomes easy to complete 
the proof of Proposition~\ref{SH} as follows: Notice that 
the left hand side of \rf{CGid} can be written as
\begin{align*}
\big[\!\big[\begin{smallmatrix}\al_1 \end{smallmatrix}\big|
\begin{smallmatrix}
\al_1 & -b\\
k_1 & \;\,ib
\end{smallmatrix}\big]\!\big]
\big[\begin{smallmatrix}
\al_3\end{smallmatrix}\big|
\begin{smallmatrix}
\al_2 & \al_1\\
k_2 & k_1+ib
\end{smallmatrix}\big]
&\;=\;e^{\pi bk_1}\,\frac{[\al_1-ik_1-b]_b}{S_b(2\al_1+b)}\,
\big[\begin{smallmatrix}
\al_3\end{smallmatrix}\big|
\begin{smallmatrix}
\al_2 & \al_1\\
k_2 & k_1+ib
\end{smallmatrix}\big]\\
&=\,
\frac{e^{\pi b k_3}}{S_b(2\al_1+b)}
\;\big(\SK_{s_\2}\ot \SE_{s_\1}\big)^t
\cdot\big[\begin{smallmatrix}
\al_3\end{smallmatrix}\big|
\begin{smallmatrix}
\al_2 & \al_1\\
k_2 & k_1 \end{smallmatrix}\big],
\end{align*}
where $\SO^t$ denotes the transpose of an operator
on $\CP_{s_\2}\ot\CP_{s_\1}$ defined by
\begin{equation}
\int dk_\2dk_\1 \;f(k_\2,k_\1)\,(\SO g)(k_\2,k_\1)\,\equiv\,
\int dk_\2dk_\1 \;(\SO^t f)(k_\2,k_\1)\,g(k_\2,k_\1).
\end{equation}
Equation \rf{CGid} may therefore be written in the form
\begin{equation}
\big(\SK_{s_\2}\ot \SE_{s_\1}\big)^t\cdot
\big[\begin{smallmatrix}
\al_3\end{smallmatrix}\big|
\begin{smallmatrix}
\al_2 & \al_1\\
k_2 & k_1 \end{smallmatrix}\big]\,=\,
\sum_{\tau =-1}^1 A_{2;s}^{\tau;s_\2 s_\1}(k_3)\,\ST_{\al_3}^{\tau}
\big[\begin{smallmatrix}\al_3\end{smallmatrix}
\big|
\begin{smallmatrix}\al_2 & \al_1\\ k_2 & k_1
\end{smallmatrix}\big],
\end{equation}
where 
\[
A_{2;s}^{\tau;s_\2 s_\1}(k_3)=e^{-\pi bk_3}S_b(2\al_1+b)
F_{\tau}\big[\begin{smallmatrix}
\al_2 & \al_1\\
\al_3 & -b
\end{smallmatrix}
\big]\;
\big[\!\big[\begin{smallmatrix}
\al_3 \end{smallmatrix}\big|
\begin{smallmatrix}\al_3+\tau b & -b\\ k_3 & ib\end{smallmatrix}
\big]\!\big].
\]
By using the explicit expressions \rf{RWres1},\rf{RWres2}
one may easily verify that the coefficients
$A_{2;s}^{t;s_\2 s_\1}$ are symmetric under the
exchange of $s_\2$ and $s_\1$, as claimed. This completes 
the proof of all the relevant statements of Proposition~\ref{SH}
for the case of the operator $\SO_2$.

In order to cover the remaining cases
let us observe that
\begin{equation}\label{EBapp}
\begin{aligned}
 \SO_2 \equiv &\  \SK_{s_\2} \otimes \SE_{s_\1} =
 \SB_{s_\1 s_\2} \, \SO_1 \, \SB_{s_\1 s_\2}^{-1} \,, \\
 \SO_3 \equiv &\ \SR_{s_\2 s_\1}^{-1} \, (\SK_{s_\2}^{-1} 
 \otimes \SE_{s_\1}) \, \SR_{s_\2 s_\1} =
 \SB^{-1}_{s_\2 s_\1 } \, \SO_1 \, \SB_{s_\2 s_\1 } \,,
\end{aligned}
\end{equation}
where $\SB = \SP \SR$ is the braiding operator. 


Therefore, invoking Proposition~\ref{TBCC}, we conclude that
$\SO_1$ and $\SO_3$ also satisfy Proposition~\ref{SH}
with coefficients 
$A_{r;s}^{\nu;s_\2s_\1}(k)$, $r=1,2,3$ being related by
\begin{equation}\label{al23}
 A_{2;s}^{\nu;s_\2 s_\1}(k) = 
 \frac{\Omega^{s\+ ib\nu}_{s_\1 s_\2}}{\Omega^{\;s}_{s_\1 s_\2}}
 A_{1;s}^{\nu;s_\1 s_\2}(k)\,, \qquad
 A_{3;s}^{\nu;s_\2 s_\1}(k)= 
 \frac{\Omega^{\;s}_{s_\2 s_\1}}{\Omega^{s\+ ib\nu}_{s_\2 s_\1}}
 A_{1;s}^{\nu;s_\1 s_\2}(k) \,,
\end{equation}
respectively. 
The proof of Proposition~\ref{SH} is complete. \hfill $\square$\\[1ex]
{\em Proof of Lemma~\ref{idlem}}. 
Our starting point is the defining relation for the
b--Racah--Wigner symbols, equation \rf{RWdef}.
Our claim will follow from \rf{RWdef} as an identity satisfied
by the residues of the meromorphic continuation of \rf{RWdef}.
We need to analyze the relevant limits step by step.\\[1ex]
\underline{$U_1\equiv \al_s+\al_0-\al_1\ra -b$:}\\[1ex]
Note that in the limit $U_1\equiv \al_s+\al_0-\al_1\ra -b$
the contour of integration in the definition of 
$\Phi^{}_4$, eqn. \rf{Phidef}, gets pinched between the poles
of the integrand $s=Q-V_4\equiv 0$ and $s=-U_1-b$ as well as between
$s=Q-V_4+b\equiv b$ and $s=-U_1$. 
This implies that $\Phi^{}_4$ has a pole when  $U_1=-b$. In order 
extract the part which gets singular in the limit under consideration
one may deform the contour of integration in \rf{Phidef} to the
sum of two circles around $s=0$ and $s=b$ plus a contour which passes
to the right of the pole at $s=b$ and which approaches the imaginary
axis at infinity. The residue is given as 
\begin{equation}
-\frac{1}{2\pi}\sin\pi b^2
\frac{S_b(U_2)S_b(U_3)S_b(U_4)}{S_b(V_1)S_b(V_2)S_b(V_3)}
\left(1+\frac{\sin\pi bU_2\sin\pi bU_3\sin\pi bU_4}
	{\sin\pi bV_1\sin\pi bV_2\sin\pi bV_3}
\right).
\end{equation}
Considering the behavior of $\{\cdots\}$ at $U_1=-b$, one finds that
the pole of $\Phi^{}_4$ at  $U_1=-b$ is canceled by the
zero from the prefactor $(S_b(\al_s+\al_0-\al_1))^{-1}$. Taken together
one obtains the following special value for $\{\cdots\}$ at 
$\al_s+\al_0-\al_1=-b:$
\begin{align}
\big\{\begin{smallmatrix}
\al_0 & \al_1\\
\al_2 & \al_3
\end{smallmatrix}\big|\begin{smallmatrix}\al_2-\al_1-b \\\al_t
\end{smallmatrix}
\big\}=&\;\frac{S_b(Q-2\al_0-b)}{S_b(2(\al_1-\al_0-b))}
\frac{S_b(\al_t+\al_2-\al_1)}{S_b(\al_3+\al_2-\al_1+\al_0+b)}
\nn\\
&\;\frac{S_b(\al_3+\al_t-\al_0)}{S_b(\al_3+\al_t-\al_0-b)}
\frac{S_b(\al_3+\al_0-\al_t)}{S_b(Q+\al_3-\al_t-\al_0-b)}
\label{spRW}
\\ &\;
S_b(\al_3+\al_0+\al_t-Q)
\left(1+\frac{\sin\pi bU_2\sin\pi bU_3\sin\pi bU_4}
	{\sin\pi bV_1\sin\pi bV_2\sin\pi bV_3} \right).\nn
\end{align}
The parameters $U_2,U_3,U_4$ and $V_1,V_2,V_3$ are now given by
\[
\begin{aligned}
U_2&=Q-2\al_0-b\\
U_3&=\al_1-\al_0-b+\al_3+\al_2-Q\\
U_4&=\al_1-\al_0-b+\al_3-\al_2
\end{aligned}
\qquad
\begin{aligned}
V_1&=Q-\al_t+\al_3-\al_0-b\\
V_2&=\al_t+\al_3-\al_0-b\\
V_3&=2(\al_1-\al_0-b).\\
\end{aligned}
\]
\underline{$k_0\ra i\al_0$:}\\[1ex]
In the same way as in the previous paragraph one may show that
the b--Clebsch--Gordan coefficients $\big[\begin{smallmatrix}
\al_s\\
k_s\end{smallmatrix}\big|
\begin{smallmatrix}
\al_1 & \al_0\\
k_1 & k_0
\end{smallmatrix}\big]
$ and $\big[\begin{smallmatrix}
\al_3 \\
k_3\end{smallmatrix}\big|
\begin{smallmatrix}\al_t & \al_0 \\ k_t & k_0\end{smallmatrix}
\big]$ develop poles,
with residues given by \rf{CGres}.\\[1ex]
\underline{Continuation to ${\rm Re}(\al_0)=-b$ with 
 ${\rm Re}(\al_3)=\frac{Q}{2}-\de$, $0<\de<b<b^{-1}$.}\\[1ex]
The left hand side of \rf{RWdef} is analytic in the range under 
consideration. In order to describe  
the analytic continuation of the right hand side let us note that 
in the continuation from  ${\rm Re}(\al_0)=Q/2$  to ${\rm Re}(\al_0)=-b$ 
exactly three poles $\al_t=\al_t^{(k)}$, $k=-1,0,1$ 
cross the contour of integration, namely 
\[ 
\begin{aligned} k=-1: \qquad & \al_t^{(k)}=\al_3+\al_0,\\
k=0: \qquad & \al_t^{(k)}=\al_3,\\
k=1: \qquad & \al_t^{(k)}=\al_3-\al_0.
\end{aligned}
\]
The analytic continuation of the right hand side of \rf{RWdef}
may therefore be represented by replacing the 
integration contour $\BS$ in \rf{RWdef} by 
$\CC=\BS\cup\bigcup_{k=-1}^1\CC_{k}$, with $\CC_t$ being a small circle
around the poles at $\al_t=\al_t^{(k)}$.\\[1ex]
\underline{Limit $\al_1\ra-b$ with ${\rm Re}(\al_3)=\frac{Q}{2}-\de$, 
$0<\de<b<b^{-1}$.}\\[1ex]
We observe that the integral over $\BS$ vanishes due to the
factor $S_b(Q-2\al_1-b)$. This is not the case for
contributions from the poles $\al_t=\al_t^{(k)}$, $k=-1,0,1$.
Our claim now follows by straightforward computations. 
\hfill$\square$

\section{Construction of the Q--operator $\SQ(u)$}\slabel{PQT}

\subsection{Preliminaries}\label{Qprel}

Let us now enter into the construction of the $\SQ$-operators. We begin
by collecting some useful preliminaries.
We will work in the Schr{\"o}dinger representation where the operators
$\sx_r$, $r=1,\ldots,\SRN$ are diagonal. We will need
operators $\SU$, $\SOmega$, and $\SJ_s$ defined in the Schr{\"o}dinger 
representation by the following integral kernels
\begin{align}
\label{wker}
 U({\bf x},{\bf x'}) &= 
	\prod_{r=1}^\SRN \, \delta(x_{r+1} -x'_{r}) \,, \qquad
 \Omega({\bf x},{\bf x'}) = 
 	\prod_{r=1}^\SRN \, \delta(x_{r} + x'_{r}) \,, \\
\label{jker}
 {}& J_s({\bf x},{\bf x'}) =  \bigl(w_b(i\fr{Q}{2} - 2s)\bigr)^\SRN  \, 
	\prod_{r=1}^\SRN \,  D_{s-\frac{i}{2}Q}(x_r -x'_r) \,.
\end{align}
$\SU$ is the cyclic shift operator defined in \rf{upar}. 
$\SOmega$~and $\SJ_s$ are products of local operators, 
\begin{equation}\label{Wwj}
 \SOmega = \prod_{r=1}^\SRN  \somega_r \,, \qquad
 \SJ_s = \prod_{r=1}^\SRN  \sj_r \,.
\end{equation}
Here $\sj_r$ is the operator which intertwines at the site $r$ the 
representations ${\cal P}_s$ and ${\cal P}_{-s}$ of $\Uq$ 
(see~\cite{PT2}), and $\somega_r$ is the operator which realizes at 
the site $r$ the parity operation:
$\somega_r \, f(x_\1,x_\2,\ldots,x_r,\ldots,x_\SRN) =
 f(x_\1,x_\2,\ldots,-x_r,\ldots,x_\SRN)$ (whence 
 $\somega_r \, \spp_r \somega_r = \theta(\spp_r)$ and 
 $\somega_r \, \sx_r \somega_r = \theta(\sx_r)$, where the automorphism 
$\theta$ is defined by~(\ref{theta})).

We will denote the standard scalar product on $L^2(\BR)$ by
$\bra f|g \ket=
 \int_{\mathbb R} d{\bf x} \, f({\bf x}) \,g({\bf x})$.
For a given operator $\SO$, its transposed $\SO^t$ and 
hermitian--conjugated $\SO^*$ are defined, respectively, by 
(the bar denotes complex conjugation)
\begin{equation}\label{trdef}
 \bra \SO^t f|g \ket=\bra f| \SO g \ket \,, \qquad 
 \bra \overline{\SO^* f} |g \ket=\bra \overline{f} | \SO g \ket \,.
\end{equation}
This definition extends to a matrix with operator--valued
coefficients as follows
\begin{equation}\label{Ltr}
 \bigl( L^t \bigr)_{ij} = \bigl( L_{ij} \bigr)^t \,, \qquad
 \bigl( L^* \bigr)_{ij} = \bigl( L_{ij} \bigr)^* \,,
\end{equation}
i.e., component--wise. If $\SO$ is represented by
the integral kernel $O({\bf x},{\bf x'})$, then the kernels
of its transposed and hermitian--conjugated are given by
\begin{equation}\label{Ot}
 O^t({\bf x},{\bf x'}) = O({\bf x'},{\bf x}) \,, \qquad
 O^*({\bf x},{\bf x'}) = \overline{O({\bf x'},{\bf x})} \,.
\end{equation}
In particular, we have
\begin{align}
\label{uw*}
 \SU^t = &\, \SU^* = \SU^{-1} \,, \qquad 
 \SOmega^t = \SOmega^* = \SOmega \,, \\
\label{j*}
{} \SJ_s^t &= \SJ_s \,, \qquad\qquad 
	\SJ_{s}^* = \SJ_{-s} = \SJ_{s}^{-1} \,.
\end{align}

In the Schr{\"o}dinger representation we have $\sx^t = \sx$, 
$\spp^t = -\spp$, $\sx^* = \sx$, $\spp^* = \spp$ and hence 
(as seen from (\ref{EFK1})):
\begin{equation}\label{bb0}
 \se^t_s = \se_{-s} \,, \qquad
 \sff^t_s = \sff_{-s} \,, \qquad
 \sk^t_s = \sk^{-1}_{-s} \,,
\end{equation}

Properties of the transfer--matrices of models in question 
with respect to the transposition and hermitian--conjugation
are described by the following statement.
\begin{lem}\label{Tt}
For the operations  defined by (\ref{trdef}) and (\ref{Ltr}) we have
\begin{align}
\label{Ttr}
 \bigl( \ST^\XXZ_s(u) \bigr)^t &= (-1)^\SRN \, \ST^\XXZ_{-s}(-u) \,, 
 {}& \bigl( \ST^\SG_s(u) \bigr)^t &= \ST^\SG_{-s}(-u) \, \\
\label{T*}
 \bigl( \ST^\XXZ_s(u) \bigr)^* &= \ST^\XXZ_s(\bar{u}) \,, 
 {}& \bigl( \ST^\SG_s(u) \bigr)^* &= \ST^\SG_s(\bar{u})  \,,
\end{align}
where $\ST_{-s}(u) \equiv \SJ_{s} \, \ST_s(u) \, \SJ_{s}^{-1}$.
\end{lem}
\begin{proof}
Taking into account (\ref{bb0}), we observe that
the L--matrices (\ref{Lxxz}) and (\ref{lSG}) satisfy
\begin{equation}\label{bb1}
\begin{aligned}
 \bigl( L_s^\XXZ(u)\bigr)^t &= - \sigma_3 \, e^{\pi b u \sigma_3} \, 
	L_{-s}^\XXZ(-u) \, e^{-\pi b u \sigma_3} \, \sigma_3 \,, \\
 \bigl( L_s^\SG(u) \bigr)^t &= \sigma_3 \, L_{-s}^\SG(-u) \, \sigma_3 \,,
\end{aligned}
\end{equation}
Substitution of these relation into
\begin{equation}\label{bb3}
 \ST^t(u) = {\rm tr}\, \bigl( 
  (L^t_1)^\kt \cdot (L^t_2)^\kt \cdot \ldots 
    \cdot (L^t_\SRN)^\kt \bigr) =
  {\rm tr}\, \bigl( L^t_\SRN \cdot \ldots 
    \cdot L^t_2 \cdot L^t_1 \bigr)  
\end{equation}
yields~(\ref{Ttr}). Relations (\ref{T*}) are derived analogously
by noticing that we have 
$\bigl(L(u)\bigr)^* = \sigma_3 \, L(\bar{u}) \, \sigma_3$ for
both models in question.
\end{proof}

A consequence of this Lemma is that it suffices to prove 
Theorem~\ref{QXXZSG} only for $\SQ^{\,\flat}_+(u)$.  Indeed,
using (\ref{Ot}) and (\ref{Dcc}), it is easy to conclude that
\begin{equation}\label{xqpm}
 \SQ^{\,\flat}_{-}(u) =  \bigl( D_{-s}(u) \bigr)^\SRN \, 
 \bigl( \SQ^{\,\flat}_{+}(\bar{u}) \bigr)^*  \,.
\end{equation}
Therefore relations 
\hbox{(\ref{defQ1}--{\rm i})}--\hbox{(\ref{defQ1}--{\rm iii})} 
for $\SQ^{\,\flat}_-(u)$ then follow immediately if we take 
(\ref{T*}) into account.  
To check the Baxter equation \hbox{(\ref{defQ1}--{\rm iv})} for
$\SQ^{\,\flat}_-(u)$, we take hermitian--conjugation of 
\hbox{(\ref{defQ1}--{\rm iv})} for $\SQ^{\,\flat}_+(u)$, using 
\hbox{(\ref{defQ1}--{\rm ii})} and the property~(\ref{Ttr}). 
After replacement of $\bar{u}$ by $u$ this yields for 
$\tilde{\SQ}^{\,\flat}_{-}(u) \equiv 
\bigl( \SQ^{\,\flat}_{+}(\bar{u}) \bigr)^*$
the following equation
\begin{equation}\label{baxtil}
 \ST^{\,\flat}(u) \cdot \tilde{\SQ}^{\,\flat}_{-}(u)  = 
 \bigl( \overline{a( \bar{u})} \bigr )^\SRN \,  
	\tilde{\SQ}^{\,\flat}_{-}(u + ib)
 + \bigl( \overline{d( \bar{u})} \bigr)^\SRN \, 
	\tilde{\SQ}^{\,\flat}_{-}(u - ib ) \,.
\end{equation}
Using relations (\ref{Dfunrel}) and (\ref{Drefl}), we observe that
\begin{equation}\label{ads}
 \overline{a( \bar{u})} = 
	\frac{D_{-s}(u+ib)}{D_{-s}(u)} \, d(u) \,, \qquad
 \overline{d( \bar{u})} = \frac{D_{-s}(u-ib)}{D_{-s}(u)} \, a(u) \,.
\end{equation}
Whence we conclude that $\SQ^{\,\flat}_-(u)$ defined by (\ref{xqpm}) 
satisfies \hbox{(\ref{defQ1}--{\rm iv})} (with the same coefficients 
$a(u)$, $d(u)$ as $\SQ^{\,\flat}_+(u)$ does).

\subsection{Construction of Q--operators}\label{QYZ}

In order to construct the Q--operators explicitly let us consider 
the following general ansatz for the Q--operator:
\begin{equation}\label{qyz}
 \SQ(u) = \SY(u) \cdot \SZ \,.
\end{equation}

We will prove Theorem~\ref{QXXZSG} for $\SQ^{\,\flat}_+(u)$ in three 
steps: first constructing a suitable solution for $\SY^{\,\flat}(u)$ by 
requiring the Baxter equation to hold, then determining the form of 
$\SZ^{\,\flat}$, and finally checking that the obtained Q--operator 
satisfies~(\ref{defQ1}). 
The first step in this proof is based on the idea to find such a gauge
transformation of the L--matrix that it becomes 
effectively upper--triangular. This approach was originally
applied by Pasquier and Gaudin~\cite{PG} to the Toda chain.
Our computation has many similarities with the modification
of this approach developed in \cite{De,DKM} for the non--compact
XXX magnet.

\begin{propn}\label{Yxxz}
Let $\ST^\flat(u)$, $\flat= \XXZ,\, \SG$ be the 
transfer--matrices corresponding to the L--matrices
(\ref{Lxxz}) and (\ref{lSG}).
Let $\SY^\flat(u)$ be defined in the Schr{\"o}dinger 
representation by the kernel 
\begin{equation}\label{y1}
 Y_u({\bf x},{\bf x'}) = \prod_{r=1}^\SRN  
 D_{\frac{1}{2}(u-\sigma)}(x_r - \ve_\flat\, x'_{r+1}) \, 
 D_{-\frac{1}{2}(u+\sigma)}(x_r - x'_{r}) \,, 
\end{equation}
where $\ve_\XXZ = 1$, $\ve_\SG = -1$.
Then $\SQ^{\flat}(u)$ of the form (\ref{qyz})
satisfies the Baxter equation~\hbox{(\ref{defQ1}--{\rm iv})} with
coefficients $a(u)$, $d(u)$ as specified in eq.~(\ref{baxxz})
of Theorem~\ref{QXXZSG}.
\end{propn}

\begin{proof}
 Let us introduce the gauge--transformed Lax operators (the
transformation depends on the site number $r$):
\begin{align}
\label{LGxxz} 
 \widetilde{L}^\flat_r(u) &= 
   G^\flat_{r+1} \cdot L^\flat_r(u) 
  \cdot \bigl( G^\flat_{r} \bigr)^{-1} \,, \qquad
 G^\flat_{r} = \bigl(\begin{smallmatrix} 1 & 0\\ \rho^\flat_r & 1
 \end{smallmatrix} \bigr) \,, \\
 \rho^\XXZ_r &= e^{\pi b (2 x'_r - u)} \,, \qquad
  \rho^\SG_r = e^{2 \pi b x'_r } \,.
\end{align}
The relevant matrix elements of the new Lax matrices
are given by
\begin{align}
 \Bigl( \widetilde{L}^\flat_r (u) \Bigr)_{21} =& \,
 4 \varkappa_\flat\, (\rho_r \rho_{r+1})^{\frac 12} \label{L21}\\
 & \times\Bigl(
 \cosh \pi b (\sx_r - \ve_\flat\, x'_{r+1} + \fr{1}{2}(\sigma -u)) \,
 \cosh \pi b (\sx_r - x'_{r} + \fr{1}{2}(\sigma +u)) \,\sk_r \nn\\
 & - \cosh \pi b (\sx_r - \ve_\flat\, x'_{r+1} - \fr{1}{2}(\sigma -u)) \,
 \cosh \pi b (\sx_r - x'_{r} - \fr{1}{2}(\sigma +u)) 
 \,\sk^{-1}_r  \Bigr) \,, \nn\\
\nonumber
\Bigl( \widetilde{L}^\flat_r (u) \Bigr)_{11} =&\,
 2 \varkappa_\flat\, e^{\pi b (x'_r -\ve_\flat\, \sx_r)} \, \Bigl( 
 e^{\frac{1}{2} \ve_\flat\, \pi b (u-\sigma)} \,
 \cosh \pi b (\sx_r - x'_{r} + \fr{1}{2}(\sigma +u)) \,\sk_r \\
\label{L11}
 & -\, e^{\frac{1}{2} \ve_\flat\, \pi b (\sigma -u)} \,
 \cosh \pi b (\sx_r - x'_{r} - \fr{1}{2}(\sigma +u)) 
 \,\sk^{-1}_r  \Bigr) \,,\\
\nonumber
\Bigl( \widetilde{L}^\flat_r (u) \Bigr)_{22} =&\,
 2 \varkappa_\flat\, e^{\pi b (x'_{r+1} -\sx_r)} \, \Bigl( 
 e^{\frac{1}{2} \pi b (\sigma +u)} \,
 \cosh \pi b (\sx_r - \ve_\flat\, x'_{r+1} +
 \fr{1}{2}(u-\sigma)) \,\sk^{-1}_r  \\
\label{L22}
 & -\,  e^{-\frac{1}{2} \pi b (u+\sigma)} \,
 \cosh \pi b (\sx_r - \ve_\flat\, x'_{r+1} + \fr{1}{2}(\sigma -u))  
  \, \sk_r  \Bigr) \,,
\end{align}
where $\varkappa_\XXZ =1$, $\varkappa_\SG = -i e^{-\pi b s}$.

In the Schr\"odinger representation, operators
$\sk_r$, $\sk^{-1}_r$ act as shifts of $x_r$ by
$\ppm \fr{i}{2}b$. Using the functional relation
(\ref{Dfunrel}), it is straightforward to 
apply (\ref{L21}) to $Y^\flat_u({\bf x},{\bf x'})$ and
verify that the condition
\begin{equation}\label{l21a}
\bigl( \widetilde{L}^\flat_r (u) \bigr)_{21} \, 
	Y^\flat_u({\bf x},{\bf x'}) = 0
\end{equation}
is satisfied for all ${\bf x'}\in {\mathbb R}^\SRN$. 
This implies that $\widetilde{L}^\flat_r(u)$ becomes upper 
triangular when acting on $Y^\flat_u({\bf x},{\bf x'})$ so 
that we can calculate the action of 
$\widetilde{\SM}^{\flat}(u)$ on 
$Y^\flat_u({\bf x},{\bf x'})$ as
\begin{equation}\label{MY1}
 \widetilde{\SM}^{\flat}(u) \, Y^\flat_u({\bf x},{\bf x'}) =
 \left( \begin{matrix} \prod_{r=1}^\SRN 
 \bigl( \widetilde{L}^\flat_r (u) \bigr)_{11} & * \\ 
 0 & \prod_{r=1}^\SRN 
 \bigl( \widetilde{L}^\flat_r (u) \bigr)_{22} \end{matrix} \right) \,
 Y^\flat_u({\bf x},{\bf x'}) \,.
\end{equation}
Hence, taking into account the periodicity
condition, $G^\flat_{\SRN+1}=G^\flat_1$, we have
\begin{equation}\label{TY1}
 \ST^{\flat}(u) \, Y^\flat_u({\bf x},{\bf x'}) \,=\,
  \Bigg(\, \prod_{r=1}^\SRN 
 \bigl( \widetilde{L}^\flat_r (u) \bigr)_{11} +
 \prod_{r=1}^\SRN 
 \bigl( \widetilde{L}^\flat_r (u) \bigr)_{22} \,\Bigg) \,
 Y^\flat_u({\bf x},{\bf x'}) \,.
\end{equation}
Applying (\ref{L11}), (\ref{L22}) to $Y^\flat_u({\bf x},{\bf x'})$
and using (\ref{l21a}), we derive
\begin{equation}\label{l11a}
\begin{aligned}
 \bigl( \widetilde{L}^\flat_r (u) \bigr)_{11} \, 
  Y^\flat_u({\bf x},{\bf x'})
 =&\ 2 \ve_\flat\, \varkappa_\flat\, e^{\pi b (x'_r - x'_{r+1})} \, 
 \sinh \pi b(u-\sigma) \, Y^\flat_{u-ib}({\bf x},{\bf x'})\,, \\
 \bigl( \widetilde{L}^\flat_r (u) \bigr)_{22} \, 
  Y^\flat_u({\bf x},{\bf x'})
 =&\ 2 \varkappa_\flat\, e^{\pi b (x'_{r+1}-x'_r)} \, 
 \sinh \pi b(u+\sigma) \, Y^\flat_{u+ib}({\bf x},{\bf x'})\,.
\end{aligned}
\end{equation}
Here we have used that 
$Y^\flat_u({\bf x},{\bf x'}) = \prod_{r=1}^\SRN Y^\flat(u,x_r)$, 
where each factor satisfies the relation
\begin{equation}\label{yy}
 \frac{Y^\flat(u,x_r \ppm i \frac{b}{2})}{Y^\flat(u \ppm ib,x_r)} 
 = \Bigl( \frac{\cosh \pi b(x_r -x'_r - \frac{1}{2}(\sigma +u))}%
 {\cosh \pi b(x_r - \ve_\flat\, x'_{r+1} + \frac{1}{2}(\sigma -u))} 
 \Bigr)^{\ppm 1} \,.
\end{equation}
Combining (\ref{TY1}) with (\ref{l11a}), we obtain
\begin{equation}\label{ty1}
\begin{aligned}
  \ST^{\flat}(u) \, Y^\flat_u({\bf x},{\bf x'}) =\;
 &\bigl(2 \varkappa_\flat\, \sinh \pi b (u+\sigma)\bigr)^\SRN \, 
   Y^\flat_{u + ib}({\bf x},{\bf x'}) \\& +
 \bigl( 2 \ve_\flat\, \varkappa_\flat\, 
 	\sinh \pi b (u-\sigma)\bigr)^\SRN \, 
   Y^\flat_{u - ib}({\bf x},{\bf x'}) \,,
\end{aligned}
\end{equation}
which implies that the Baxter equation \hbox{(\ref{defQ1}--{\rm iv})}
holds with the coefficients $a(u)$, $d(u)$ as specified
in eq.~(\ref{baxxz}). 
\end{proof}

The possible form of~$\SZ$ can be found from the 
requirement that \hbox{(\ref{defQ1}--{\rm iii})} holds.

\begin{propn}\label{ZXXZ}
Let $\SY^\flat(u)$ be chosen as in Proposition~\ref{Yxxz}.
Then the commutativity condition
\begin{equation}\label{qt1}
 \SQ^{\flat}(u) \, \ST^{\flat}(u) = \ST^{\flat}(u) \, \SQ^{\flat}(u)
\end{equation}
holds for $ \SQ^{\flat}(u)$ of the form (\ref{qyz}) provided that 
the corresponding operator $\SZ^\flat$ satisfies the following relation
\begin{equation}\label{zz}
 \SZ^\flat \, \ST_s^{\flat} (u)  = 
  \ST_{-s}^{\flat}(u) \, \SZ^\flat  \,.
\end{equation}
\end{propn}

\begin{proof}
In order to treat both models in a uniform way, let us 
introduce the operator 
\begin{equation}\label{Wb}
   \SOmega^\flat = \begin{cases} {\sf 1}\,, & \flat = \XXZ \\
    \SOmega\,, & \flat = \SG \end{cases}\,,
\end{equation}
where the parity operator was defined in~(\ref{wker}). Then,
using the explicit expressions (\ref{y1}) and taking (\ref{Dpar}) 
into account, it is easy to verify that (the subscript $\bf x$ or 
$\bf x'$ of an operator specifies the argument on which it acts) 
\begin{equation}\label{bb7}
 \SOmega^\flat_{\bf x} \, Y^\flat_u({\bf x},{\bf x'}) =  
 \SOmega^\flat_{\bf x'} \, Y^\flat_u({\bf x},{\bf x'}) =
 \SU_{\bf x}^{-1} \, Y^\flat_{-u}({\bf x'},{\bf x}) \,.
\end{equation}
Now we derive 
\begin{equation}
\begin{aligned}
\nonumber
{} & \ST_{s;\bf x}^{\,\flat}(u) \, Y^\flat_u({\bf x},{\bf x'}) 
 \={ty1} \bigl( a(u) \bigr)^\SRN \, 
   Y^\flat_{u - ib}({\bf x},{\bf x'}) +
 \bigl( d(u) \bigr)^\SRN \, 
   Y^\flat_{u + ib}({\bf x},{\bf x'})  \\
{} & \={bb7} (-\ve_\flat)^\SRN \, \SOmega^\flat_{\bf x} \, 
 \SU_{\bf x}^{-1} \, \Bigl( \bigl( d(-u) \bigr)^\SRN \, 
   Y^\flat_{-u + ib}({\bf x'},{\bf x}) 
 + \bigl( a(-u) \bigr)^\SRN \,
 Y^\flat_{-u - ib}({\bf x'},{\bf x}) \Bigr) \nn \\
{} & \={ty1} (-\ve_\flat)^\SRN \, \SOmega^\flat_{\bf x} \,\SU_{\bf x}^{-1} \,
 \ST_{s;\bf x'}^{\,\flat}(-u) \, Y^\flat_{-u}({\bf x'},{\bf x}) 
 \={bb7} (-\ve_\flat)^\SRN \, \ST_{s;\bf x'}^{\,\flat}(-u) \, 
  Y^\flat_u({\bf x},{\bf x'}) \\
{} & = (-\ve_\flat)^\SRN \, Y^\flat_u({\bf x},{\bf x'}) \, 
	\bigl( \ST_{s;\bf x'}^{\SG}(-u) \bigr)^t
 \={Ttr} Y^\flat_u({\bf x},{\bf x'}) \,  \ST_{-s;\bf x'}^{\,\flat}(u) \,.
\end{aligned}
\end{equation}
Thus, we verified that $\ST_{s}^{\flat}(u) \, \SY(u) = 
 \SY(u) \, \ST_{-s}^{\flat}(u) $, which is equivalent to (\ref{qt1})
if relation (\ref{zz}) is satisfied.
\end{proof}

Proposition~\ref{ZXXZ} implies that we can choose 
\begin{equation}\label{zzj}
   \SZ^{\flat} =  \SJ_s \,,
\end{equation}
where $\SJ_s$ was defined in~(\ref{jker}). Then we compute the 
integral kernel of $\SQ^{\flat}_+(u)$:
\begin{align} \label{qker1}
 & Q_{+;u}^{\,\flat} ({\bf x},{\bf x'}) = 
  \bigl(w_b(i\fr{Q}{2} - 2s) \bigr)^\SRN 
 \int\limits_{\BR^{\SRN}}dz_1\dots dz_\SRN\,  \prod_{r=1}^\SRN  \,
 D_{\frac{u-\sigma}{2}}(z_{r} - \ve_\flat\, x_{r-1}) \,  \\[-1ex] 
{} & \hspace{7.3cm}\times
 D_{-\frac{u+\sigma}{2}}(z_{r} - x_{r}) \,
 D_{\bar\sigma}(z_{r} - x'_{r}) \nn \\
\label{qker2}
 & \quad= \bigl( D_{-s}(u) \bigr)^\SRN\, \prod_{r=1}^\SRN
 D_{\frac{\bar\sigma -u}{2}}(x_{r} -  x'_{r}) \,
 D_{\frac{\bar\sigma +u}{2}}(x_{r-1} - \ve_\flat\, x'_{r}) \,
 D_{-s}(x_{r} - \ve_\flat\, x_{r-1}) \,.
\end{align}
Equivalence of (\ref{qker1}) and (\ref{qker2}) is due to the 
identity~(\ref{D3}). 

\begin{rem}
The Baxter equation \hbox{(\ref{defQ1}--{\rm iv})} along
with the self--commutativity \hbox{(\ref{defQ1}--{\rm ii})}
relation, which will be proven below, imply that
\rf{qt1} extends to commutativity
of $\SQ_+^{\flat}(u)$ with $\ST^{\flat}(v)$ for those values of $v$, 
where $\SQ_+^{\flat}(v)$ is invertible. 
\end{rem}

\subsection{Proof of commutativity relations}

To complete the proof of Theorem~\ref{QXXZSG},
we have to establish relations \hbox{(\ref{defQ1}--{\rm i})} 
and \hbox{(\ref{defQ1}--{\rm ii})}.

\begin{lem}\label{YYB} 
Let $\SY^\flat(u)$ be chosen as in Proposition~\ref{Yxxz}. 
Then the following identities hold
\begin{align}
\label{yyb1}
 \bigl( \SY^\flat(\bar{u}) \bigr)^* \cdot \SY^\flat(v) &=
 \bigl( D_{s}(u) \, D_{-s}(v) \bigr)^\SRN \, 
 \bigl( \SY^\flat(\bar{v}) \bigr)^* \cdot \SY^\flat(u) \,, \\
\label{yyb2}
 \SY^\flat(u) \cdot \bigl( \SY^\flat(\bar{v}) \bigr)^* &= 
  \bigl( D_{s}(v) \, D_{-s}(u) \bigr)^\SRN \,
 \SY^\flat(v) \cdot \bigl( \SY^\flat(\bar{u}) \bigr)^* \,.
\end{align}
\end{lem}
\begin{proof} These identities are just particular cases of
the integral identity~(\ref{D4}). Indeed, let us denote
$\alpha_u \equiv \fr{1}{2}(u-\sigma)$,
$\beta_u \equiv -\fr{1}{2}(u+\sigma)$.
We will also use the notation $\alpha^\star\equiv -\fr{i}{2}Q -\alpha$.
Let us consider the operator
$\SV^\flat(u,v) = 
	\bigl( \SY^\flat(\bar{u}) \bigr)^* \cdot \SY^\flat(v)$.
Its kernel is given by 
\begin{align}\label{mker3}
V{\,^\flat}_{u,v}({\bf x},{\bf x'}) 
 = \int\limits_{\BR^{\SRN}}dz_1\dots dz_\SRN 
 \; \prod_{r=1}^\SRN \,& D_{\alpha^\star_u}(z_r - \ve_\flat \, x_{r+1}) \,
 D_{\beta^\star_u}(z_{r} -  x_{r}) \\[-1ex] 
 \times & D_{\alpha_v}(z_r - \ve_\flat \, x'_{r+1}) \,
 D_{\beta_v}(z_{r} - x'_r) \,, \nn
\end{align}
Now we can apply identity (\ref{D4}) choosing $\alpha=\alpha^\star_u$,
$\beta=\alpha_v$, $\gamma=\beta^\star_u$, $\omega=\beta_v$, and
$u= \ve_\flat \, x_{r+1}$, $v= \ve_\flat \, x'_{r+1}$, 
$w= x_{r}$, $z=x'_r$. This yields
\begin{align}
\nonumber
V^{\,\flat}_{u,v}({\bf x},{\bf x'}) 
 =&\, \bigl(A(\alpha^\star_u,\alpha_v,\beta^\star_u,\beta_v)\bigr)^\SRN \
  \prod_{r=1}^\SRN  
  { D_{\frac{v-u}{2}} \left( \ve_\flat ( x_{r+1} - x'_{r+1}) \right) }
  { D_{\frac{u-v}{2}} ( x_{r}-x'_r ) } \\
\nonumber
{} &\times  \int\limits_{\BR^{\SRN} }dz_1\dots dz_\SRN 
{}  \prod_{r=1}^\SRN  \,
 D_{\alpha_u}(z_r- \ve_\flat \, x'_{r+1}) \, D_{\beta_u}(z_{r} - x'_r) \, 
 \\{} & \hspace{28mm} \times 
 D_{\alpha^\star_v}(z_r- \ve_\flat \, x_{r+1}) \, 
 D_{\beta^\star_v}(z_{r} - x_{r}) \nn \\
\label{Wker1}
  =&\, \bigl( D_{s}(u) \, D_{-s}(v) \bigr)^\SRN \ 
 V^{\,\flat}_{v,u}({\bf x},{\bf x'}) \,.
\end{align}
Here we used (\ref{Dpar}), the definition (\ref{Adef}) of the function 
$A(\alpha_1,\alpha_2,\ldots)$, and took into account the periodic 
boundary conditions. 

Identity (\ref{yyb2}) can be proven absolutely analogously.
\end{proof}

\begin{propn}\label{QQXXZ}
The operators $\SQ_+^{\flat}(u)$ and $\SQ_-^{\flat}(u)$
with the kernels given in Theorem~\ref{QXXZSG} by eqs.
(\ref{Qxxz+}) and (\ref{Qxxz-}), respectively, satisfy
the following commutativity and exchange relations
\begin{align}
 \label{qq3a}
{} \SQ_+^{\flat}(u) \, \SQ_+^{\flat}(v) =  
  \SQ_+^{\flat}(v) \, \SQ_+^{\flat}(u) \,, &\qquad
  \SQ_-^{\flat}(u) \, \SQ_-^{\flat}(v) =  
  \SQ_-^{\flat}(v) \, \SQ_-^{\flat}(u) \,, \\
 \label{qq3b}
{}  \SQ_+^{\flat}(u) \, \SQ_-^{\flat}(v) =  
  \SQ_-^{\flat}(u) \, \SQ_+^{\flat}(v) &=
  \SQ_+^{\flat}(v) \, \SQ_-^{\flat}(u) =
  \SQ_-^{\flat}(v) \, \SQ_+^{\flat}(u) \,.
\end{align}
for all $u,v \in \BC$.
\end{propn}

\begin{proof}
Observe that, using (\ref{Ot}) and (\ref{Dcc}), the equality of 
(\ref{qker1}) and (\ref{qker2}) can be written in the following 
operator form:
\begin{equation}\label{qqa1}
 \SQ_+^{\flat}(u) = 
  \SY^\flat(u) \cdot \SJ_s = \bigl( D_{-s}(u) \bigr)^\SRN\, 
  \bigl( \SX^\flat_{s})^{-1} \cdot \SOmega^\flat \cdot 
  \SU^{-1} \cdot \bigl( \SY^\flat(\bar{u}) \bigr)^*  \,,
\end{equation}
where we used the notation (\ref{Wb}) and introduced
\begin{equation}\label{Xs}
 \SX^\flat_{s} = \prod_{r=1}^\SRN 
 	D_{s}(\sx_r - \ve_\flat \sx_{r-1}) \,.
\end{equation}
Using Lemma~\ref{YYB}, we can write down the product of two 
such Q--operators as follows
\begin{equation}\label{qq1}
\begin{aligned} 
 {}& \SQ_+^{\flat}(u) \, \SQ_+^{\flat}(v) = 
 \bigl(D_{-s}(u) \bigr)^\SRN \, \bigl(\SX^\flat_{s}\bigr)^{-1} 
 \cdot \SOmega^\flat \cdot \SU^{-1} \cdot
 \bigl( \SY^\flat(\bar{u}) \bigr)^* \cdot \SY^\flat(v) \cdot \SJ_s \\
 {} & \={yyb1} \bigl(D_{-s}(v) \bigr)^\SRN \, 
 \bigl(\SX^\flat_{s}\bigr)^{-1} \cdot \SOmega^\flat\cdot \SU^{-1} 
 \cdot \bigl( \SY^\flat(\bar{v}) \bigr)^* \cdot \SY^\flat(u) 
 \cdot \SJ_s  = \SQ_+^{\flat}(v) \, \SQ_+^{\flat}(u) \,.
\end{aligned}
\end{equation}
This proves the first relation in (\ref{qq3a}) and hence
\hbox{(\ref{defQ1}--{\rm ii})} for ~$\SQ_+^{\flat}(u)$. As was 
explained in Subsection~\ref{Qprel}, relation 
\hbox{(\ref{defQ1}--{\rm ii})} for ~$\SQ_-^{\flat}(u)$ (i.e., the 
second relation in~(\ref{qq3a})) follows then as a consequence of the 
relation (\ref{xqpm}) between $\SQ_+^{\flat}(u)$ and~$\SQ_-^{\flat}(u)$. 
By the same token, relation \hbox{(\ref{defQ1}--{\rm i})} is equivalent 
to~(\ref{qq3b}). To prove the latter relation, we substitute (\ref{qqa1}) 
into (\ref{xqpm}) and use (\ref{uw*})--(\ref{j*}). This yields
the operator~$\SQ^\flat_-(u)$ in the following form:
\begin{equation}\label{qqa3x}
 \SQ_-^{\flat}(u) = \bigl( D_{-s}(u) \bigr)^\SRN \,
 \SJ_{s}^{-1} \cdot \bigl( \SY^\flat(\bar{u}) \bigr)^* =
 \SY^\flat(u) \cdot \SU \cdot \SOmega^\flat \cdot \SX^\flat_s \,. 
 \end{equation}
As seen from \rf{qqa1} and \rf{qqa3x}, the two expressions on the 
l.h.s. of \rf{qq3b} are just two ways to write down 
$\bigl( D_{-s}(v) \bigr)^\SRN \,
	\SY^\flat(u) \cdot \bigl( \SY^\flat(\bar{v}) \bigr)^*$. 
Analogously, the two expressions on the r.h.s. of \rf{qq3b} 
are two ways to write down 
$\bigl( D_{-s}(u) \bigr)^\SRN \, 
  \SY^\flat(v) \cdot \bigl( \SY^\flat(\bar{u}) \bigr)^*$. 
The middle equality in \rf{qq3b} is due to the identity (\ref{yyb2}) 
in Lemma~\ref{YYB}.
\end{proof}


\begin{thebibliography}{11}
\newcommand{\CMP}[3]{{ Commun. Math. Phys. }{\bf #1} (#2) #3}
\newcommand{\LMP}[3]{{ Lett. Math. Phys. }{\bf #1} (#2) #3}
\newcommand{\IMP}[3]{{ Int. J. Mod. Phys. }{\bf A#1} (#2) #3}
\newcommand{\NP}[3]{{ Nucl. Phys. }{\bf B#1} (#2) #3}
\newcommand{\PL}[3]{{ Phys. Lett. }{\bf B#1} (#2) #3}
\newcommand{\MPL}[3]{{ Mod. Phys. Lett. }{\bf A#1} (#2) #3}
\newcommand{\PRL}[3]{{ Phys. Rev. Lett. }{\bf #1} (#2) #3}
\newcommand{\AP}[3]{{ Ann. Phys. (N.Y.) }{\bf #1} (#2) #3}
\newcommand{\LMJ}[3]{{ Leningrad Math. J. }{\bf #1} (#2) #3}
\newcommand{\FAA}[3]{{ Funct. Anal. Appl. }{\bf #1} (#2) #3}
\newcommand{\TMP}[3]{{ Theor. Math. Phys. }{\bf #1} (#2) #3}
\newcommand{\PTPS}[3]{{ Progr. Theor. Phys. Suppl. }{\bf #1} (#2) #3}
\newcommand{\LMN}[3]{{ Lecture Notes in Mathematics }{\bf #1} (#2) #2}
\small  \setlength{\itemsep}{-3pt}

\bibitem[Ba]{Ba} E.W.~Barnes: {\em Theory of the double gamma function}, 
 Phil. Trans. Roy. Soc. {\bf A196} (1901) 265--388

\bibitem[BL]{BL} V.~Brazhnikov and S.~Lukyanov:
{\em Angular quantization and form factors in massive integrable models}
\NP{512}{1998}{616--636}

\bibitem[BD]{BD} A.G.~Bytsko and A.~Doikou:
{\em Thermodynamics and conformal properties of XXZ chains 
with alternating spins},
{J.\ Phys.} {\bf A37} (2004) 4465--4492

\bibitem[BT]{BT} A.~Bytsko and J.~Teschner: {\em R--operator, 
co--product and Haar--measure for the modular double of $\Uq$},
\CMP{240}{2003}{171--196} 

\bibitem[De]{De} S.E.~Derkachov: 
{\em Baxter's Q--operator for the homogeneous XXX spin chain},
J.~Phys.\ {\bf A32} (1999) 5299--5316

\bibitem[DKK]{DKK} 
S.E.~Derkachov, D.~Karakhanyan, and R.~Kirschner: 
{\em Baxter Q--operators of the XXZ chain and 
R--matrix factorization}, 
Nucl. Phys. {\bf B738} (2006) 368--390

\bibitem[DKM]{DKM}
S.E.~Derkachov, G.P.~Korchemsky, and A.~N.~Manashov:
{\em Noncompact Heisenberg spin magnets from high--energy QCD.
 I. Baxter Q--operator and separation of variables},  
Nucl.\ Phys.\ {\bf B617} (2001) 375--440;
{\em Separation of variables for the quantum $SL(2,\BR)$ spin chain},
JHEP {\bf 0307} (2003) 047

\bibitem[F1]{F1} 
L.D.~Faddeev: {\em How algebraic Bethe ansatz works for 
integrable model}. In: {\em Sym\'{e}tries quantiques}
 (North-Holland, 1998), 149--219 [hep-th/9605187]

\bibitem[F2]{F2} L.D.~Faddeev: {\em Discrete Heisenberg--Weyl group 
and modular group}, \LMP{34}{1995}{249--254} 

\bibitem[F3]{F3} L.D.~Faddeev: {\em Modular double of a quantum group}, 
 Math.\ Phys.\ Stud.\ {\bf 21} (2000) 149--156 

\bibitem[FK2]{FK2}
L.D.~Faddeev and R.M.~Kashaev: {\em Quantum dilogarithm}, 
Mod.\ Phys.\ Lett.\ {\bf A9} (1994) 427--434

\bibitem[FKV]{FKV}
L.D.~Faddeev, R.M.~Kashaev, and A.Yu.~Volkov: {\em Strongly 
coupled quantum discrete Liouville theory. I: Algebraic  
approach and duality}, \CMP{219}{2001}{199--219}

\bibitem[FK1]{FK1} L.D.~Faddeev and G.P.~Korchemsky: {\em High energy
 QCD as a completely integrable model}, Phys.\ Lett.\
 {\bf B342} (1995) 311--322 

\bibitem[FST]{FST}
L.D.~Faddeev, E.K.~Sklyanin, and L.A.~Takhtajan:
{\em Quantum inverse problem method.~I},
 Theor.\ Math.\ Phys. {\bf 40} (1979) {688--706}

\bibitem[FTT]{FTT}
L.D.~Faddeev, V.O.~Tarasov, and L.A.~Takhtajan: {\em Local 
Hamiltonians for integrable quantum models on a lattice},
\TMP{57}{1983}{1059--1073}

\bibitem[FT]{FT}
L.D.~Faddeev and O.~Tirkkonen: {\em Connections of the 
Liouville model and XXZ spin chain},
\NP{453}{1995}{647--669}

\bibitem[FV]{FV} L.D.~Faddeev and A.Yu.~Volkov:
 {\em  Yang--Baxterization of the quantum dilogarithm},
 Zapiski nauch.\ semin.\ POMI {\bf 224} (1995) 146--154
 [Engl.\ transl.: J.~Math.\ Sci.\ {\bf 88} (1998) {202--207}]

\bibitem[FMS]{FMS}
A.~Fring, G.~Mussardo, and P.~Simonetti:
{\em Form--factors for integrable Lagrangian field theories, 
 the sinh--Gordon theory}, Nucl.\ Phys.\ {\bf B393} (1993) 413--441

\bibitem[IK]{IK}
A.G.~Izergin and V.E.~Korepin: {\em Lattice versions of 
quantum field theory models in two dimensions},
\NP{205}{1982}{401--413}

\bibitem[Ji]{Ji} 
M.~Jimbo: {\em A $q$--difference analogue of $U(gl(N+1))$ and the
 Yang-Baxter equations}, {Lett.\ Math.\ Phys.} {\bf 10} (1985) 63--69

\bibitem[K1]{K1}
R.M.~Kashaev: {\em The non--compact quantum dilogarithm and the 
Baxter equations}, J.\ Stat.\ Phys.\ {\bf 102} (2001) 923--936

\bibitem[K2]{K2}
R.M.~Kashaev: {\em The quantum dilogarithm and Dehn
twists in quantum Teichm{\"u}ller theory}, In:
Integrable structures of exactly solvable two--dimensional
models of quantum field theory (Kiev, 2000), 211--221
(NATO Sci.Ser.II Math.Phys.Chem., {\bf 35}, Kluwer Acad.\
Publ., Dordrecht, 2001)

\bibitem[KL]{KL} S.~Kharchev, D.~Lebedev:
 {\em Integral representation for the eigenfunctions 
      of quantum periodic Toda chain},
  Lett. Math. Phys. {\bf 50} (1999) 53--77

\bibitem[KLS]{KLS}
S.~Kharchev, D.~Lebedev, and M.~Semenov--Tian--Shansky:
{\em Unitary representations of $\Uq$, 
the modular double, and the multiparticle $q$--deformed Toda chains},
\CMP{225}{2002}{573--609} 

\bibitem[KM]{KM}
M.~Kirch and A.N.~Manashov: {\em Noncompact $SL(2,\BR)$ 
spin chain}, JHEP {\bf 0406} (2004) 035

\bibitem[KBI]{KBI} V.~Korepin, N.~Bogoliubov, and A.~Izergin:
{\em Quantum inverse scattering method and correlation functions}
 (Cambridge U.\ Press, 1993)

\bibitem[KMu]{KMu} A.~Koubek and G.~Mussardo: 
{\em On the operator content of the sinh--Gordon model},
\PL{311}{1993}{193--201}

\bibitem[Le]{Le} G. Lechner: {\em An existence proof for interacting
  quantum field theories with a factorising S--matrix}, 
  math-ph/0601022

\bibitem[Li]{Li} L.N.~Lipatov: {\em High energy asymptotics of 
 multi--colour QCD and exactly solvable lattice models},
 JETP Lett.\ {\bf 59} (1994) 596--599

\bibitem[Lu]{Lu} S.~Lukyanov: {\em Finite temperature expectation
 values of local fields in the sinh--Gordon model},
 Nucl.\ Phys.\ {\bf B612} (2001) 391--412

\bibitem[PG]{PG} V.~Pasquier and M.~Gaudin: {\em The periodic
Toda chain and a matrix generalization of the Bessel function},
J.~Phys.\ {\bf A25} (1992) 5243--5252

\bibitem[PT1]{PT1}
B.~Ponsot and J.~Teschner: {\em Liouville bootstrap via 
harmonic analysis on a non--compact quantum group},
 hep-th/9911110

\bibitem[PT2]{PT2} B.~Ponsot and J.~Teschner: {\em Clebsch--Gordan 
and Racah--Wigner coefficients for a continuous series of 
representations of $\Uq$}, 
\CMP{224}{2001}{613--655} 

\bibitem[Ru]{Ru} S.N.M.~Ruijsenaars: {\em First order analytic 
 difference equations and integrable quantum systems},
J.~Math.\ Phys.\ {\bf  38} (1997) 1069--1146

\bibitem[S]{S} K.~Schm\"udgen: {\em Integrable operator representations 
of ${\BR}^2_q$, $X_{q,\gamma}$ and ${\rm SL}_q(2,{\BR})$}, 
\CMP{159}{1994}{217--237}

\bibitem[Sh]{Sh} T.~Shintani: {\em On a Kronecker limit formula for real 
 quadratic fields}, J.~Fac.\ Sci.\ Univ.\
 Tokyo Sect.\ 1A Math.\ {\bf 24} (1977) 167--199

\bibitem[S1]{S1}
E.K.~Sklyanin: {\em Exact quantization of the sinh--Gordon 
model}, \NP{326}{1989}{719--736}

\bibitem[S2]{Sk2}
E.K.~Sklyanin: {\em The quantum Toda chain},
 Lect.\ Notes Phys.\ {\bf 226} (1985) 196--233

\bibitem[S3]{Sk3}
E.K.~Sklyanin: {\em Quantum inverse scattering method. 
 Selected topics}. In: {\em Quantum groups and quantum
integrable systems} (World Scientific, 1992) 63--97
[hep-th/9211111]; {\em Separation of variables -- new trends},
Prog.\ Theor.\ Phys.\ Suppl.\  {\bf 118} (1995) 35--60

\bibitem[Sm1]{Sm} F.A.~Smirnov: {\em Quasi--classical study of 
form factors in finite volume}. In: {\em L.D.~Faddeev's seminar on 
 mathematical physics} (AMS Transl. Ser. 2, 201, AMS, 
  Providence, RI, 2000) 283--307 [hep-th/9802132]

\bibitem[Sm2]{Sm2} F.A.~Smirnov: {\em 
   Baxter equations and deformation of abelian differentials},
   Int. J. Mod. Phys. {\bf A19S2} (2004) 396--417.

\bibitem[Ta]{Ta}
V.O.~Tarasov: {\em Irreducible monodromy matrices for 
the R matrix of the XXZ model and local lattice quantum 
Hamiltonians}, \TMP{63}{1985}{440--454}


\bibitem[T1]{T1} J.~Teschner: {\em On structure constants and fusion 
 rules in the $SL(2,\BC)/SU(2)$--WZNW model}, \NP{546}{1999}{390--422};
   {\em Operator product expansion and factorization in the 
   $H_3^+$--WZNW model}, \NP{571}{2000}{555--582};
   {\em Crossing symmetry in the $H_3^+$ WZNW model},
   Phys.\ Lett.\ {\bf B521} (2001) 127--132

\bibitem[T2]{T2} J.~Teschner:
  {\em Liouville theory revisited},
  {Class.\ Quant.\ Grav.} {\bf 18} (2001)  R153--R222; 
  {\em A lecture on the Liouville vertex operators},
  Int. J. Mod. Phys. {\bf A19S2} (2004) 436--458
 
\bibitem[Ti]{Ti} E.C.~Titchmarsh: {\em The theory of functions},
  2nd ed. (Oxford Univ. Press, Oxford, 1975)

\bibitem[VG]{VG} S.N.~Vergeles and V.M.~Gryanik:
 {\em Two--dimensional quantum field theories which admit exact 
 solutions},  Sov. J. Nucl. Phys. {\bf 23} (1976) 704--709

\bibitem[V1]{V1}
A.Yu.~Volkov: {\em Quantum Volterra model}, 
 Phys.\ Lett.\ {\bf A167} (1992) {345--355} 

\bibitem[V2]{V2}
A.Yu.~Volkov: {\em Noncommutative hypergeometry}, 
 \CMP{258}{2005}{257--273} 


\bibitem[Wo]{Wo}
S.L.~Woronowicz: {\em Quantum exponential function}, 
Rev.\ Math.\ Phys.\ {\bf 12} (2000) 873--920

\bibitem[Za]{Za} Al.~Zamolodchikov: {\em On the thermodynamic Bethe
Ansatz equation in sinh--Gordon model}, hep-th/0005181

\bibitem[ZZ]{ZZ} 
A.B.~Zamolodchikov and Al.B.~Zamolodchikov: {\em 
Structure constants and conformal bootstrap in Liouville field theory},
 Nucl.\ Phys.\ {\bf B477} (1996) 577--605
 
\end{thebibliography}
\end{document}